\documentclass[twocolumn,amsmath,amssymb,a4paper,prb,superscriptaddress,floatfix]{revtex4-1}
\usepackage[dvipdfmx]{graphicx}
\usepackage{natbib}
\usepackage{multirow}
\usepackage{amsmath}
\usepackage{bm}
\usepackage{mathrsfs}
\usepackage{url}
\usepackage{color}
\usepackage{ulem}
\begin{document}

\title{Distribution of phonon lifetime in Brillouin zone}

\author{Atsushi Togo}
\email{togo.atsushi@gmail.com}
\affiliation{Center for Elements Strategy Initiative for Structural
Materials, Kyoto University, Sakyo, Kyoto 606-8501, Japan}

\author{Laurent Chaput}
\affiliation{Institut Jean Lamour, UMR CNRS 7198, Universit\'{e} de
Lorraine, Boulevard des Aiguillettes, BP 70239, 54506 Vandoeuvre Les
Nancy Cedex, France}

\author{Isao Tanaka}
\affiliation{Center for Elements Strategy Initiative for Structural
Materials, Kyoto University, Sakyo, Kyoto 606-8501, Japan}
\affiliation{Department of Materials Science and
Engineering, Kyoto University, Sakyo, Kyoto 606-8501, Japan}
\affiliation{Nanostructures Research Laboratory, Japan Fine Ceramics
Center, Atsuta, Nagoya 456-8587, Japan}

\begin{abstract}
Lattice thermal conductivities of zincblende- and wurtzite-type
compounds with 33 combinations of elements are calculated with the
single-mode relaxation-time approximation and linearized phonon
Boltzmann equation from first-principles anharmonic lattice dynamics
calculations. In 9 zincblende-type compounds, distributions of phonon
linewidths (inverse phonon lifetimes) are discussed in detail. The
phonon linewidths vary non-smoothly with respect to wave vector, which
is explained from the imaginary parts of the self energies. It is
presented that detailed combination of phonon-phonon interaction
strength and three phonon selection rule is critically important to
determine phonon lifetime for these compounds. This indicates difficulty
to predict phonon lifetime quantitatively without anharmonic force
constants. However it is shown that joint density of
states weighted by phonon numbers, which is calculated only from
harmonic force constants, can be potentially used for a screening of the
lattice thermal conductivity of materials.
\end{abstract}

\maketitle

\section{Introduction}
At temperatures near or above room temperature, the thermal conductivity
of non-metallic solids is governed by phonon-phonon collisions, which is
an anharmonic phenomenon. Recent progress of computers, numerical
algorithms, and methods of first-principles calculations have enabled us
to predict anharmonic force constants quantitatively, and it has even
become possible to reproduce well the lattice thermal conductivities of
compounds using those anharmonic force
constants.\cite{Laurent-LBTE-2013,Broido-kappa-2007,Ward-kappa-2008,
Ward-kappa-2009,Ward-kappa-2010,Esfarjani-kappa-2011,Shiomi-kappa-2011,
Tian-kappa-2012,Shiga-kappa-2012,Lindsay-kappa-2012,Lindsay-kappa-2013-1,
Lindsay-kappa-2013-2,Lindsay-kappa-2013-3} However a detailed analysis
of anharmonicity which is at work to produce the lattice thermal
resistivity is still not well investigated.

In this work we performed systematic lattice thermal conductivity
calculations for a class of compounds with the zincblende- and
wurtzite-type structures for 33 different combinations of chemical
elements. These two crystal structures are equivalent in local
coordinations and different in their stacking orders. This is similar to
the difference between the face-centered cubic and hexagonal close-packed
structures.

\begin{figure}[ht]
 \begin{center}
  \includegraphics[width=0.90\linewidth]{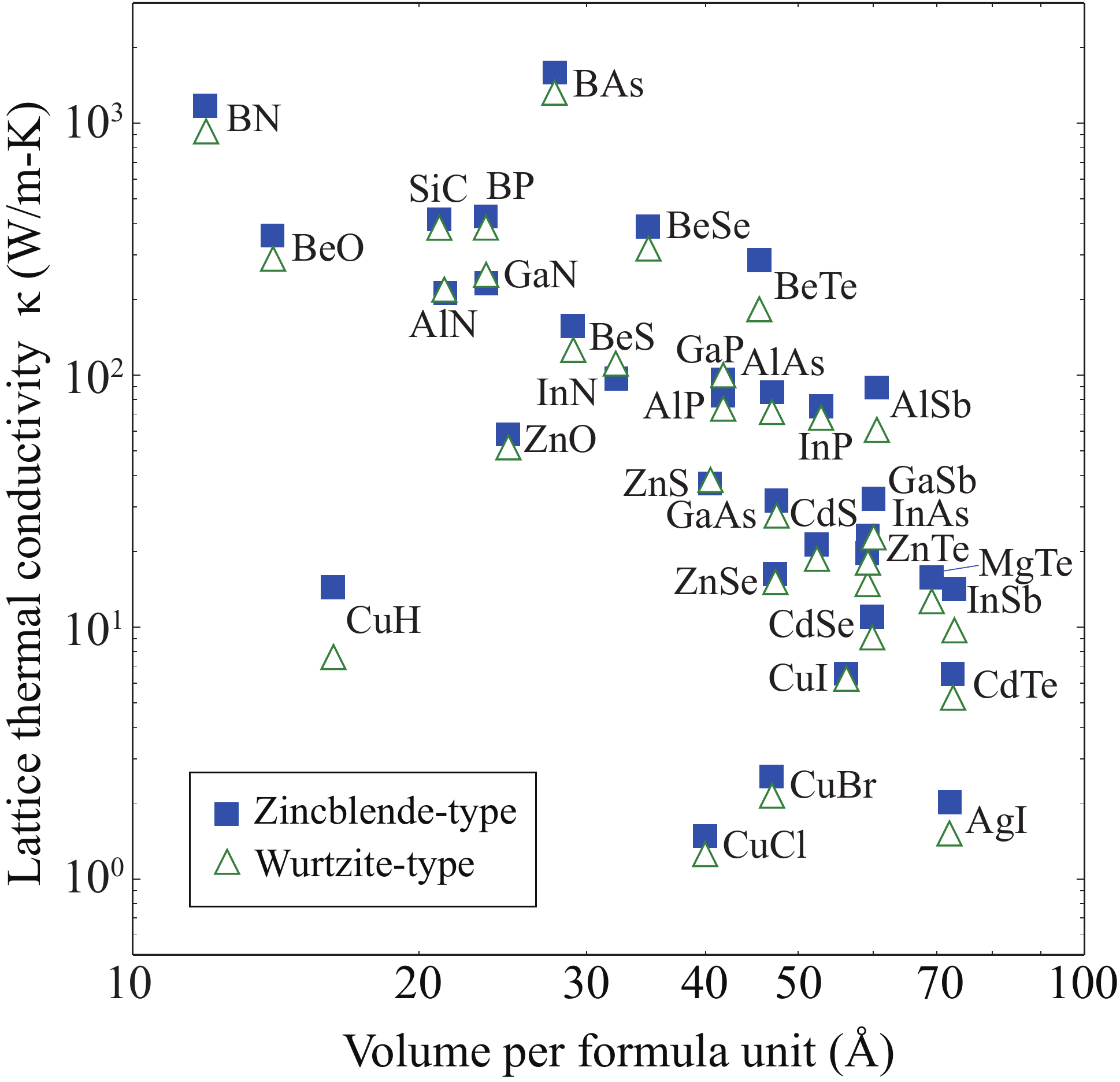}
  \caption{(color online) Lattice thermal conductivities calculated at
  300K with respect to volume per formula unit. \label{fig:vol-kappa} }
 \end{center}
\end{figure}

\begin{figure}[ht]
 \begin{center}
  \includegraphics[width=0.95\linewidth]{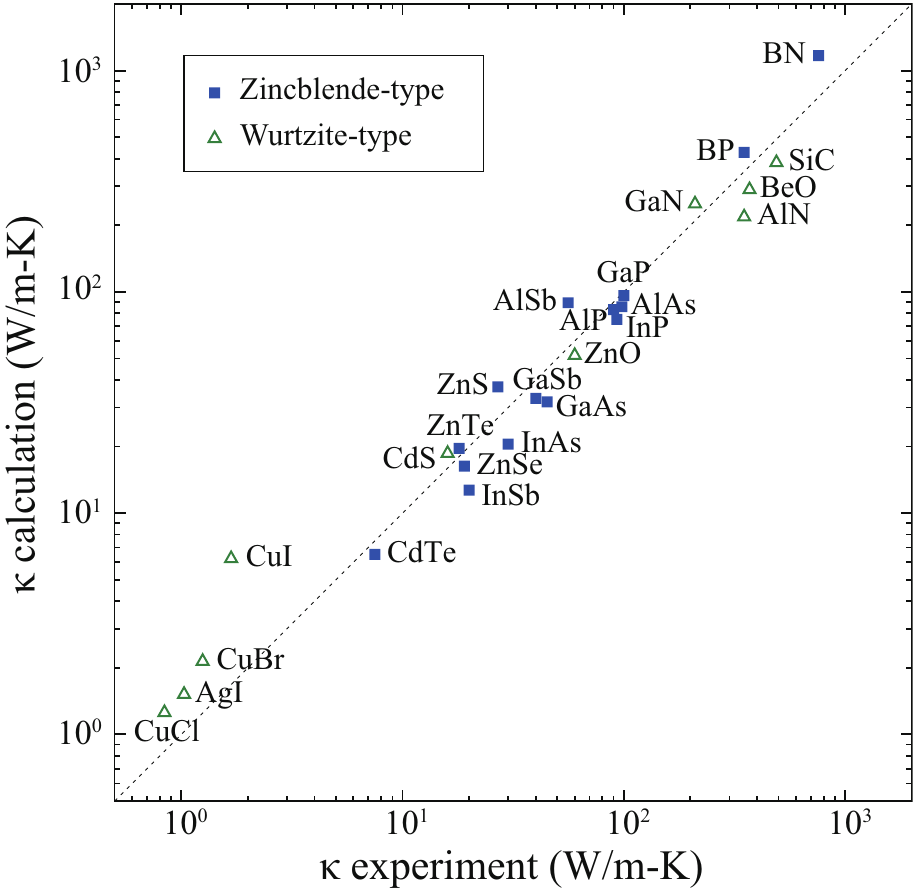}
  \caption{(color online) Comparison in lattice thermal conductivities
  at room temperature between experiments and calculations with RTA and
  witout isotope scattering. \label{fig:kappa-exp-calc} }
 \end{center}
\end{figure}

We mainly focused on the phonon lifetime since we believe that it is the
most critical physical property to seek for a mechanism of lattice
thermal conduction. In fact the lattice thermal conductivity and phonon
lifetimes are related straightforwardly (see
Sec.~\ref{sec:lattice-thermal-conductivity}) in the single-mode
relaxation-time (SMRT) method, being merely proportional to each
others. The results of these calculations are presented in
Fig.~\ref{fig:vol-kappa} and Fig.~\ref{fig:kappa-exp-calc}. 

In Fig.~\ref{fig:vol-kappa} it is found that the differences of the two
crystal structures, i.e., the difference in their stacking orders,
affects little the lattice thermal conductivities.  The same behavior is
also observed for the phonon linewidth distributions and phonon density
of states which are presented in Appendix~\ref{apx:bands}. For the ease
of presentation this allows us to focus our discussions only on the
zincblende-type compounds. Moreover, since the zincblende-type compound
has cubic symmetry, the phonon linewidth distribution in the Brillouin
zone will be considered to be nearly isotropic. Therefore the phonon
linewidth distributions will mostly be investigated along $\Gamma$-L
direction.

Fig.~\ref{fig:kappa-exp-calc} shows that a good agreement is obtained with experiments, and evidence the lifetime as an important, controlling, parameter. The lattice thermal conductivities have also be computed from a full
solution of the linearized phonon Boltzmann equation
(LBTE).\cite{Laurent-LBTE-2013} As presented in
Appendix~\ref{apx:kappa}, the numerical results, with and without
isotope scattering, are not very much different from those presented in
Fig.~\ref{fig:vol-kappa} and  Fig.~\ref{fig:kappa-exp-calc}. This confirms our claim that the anharmonic
lifetime should be understood carefully. Effect of structural disorder,
which can vary phonon frequency, linewidth, and lattice thermal
conductivity,\cite{van-de-Walle-disorder-2002,
Alam-disorder-2004,Alam-disorder-2005,Alam-disorder-kappa-2005} was not
considered in this study since the effect is expected to be small for
most of these compounds.

The paper is organized as follow. In section \ref{sec:phonon-lifetime}
the method of calculation is reviewed, with the most technical aspects
described in Appendix \ref{apx:supercell-approach} to
\ref{apx:LBTE}. The results are discussed in section \ref{sec:results},
which is separated in two parts. In the first part we study the
dynamical aspects of the scattering from the $(\omega, \mathbf{q})$ maps
of the self energy, and in the second part the average value of the
lifetime over the Brillouin zone. The numerical results are collected in
Appendix \ref{apx:lattice-parameters} to \ref{apx:kappa} and compared to
experiments in Table \ref{table:kappa-isotope}.

\section{Method of calculation}

\subsection{Phonon lifetime}
\label{sec:phonon-lifetime} The phonon lifetime that we discuss in this
study is computed from the imaginary part of phonon self energy. The
crystal potential is expanded with respect to atomic displacements. The
coefficients of the third order which contain the anharmonicity are used
to compute the imaginary part of the self energy. For the Hamiltonian
$\mathscr{H}$ the expansion can be written as
\begin{align}
 \label{eq:hamiltonian-expansion}
 \mathscr{H} = \Phi_0 + \mathscr{T} + \mathscr{H}_\mathrm{2} +
 \mathscr{H}_\mathrm{3} + \cdots
 \end{align}
with
\begin{align}
 \mathscr{T} =& \frac{1}{2}\sum_{l\kappa\alpha}m_\kappa
 \left[\dot{u}_\alpha(l\kappa)\right]^2, \\
 \mathscr{H}_\mathrm{2} =& \frac{1}{2} \sum_{l\kappa\alpha}\sum_{l'\kappa'\beta} 
 \Phi_{\alpha\beta}(l\kappa,l'\kappa')u_\alpha(l\kappa)u_\beta(l'\kappa'), \\
 \label{eq:third-order-hamiltonian}
 \mathscr{H}_\mathrm{3} =& \frac{1}{6} \sum_{l\kappa\alpha}\sum_{l'\kappa'\beta}\sum_{l''\kappa''\gamma}
  \Phi_{\alpha\beta\gamma}(l\kappa,l'\kappa',l''\kappa'') 
 \nonumber \\
 & \times u_\alpha(l\kappa)u_\beta(l'\kappa')u_\gamma(l''\kappa''),
\end{align}
where $\Phi_0$, $\mathscr{T}$, and $\mathscr{H}_n$ correspond to the
constant potential, kinetic energy, and $n$-body crystal potential
terms, respectively. $\mathbf{u}(l\kappa)$ is the atomic displacement of
the $\kappa$-th atom in the $l$-th unit cell, $m_\kappa$ is the atomic
mass of type $\kappa$, and $\alpha$, $\beta$, and $\gamma$ are the
Cartesian indices.  $\Phi_{\alpha\beta}$ and $\Phi_{\alpha\beta\gamma}$
denote the harmonic and cubic anharmonic force constants,
respectively. The harmonic Hamiltonian is defined as
$\mathscr{H}_\mathrm{H} = \mathscr{T} + \mathscr{H}_\mathrm{2}$.

An atomic displacement operator may be written as
~\cite{Thermodynamics-of-crystals}
\begin{align}
 \label{eq:displacement-operator}
 & u_\alpha(l\kappa)=  \nonumber \\
 & \left(\frac{\hbar}{2Nm_\kappa}\right)^{\frac{1}{2}}
 \sum_{\mathbf{q}j}\omega_{\mathbf{q}j}^{-\frac{1}{2}}
 \left[\hat{a}_{\mathbf{q}j}+\hat{a}^\dagger_{-\mathbf{q}j}\right]
 e^{i\mathbf{q}\cdot\mathbf{r}(l\kappa)}
 W_{\alpha}(\kappa,\mathbf{q}j) 
\end{align}
where $N$ is the number of unit cells in the crystal and $\hbar$ is the
reduced Planck constant. $\mathbf{r}(l\kappa)$ is the equilibrium atomic
position and $\hat{a}^\dagger_{\mathbf{q}j}$ and $\hat{a}_{\mathbf{q}j}$
are the phonon creation and annihilation operators of the normal mode of
band index $j$ and wave vector $\mathbf{q}$. The harmonic frequency
$\omega_{\mathbf{q}j}$ and polarization vector
$\mathbf{W}(\kappa,\mathbf{q}j)$ are obtained from the eigenvalue
problem of a dynamical matrix $\mathbf{D}(\mathbf{q})$,
\begin{equation}
 \label{eq:eigenvalue-problem}
 \sum_{\kappa'\beta} D_{\alpha\beta}(\kappa\kappa',\mathbf{q})
  W_\beta({\kappa'},\mathbf{q}j) =
 \omega_{\mathbf{q}j}^2 W_\alpha(\kappa,\mathbf{q}j),
\end{equation}
with
\begin{align}
 \label{eq:dynamical-matrix-ti}
 D_{\alpha\beta}(\kappa\kappa', \mathbf{q}) = 
 \frac{1}{\sqrt{m_\kappa m_{\kappa'}}}
 \sum_{l'} \Phi_{\alpha\beta}(0\kappa,l'\kappa')
  e^{i\mathbf{q}\cdot\left[\mathbf{r}(l'\kappa')-\mathbf{r}(0\kappa)\right]}.
\end{align}
Using Eq.~(\ref{eq:displacement-operator}) the hamiltonian
$\mathscr{H}$ is easily expressed using creation and annihilation
operators. We obtain the harmonic part as a sum of harmonic oscillators,
\begin{equation}
 \mathscr{H}_\mathrm{H} = \sum_{\mathbf{q}j} \hbar\omega_{\mathbf{q}j}
  \left(\frac{1}{2}+\hat{a}_{\mathbf{q}j}^\dagger \hat{a}_{\mathbf{q}j}\right),
\end{equation}
and the third-order potential as a sum of 3-phonon collisions
\begin{equation}
 \label{eq:hamiltonian3}
 \mathscr{H}_3 = \sum_{\lambda\lambda'\lambda''}
  \Phi_{\lambda\lambda'\lambda''}
  (\hat{a}_{\lambda} + \hat{a}_{-\lambda}^\dagger) (\hat{a}_{\lambda'} +
  \hat{a}_{-\lambda'}^\dagger) (\hat{a}_{\lambda''} + \hat{a}_{-\lambda''}^\dagger).
\end{equation}
Here the phonon modes $(\mathbf{q},j)$ and $(-\mathbf{q},j)$ have been
abbreviated by $\lambda$ and $-\lambda$, respectively.
$\Phi_{\lambda\lambda'\lambda''}$ represent the strength of interaction
between the three phonon $\lambda,\lambda'$ and $\lambda'$ involved
in the scattering. $\Phi_{\lambda\lambda'\lambda''}$ is explicitly given
by
\begin{widetext}
\begin{align}
 \label{eq:ph-ph-strength}
 \Phi_{\lambda\lambda'\lambda''} &= 
\frac{1}{ \sqrt{N}} \frac{1}{3!} 
 \sum_{\kappa\kappa'\kappa''}\sum_{\alpha\beta\gamma}
W_{\alpha}(\kappa,\lambda)
  W_{\beta} ({\kappa'},\lambda')
  W_{\gamma}({\kappa''},\lambda'')    \sqrt{\frac{\hbar}{2m_{\kappa}  \omega_{\lambda}  }} \sqrt{\frac{\hbar}{2m_{\kappa'}  \omega_{\lambda'}  }}  \sqrt{\frac{\hbar}{2m_{\kappa''}  \omega_{\lambda''}  }}\nonumber \\
  &\times \sum_{l'l''}
  \Phi_{\alpha\beta\gamma}(0\kappa,l'\kappa',l''\kappa'')
  e^{i\mathbf{q}' \cdot
  [\mathbf{r}(l'\kappa')-\mathbf{r}(0\kappa)]} e^{i\mathbf{q}''
  \cdot [\mathbf{r}(l''\kappa'')-\mathbf{r}(0\kappa)]}   
  \times  e^{i(\mathbf{q}+\mathbf{q}'+\mathbf{q}'')\cdot
  \mathbf{r}(0\kappa)} \Delta(\mathbf{q}+\mathbf{q}'+\mathbf{q}'')   
\end{align}
\end{widetext}
 where $\Delta(\mathbf{q}+\mathbf{q}'+\mathbf{q}'') = 1$ when
$\mathbf{q}+\mathbf{q}'+\mathbf{q}''$ is a reciprocal lattice vector,
and is zero otherwise. In Eqs.~(\ref{eq:dynamical-matrix-ti}) and
~(\ref{eq:ph-ph-strength}) we have used the translational invariance
condition to simplify the more symmetric equations which appears in
Ref.~\onlinecite{Thermodynamics-of-crystals}.

The imaginary part of the self energy $\Gamma_\lambda(\omega)$ can now
be computed up to second order in $ \mathscr{H}_\mathrm{3}$ using many
body perturbation theory. It takes a form analogous to the Fermi's golden
rule,\cite{Maradudin-1962}
\begin{widetext}
\begin{align}
 \label{eq:imag-selfenergy}
 \Gamma_\lambda(\omega) = \frac{18\pi}{\hbar^2}
  \sum_{\lambda' \lambda''}
  \bigl|\Phi_{-\lambda\lambda'\lambda''}\bigl|^2 
  \left\{(n_{\lambda'}+ n_{\lambda''}+1) 
   \delta(\omega-\omega_{\lambda'}-\omega_{\lambda''}) \right.
   + (n_{\lambda'}-n_{\lambda''})
  \left[\delta(\omega+\omega_{\lambda'}-\omega_{\lambda''})
 - \left. \delta(\omega-\omega_{\lambda'}+\omega_{\lambda''})
 \right]\right\},
\end{align}
\end{widetext}
where $n_{\lambda}$ is the phonon occupation number at the equilibrium,
\begin{equation}
n_{\lambda} = \frac{1}{\exp(\hbar\omega_\lambda/\mathrm{k_B}T)-1}.
\end{equation}
The double Brillouin zone summation in Eq.~(\ref{eq:imag-selfenergy}) is
reduced to a single summation due to
$\Delta(-\mathbf{q}+\mathbf{q}'+\mathbf{q}'')$ in the
calculation. 

$2\Gamma_\lambda(\omega_\lambda)$ corresponds to the phonon linewidth of
the phonon mode $\lambda$ and its reciprocal is known as the phonon
lifetime,\cite{Maradudin-1962}
\begin{equation}
 \label{eq:phonon-lifetime}
\tau_\lambda=\frac{1}{2\Gamma_\lambda(\omega_\lambda)}.
\end{equation}

\subsection{Lattice thermal conductivity}
\label{sec:lattice-thermal-conductivity}

When the LBTE is solved under the SMRT method, the lattice thermal
conductivity tensor can be written in a closed
form,\cite{Physics-of-phonons}
\begin{align}
 \label{eq:SMRT-kappa}
  \kappa = \frac{1}{NV_0} \sum_\lambda C_\lambda \mathbf{v}_\lambda \otimes
  \mathbf{v}_\lambda \tau_\lambda^{\mathrm{SMRT}},
\end{align}
where $V_0$ is the volume of a unit cell, $\mathbf{v}_\lambda$
and $\tau_\lambda^{\mathrm{SMRT}}$ are the group velocity and
SMRT of the phonon mode $\lambda$,
respectively. $C_\lambda$ is the mode dependent heat capacity defined as
\begin{equation}
C_\lambda = k_\mathrm{B}
 \left(\frac{\hbar\omega_\lambda}{k_\mathrm{B} T} \right)^2
 \frac{\exp(\hbar\omega_\lambda/k_\mathrm{B}
 T)}{[\exp(\hbar\omega_\lambda/k_\mathrm{B} T)-1]^2}.
\end{equation}
 The group velocity can be obtained directly from the eigenvalue
 equation~(\ref{eq:eigenvalue-problem}),
\begin{align}
 \label{eq:group-velocity}
 v_\alpha(\lambda) &\equiv \frac{\partial \omega_\lambda}{\partial q_{\alpha}} \\
  % = \frac{1}{2\omega_\lambda}\frac{\partial \omega_\lambda^2}{\partial \mathbf{q}}
  &= \frac{1}{2\omega_\lambda} \sum_{\kappa\kappa'\beta\gamma}
  W_\beta(\kappa,\lambda)\frac{\partial
  D_{\beta\gamma}(\kappa\kappa',\mathbf{q})}{\partial
  q_\alpha}W_\gamma(\kappa',\lambda).
\end{align}
In this study it is further assumed that the phonon relaxation time is
given by the phonon lifetime of Eq.~(\ref{eq:phonon-lifetime}),
\begin{equation}
\tau_\lambda^{\mathrm{SMRT}} \equiv \tau_\lambda.
\end{equation}
This is an approximation. Therefore we have also computed the lattice
thermal conductivity from a full solution of the LBTE using the method
of Ref.~\onlinecite{Laurent-LBTE-2013} which is summarized in Appendix
\ref{apx:LBTE}. Those calculations are much more expensive than the
closed expression~(\ref{eq:SMRT-kappa}). The results presented in the
Appendix \ref{apx:kappa} show however that they give thermal
conductivities which are numerically quite similar to those with the
SMRT approximation for the class of compounds we consider.

\subsection{Computational details}
\label{computational-details}

Second- and third-order force constants were obtained by the supercell
approach with finite atomic displacements of 0.03 \AA. The details of
the method are written in Appendix~\ref{apx:supercell-approach} and in
Ref.~\onlinecite{Laurent-phph-2011}. Non-analytical term
correction~\cite{Pick-1970} was applied to the second-order force
constants to take into account the long range Coulomb forces present in
ionic crystals. The technical details to incorporate the non-analytical
correction are presented in
Appendix~\ref{apx:non-analytical-term-correction} and in
Ref.~\onlinecite{Wang-2010}.

For the first-principles calculations, we employed the plane-wave basis
projector augmented wave method~\cite{PAW-Blochl-1994} within the
framework of density functional theory as implemented in the VASP
code.\cite{VASP-Kresse-1995,VASP-Kresse-1996,VASP-Kresse-1999} The
generalized gradient approximation (GGA) of Perdew, Burke, and
Ernzerhof~\cite{Perdew-PBE-1996} was used for the exchange correlation
potential. A plane-wave energy cutoff of 500 eV was employed.
Reciprocal spaces of conventional unit cells of the zincblende-type
structures and unit cells of the wurtzite-type structures were sampled
by $4\times 4\times 4$ and $6\times 6\times 4$ meshes, respectively.
The sampling meshes were shifted from $\Gamma$-point by $(1/2, 1/2,
1/2)$ and $(0, 0, 1/2)$ of neighboring mesh grid points along axes,
respectively. To obtain atomic forces, the total energies were minimized
until the energy convergences to be less than $10^{-8}$ eV. The lattice
parameters of the compounds were optimized under the calculations at
zero pressure condition. The results are summarized in
Appendix~\ref{apx:lattice-parameters}.

The second-order force constants were computed using $4\times 4\times 4$
supercells of the conventional unit cell (512 atoms) for the
zincblende-structures and $5\times 5\times 3$ supercells for the
wurtzite-type structures (300 atoms). Reciprocal spaces of the
supercells of the zincblende- and wurtzite-type structures were sampled
by $1\times 1\times 1$ of $\Gamma$-centered meshes and $1\times 1\times
2$ meshes shifted by $(0, 0, 1/2)$ with respect to the grid spacing,
respectively. For zincblende-type GaSb, InN, InAs, and InSb, the
sampling meshes were shifted by $(1/2, 1/2, 1/2)$ to avoid sampling
their $\Gamma$-points since their valence and conduction bands slightly
touch around the $\Gamma$-points due to the underestimation of band gaps
by the GGA.
Static dielectric constant tensors and Born effective charge tensors
were obtained from density functional perturbation theory as implemented
in the VASP code.\cite{Gajdos-2006,Wu-2005} These tensors were
symmetrized and a sum rule was applied to the Born effective charge
tensors following Ref.~\onlinecite{Gonze-1997}. The phonon band
structures and densities of states obtained from those harmonic
calculation are presented in Appendix~\ref{apx:bands} for all the
compounds considered in this study.

For the third-order force constants, $2\times2\times 2$ supercells were
built for the zincblende-structures (64 atoms) and $3\times 3\times 2$
for the wurtzite-type structures (72 atoms). Reciprocal spaces of the
supercells of the zincblende- and wurtzite-type structures were sampled
by $2\times 2\times 2$ meshes shifted by $(1/2, 1/2, 1/2)$ and $(0, 0,
1/2)$ with respect to the grid spacing, respectively. These anharmonic
calculations allow to obtain the lifetimes according to
Eq.~(\ref{eq:imag-selfenergy}).  The lifetimes for the phonon states of
the band structures are shown in Appendix \ref{apx:bands} for all the
compounds considered in this study. To obtain those results Brillouin zones were sampled using
the $96\times 96\times 96$ and $80\times 80\times 48$ sampling meshes
for the reciprocal spaces of the primitive cells of the zincblende- and
wurtzite-type structures, respectively.

Finally, to compute lattice thermal conductivities, the reciprocal
spaces of the primitive cells of the zincblende- and wurtzite-type
structures were sampled using the $19\times 19\times 19$ and $19\times
19\times 15$ meshes, respectively. In Brillouin zone integrations,
crystal symmetry was used to improve numerical accuracy and to reduce
the computational cost. A tetrahedron method was employed in the
calculation of the imaginary parts of the self
energies.\cite{MacDonald-tetrahedron-1979,Blochl-tetrahedron-1994,Tadano-2014}
The technical details on the Brillouin zone integration are summarized
in Appendix~\ref{apx:brillouin-zone-integration}. The computed lattice
thermal conductivities are presented in Appendix \ref{apx:kappa}.

\section{Results and discussion \label{sec:results}}
\begin{figure}[ht]
 \begin{center}
  \includegraphics[width=0.60\linewidth]{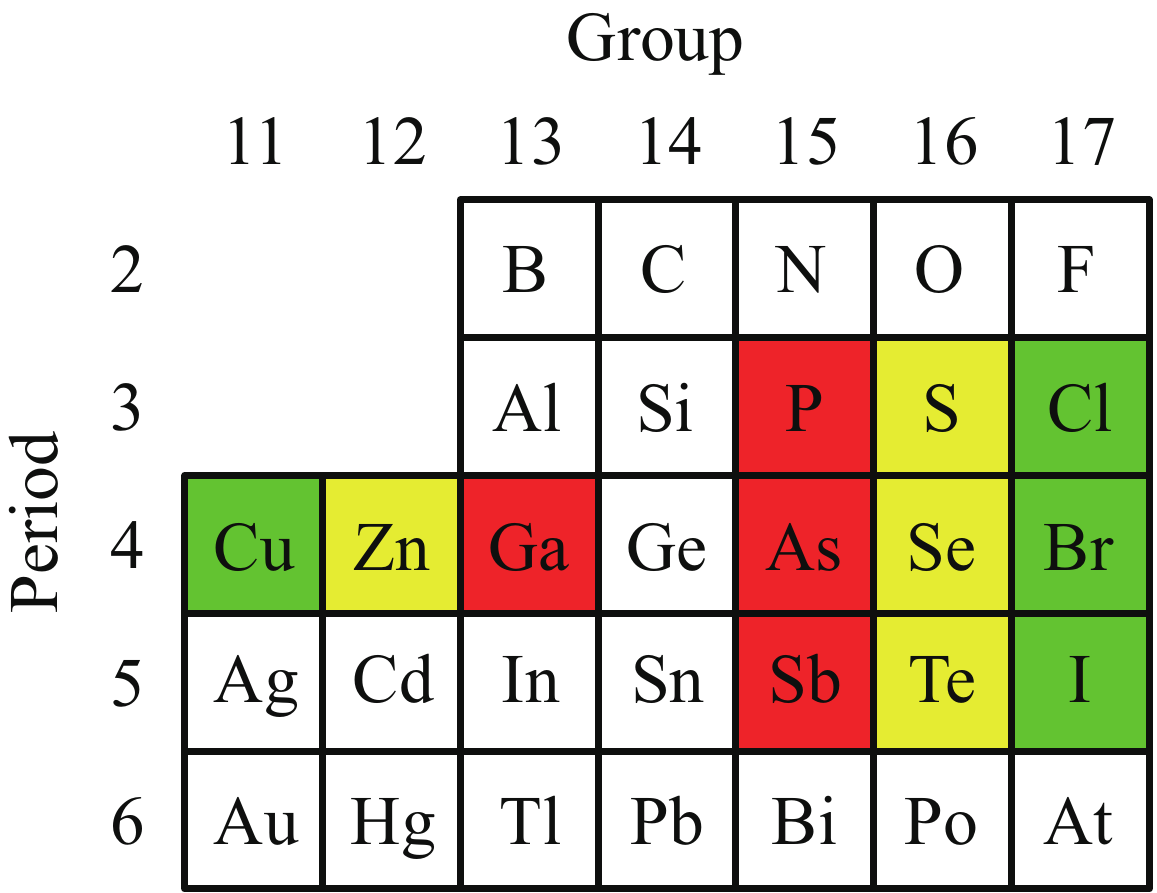}
  \caption{(color online) Positions of elements in Cu$X$, Zn$X$, and
  Ga$X$ on the periodic table. \label{fig:periodic-table}}
 \end{center}
\end{figure}

In this section, we attempt to capture the features how phonons interact
in Brillouin zone schematically by limiting our discussion to selected
compounds. In Fig.~\ref{fig:vol-kappa}, we see a trend that the
compounds with larger unit-cell volumes have smaller lattice thermal
conductivities. However the copper compounds, CuCl, CuBr, CuI (Cu$X$),
clearly ignore this trend. Therefore, to understand this behavior, as
shown in Fig.~\ref{fig:periodic-table}, we focus on Cu$X$ and their
neighbors in the periodic table, ZnS, ZnSe, ZnTe (Zn$X$) and GaP, GaAs,
GaAs (Ga$X$), and compare them to each others to find the
characteristics they share. Figures similar to those found in this
section are presented in Appendix \ref{apx:bands} and \ref{apx:self} for
all the zincblende-type compounds considered in this study.

\begin{figure*}[ht]
 \begin{center}
  \includegraphics[width=0.70\linewidth]{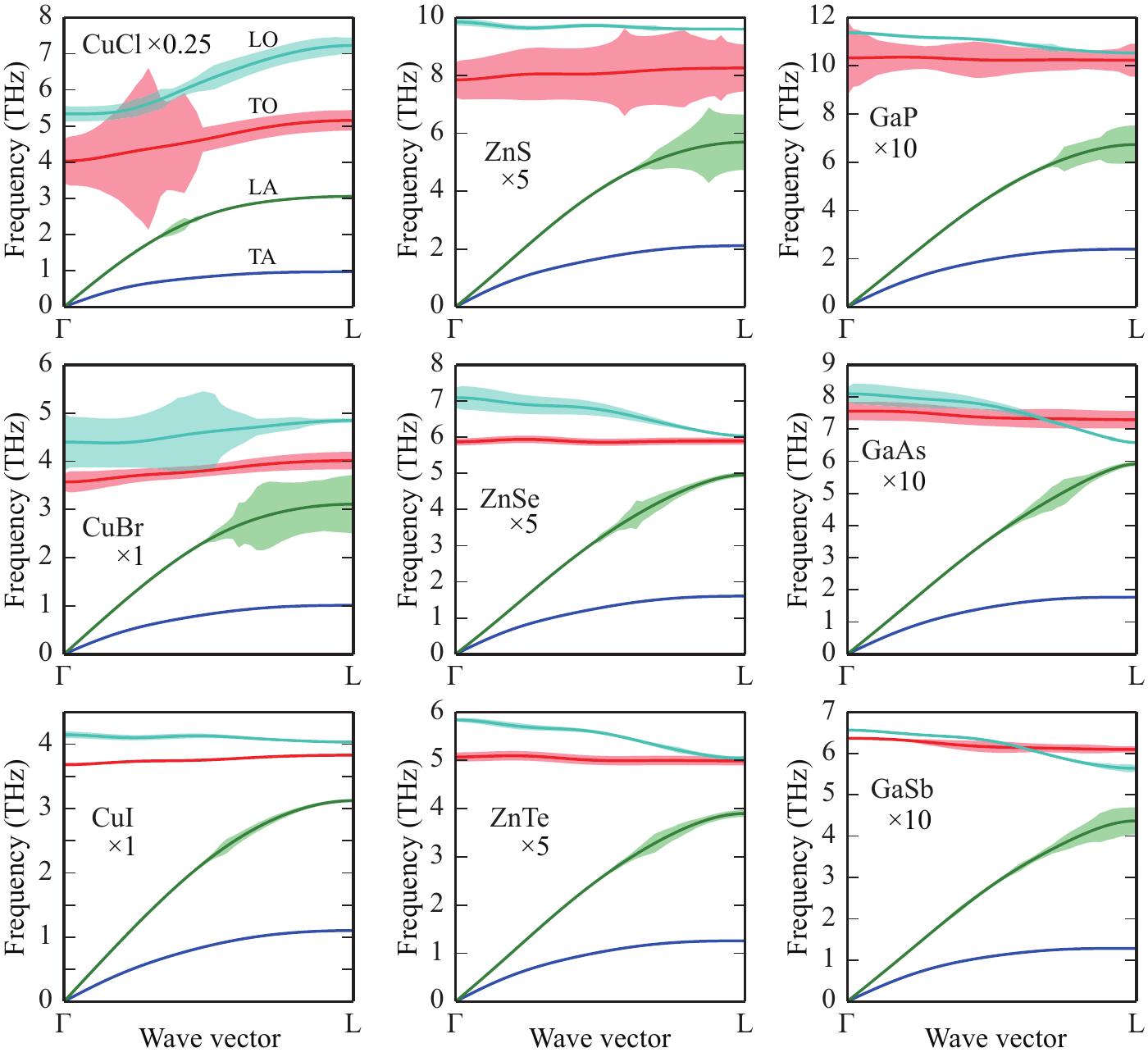}
  \caption{(color online) \label{fig:band-CuZnGaX} Phonon linewidth
  distributions of Cu$X$, Zn$X$, and Ga$X$ of the zincblende-type
  structure which are drawn on the phonon band structures between the
  $\Gamma$ and L points. The scales of the phonon linewidths and phonon
  frequencies are shared in THz, but to make the phonon linewidths
  visible, their magnitudes are adjusted by scale factors ($\times
  0.25$, $\times 1$, $\times 5$, and $\times 10$) that are shown close
  to their compound names. Four phonon branches correspond to TA (blue),
  LA (green), TO (red), and LO (pale blue) modes in order of increasing
  phonon frequency near $\Gamma$-point, respectively.}
 \end{center}
\end{figure*}

The phonon band structures between $\Gamma$ and L points for Cu$X$,
Zn$X$, and Ga$X$ are shown in Fig.~\ref{fig:band-CuZnGaX} from the left
column to the right column, respectively. The panels are aligned in
relation to the periodic table in Fig.~\ref{fig:periodic-table}. The
compounds at the lower panels have heavier anion atoms, therefore as it
is expected from Eqs.~(\ref{eq:eigenvalue-problem}) and
(\ref{eq:dynamical-matrix-ti}), phonon frequencies become systematically
lower in the lower panels. Looking at the panels from left to right,
phonon frequencies increase gradually. Therefore, except for CuCl and
CuBr, the compounds show rather similar band structures, at different
scales. This means that in Eq.~(\ref{eq:imag-selfenergy}) the three
phonon collisions and decay processes allowed by the conservation laws
for energy and momentum will be about the same. Then, if the
phonon-phonon interaction strength can be approximated by a function
with a simple form, we can expect to analyze those compounds
systematically using a small number of parameters.

CuCl and CuBr reveal an interesting behavior. For all compounds there
are four phonon branches between $\Gamma$ and L points. They are labeled
as TA (transverse acoustic), LA (longitudinal acoustic), TO (transverse
optical), and LO (longitudinal optical) branches in ascending order
along vertical axis near $\Gamma$ point. Usually, acoustic branches have
larger dispersion and group velocity than optical branches and the
phonon mode frequency of an optical branch is weakly dependent on wave
vector. However CuCl shows relatively large dispersions in the optical
branches.
In non-metallic crystals, atomic displacements of LO-like phonon modes
near $\Gamma$ point induce a macroscopic electric field. As a result, the
phonon mode frequency of an LO branch is lifted up and LO and TO
branches split near $\Gamma$ point. Therefore phonon mode frequency of
an LO branch near $\Gamma$ is higher than that near Brillouin zone
boundary in the most compounds. Clear LO-TO splitting is observed in
CuCl and CuBr, however the slopes of the LO branches are inverted. 

Inverse phonon lifetimes, or linewidths, given in
Eq.~(\ref{eq:phonon-lifetime}) are drawn on the phonon band structures
as the vertical widths of the shaded areas in
Fig.~\ref{fig:band-CuZnGaX}. Larger linewidth indicates larger phonon
scattering and shorter phonon lifetime. The physical unit of THz is
shared with phonon mode frequency. To well visualize distributions of
the linewidths, the scales of the linewidths are adjusted with
multiplying factors. It has to be noticed that phonon lifetimes can be
different in logarithmic scale among compounds, which is presumed from
the fact that, as shown Fig.~\ref{fig:vol-kappa}, lattice thermal
conductivities also range in logarithmic scale. It is found that
linewidths vary non-smoothly with respect to wave vector. Partly, this
come from the conservation laws for energy and momentum which select
regions where more scattering mechanism are possible. As mentioned
before, this would also explain the similarities in lifetime
distribution for compounds with similar band structures. This will be
described in more details.

\subsection{$(\mathbf{q}, \omega)$ map of self energy}

To find common features, linewidth distributions in
Fig.~\ref{fig:band-CuZnGaX} are observed: linewidths of the TA branches
are small compared with the other branches, knot-like distributions are
found on the LA branches, and, except for CuCl, the TO branch linewidths
are roughly constant. To understand how this variety of the linewidth
distributions is obtained, we investigated the frequency and wave vector
dependencies of the imaginary parts of the self energies. The result of
these calculations for all the zincblende-type compounds considered in
this study are collected in Appendix \ref{apx:self}.

\begin{figure*}[ht]
 \begin{center}
  \includegraphics[width=0.7\linewidth]{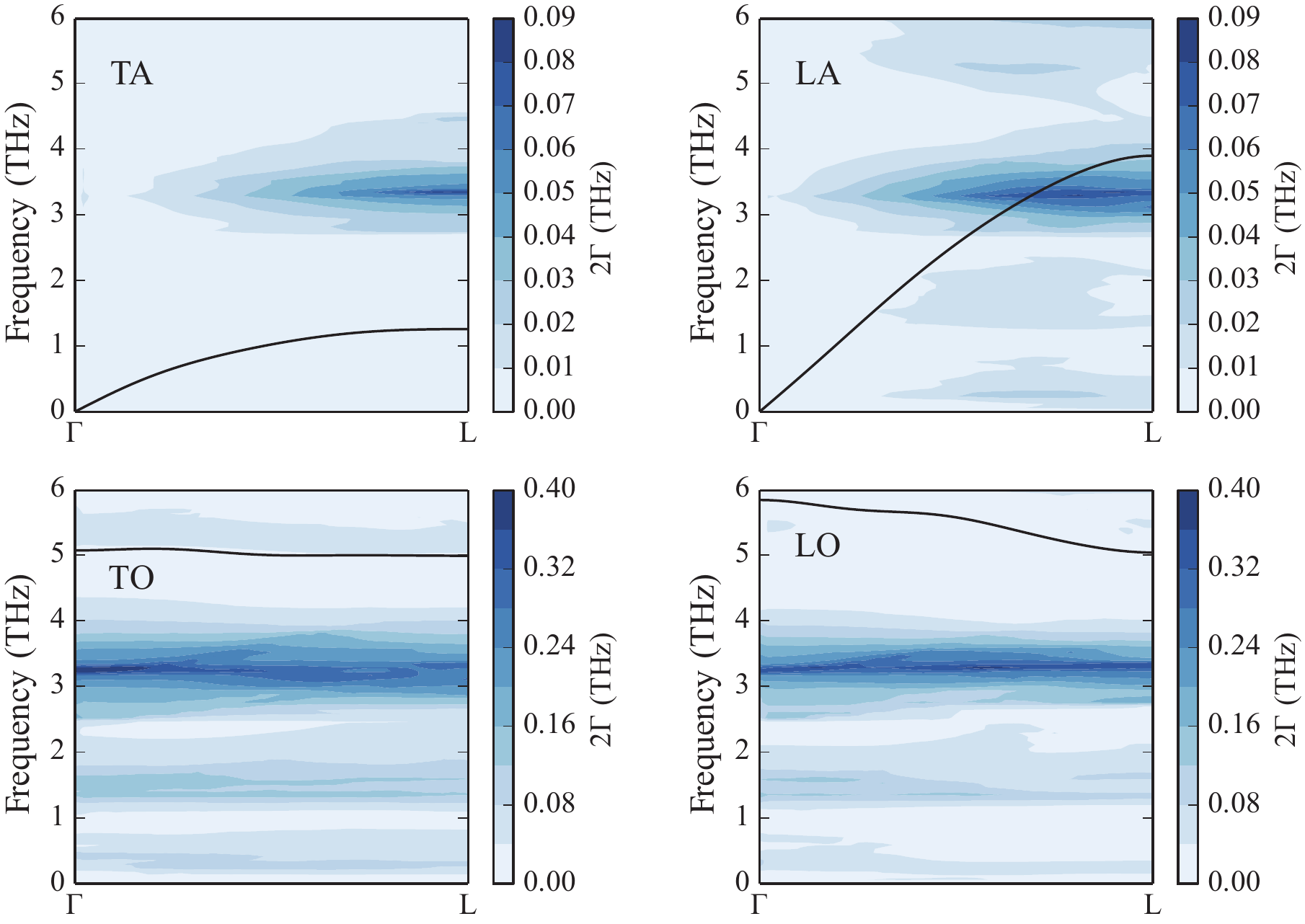}
  \caption{(color online) \label{fig:imag-self-energy-ZnTe} Imaginary
  parts of self energies of ZnTe as functions of wave vector between
  $\Gamma$ and L points and frequency at 300 K for TA, LA, TO, and LO
  branches. Phonon dispersion curves of the corresponding branches are
  superimposed.}
 \end{center}
\end{figure*}
Taking ZnTe as an example, we draw $(\mathbf{q}, \omega)$ maps of the
imaginary parts of the self energies for the TA, LA, TO, and LO
branches.  As shown in Fig.~\ref{fig:imag-self-energy-ZnTe}, they are
drawn between $\Gamma$ and L points for $\omega < 6$ THz. The four
branches of phonon band structure are superimposed on the corresponding
$(\mathbf{q}, \omega)$ maps as solid curves. The linewidth distribution
of ZnTe in Fig.~\ref{fig:band-CuZnGaX} is reproduced by collecting
$2\Gamma_\lambda(\omega)$ values on the solid curves where
$\omega=\omega_\lambda$, i.e. $2\Gamma_\lambda(\omega_\lambda)$. As a
general tendency, it is found that the imaginary part of the self energy
is weakly dependent on wave vector, however it varies vividly along
frequency.

\begin{figure}[ht]
 \begin{center}
  \includegraphics[width=1.0\linewidth]{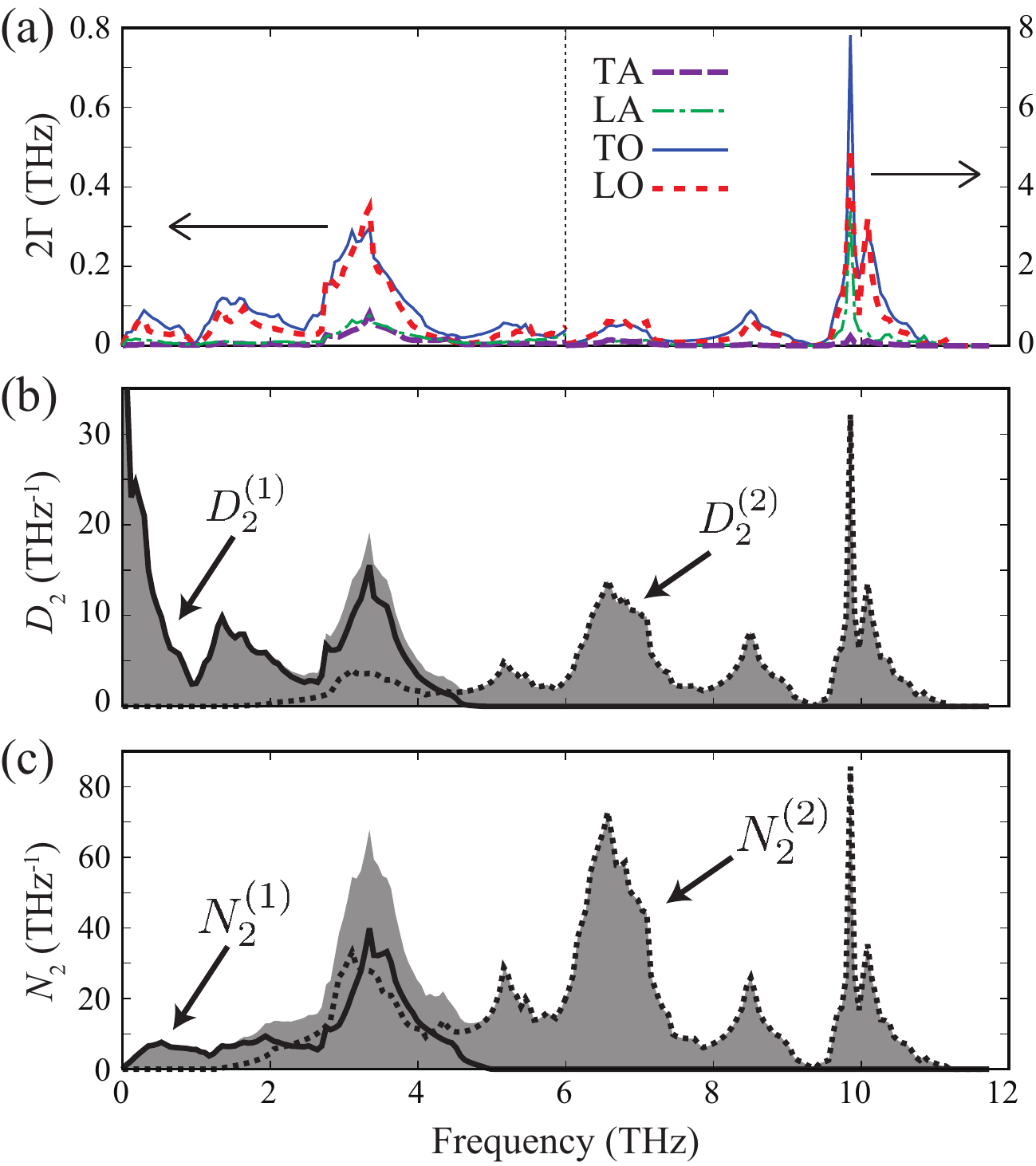}
  \caption{(color online) \label{fig:gamma-jdos-ZnTe-spectra} Imaginary
  parts of self energies ($2\Gamma$), JDOS ($D_2$), and w-JDOS ($N_2$)
  are drawn for ZnTe as a function of phonon frequency at L point.  (a)
  Imaginary parts of self energies 2$\Gamma$ of TA, LA, TO, and LO
  branches at 300 K are depicted by dashed, dashed-dotted, solid, and
  dotted curves, respectively. Note that the scale for the vertical axis
  is changed below and above 6 THz.  (b) JDOS $D_2^{(1)}$ and
  $D_2^{(2)}$ are drawn by the solid and dotted curves, respectively,
  and the sum $D_2 = D_2^{(1)} + D_2^{(1)}$ is depicted as the shaded
  region.  (c) w-JDOS $N_2^{(1)}$ and $N_2^{(2)}$ at 300 K are presented
  by the solid and dotted curves, respectively, and the sum $N_2 =
  N_2^{(1)} + N_2^{(1)}$ is depicted as the shaded region.}
 \end{center}
\end{figure}

To see their frequency dependency in detail, the profiles of the
$(\mathbf{q}, \omega)$ maps at L point are plotted in
Fig.~\ref{fig:gamma-jdos-ZnTe-spectra} (a). They are drawn up to $\omega
\sim 12$ THz since $2\Gamma_\lambda(\omega)$ can be nonzero up to twice
the maximum phonon mode frequency due to the term
$\delta(\omega-\omega_{\lambda'}-\omega_{\lambda''})$ in
Eq.~(\ref{eq:imag-selfenergy}). Note that 10 times different scales are
used to plot $2\Gamma_\lambda(\omega)$ below and above 6 THz in
Fig.~\ref{fig:gamma-jdos-ZnTe-spectra} (a). The profiles in the lower
frequency range are much smaller than those in the high frequency range.
For each profile, the strongest peak appears at $\omega \sim 10$ THz
except for the TA branch. The peaks at $\omega \sim 3.5$ THz are
relatively large in the frequency range where the phonon modes exist,
$\omega < 6$ THz. The phonon band structure of the LA branch crosses
this peak frequency as can be seen clearly in
Fig.~\ref{fig:imag-self-energy-ZnTe}. This causes the knot-like shape of
the LA branch linewidth.

These peaks of the profiles are mainly brought about by the three phonon
selection rules. In Eq.~(\ref{eq:imag-selfenergy}), the energy
conservation among three phonons is represented by the $\delta$
functions and the conservation of momentum appears in
Eq.~(\ref{eq:ph-ph-strength}) with the $\Delta$ function. To understand
the role of the selection rules to imaginary part of self energy,
the joint density of states (JDOS) is employed. This function is
temperature independent. There are two classes of three phonon
scattering events. The collision processes, which correspond to that
$\delta(\omega-\omega_{\lambda_1} + \omega_{\lambda_2})$ or
$\delta(\omega + \omega_{\lambda_1} - \omega_{\lambda_2})$ in
Eq.~(\ref{eq:imag-selfenergy}), and the decay processes which correspond
to $\delta(\omega-\omega_{\lambda_1}-\omega_{\lambda_2})$.  In
Ref.~\onlinecite{Physics-of-phonons} they are respectively named class 1
and class 2 events, and we will keep this terminology in the following
to label different quantities.  For example the JDOS $D_2(\mathbf{q},
\omega)$ is defined by the sum of the JDOS of the two classes
$D_2^{(1)}(\mathbf{q}, \omega)$ and $D_2^{(2)}(\mathbf{q}, \omega)$ as
\begin{equation}
 \label{eq:D2}
  D_2(\mathbf{q}, \omega) = D_2^{(1)}(\mathbf{q}, \omega)  +
  D_2^{(2)}(\mathbf{q}, \omega)
\end{equation}
where
\begin{align*}
D_2^{(1)} (\mathbf{q}, \omega) 
&= \frac{1}{N} \sum_{\lambda'\lambda''}
\Delta(-\mathbf{q}+\mathbf{q}'+\mathbf{q}'') \\ \nonumber
& \times \left[
\delta(
 \omega + \omega_{\lambda'} - \omega_{\lambda''}) + 
\delta(
 \omega - \omega_{\lambda'} + \omega_{\lambda''})\right],
\end{align*}
for class 1 events and
\begin{align*}
 D_2^{(2)} (\mathbf{q}, \omega) 
& =  \frac{1}{N}\sum_{\lambda'\lambda''}
\Delta(-\mathbf{q}+\mathbf{q}'+\mathbf{q}'') \\ \nonumber
& \times \delta(
\omega - \omega_{\lambda'} - \omega_{\lambda''}),
\end{align*}
for class 2 events. $D_2(\mathbf{q}, \omega)$, $D_2^{(1)}(\mathbf{q},
\omega)$, and $D_2^{(2)}(\mathbf{q}, \omega)$ are shown in
Fig.~\ref{fig:gamma-jdos-ZnTe-spectra} (b) for point L. The peak
positions of the profiles coincide with those of the imaginary parts of
the self energies.  Moreover $D_2^{(1)}(\mathbf{q}, \omega)$ and
$D_2^{(2)}(\mathbf{q}, \omega)$ cover the low and high frequency ranges,
respectively. Therefore it is expected that the class 1 (2) event tends
to contribute to the phonon lifetimes of the acoustic (optical)
branches.

Comparing Figs.~\ref{fig:gamma-jdos-ZnTe-spectra} (a) and (b), relative
intensities of the profiles are clearly different since phonon-phonon
interaction strength and phonon occupation numbers are ignored in the
JDOS. If we assume that the scattering of the phonon $\lambda$ with
$\lambda'$ and $\lambda''$ can be averaged over $\lambda'$ and
$\lambda''$ then the imaginary part of self energy can be approximated
as
\begin{align}
 \label{eq:imag-selfenergy-ave}
 \widetilde{\Gamma}_{\mathbf{q} j}(\omega)& = \frac{18\pi}{\hbar^2} P_{-\mathbf{q} j}  N_2(\mathbf{q}, \omega)
\end{align}
with 
\begin{align}
 \label{eq:defPeierls}
&P_{\mathbf{q} j}= \frac{1}{(3n_a)^2} \sum_{\lambda' \lambda''} \bigl|\Phi_{\lambda \lambda' \lambda''}\bigl|^2, \\
&N_2(\mathbf{q}, \omega) = N_2^{(1)}(\mathbf{q}, \omega) + N_2^{(2)}(\mathbf{q}, \omega) 
\end{align}
and
\begin{widetext}
\begin{align}
&N_2^{(1)}(\mathbf{q}, \omega) =  \frac{1}{N} \sum_{\lambda'\lambda''} \Delta(-\mathbf{q}+\mathbf{q}'+\mathbf{q}'') (n_{\lambda'} - n_{\lambda''}) [ \delta( \omega + \omega_{\lambda'} - \omega_{\lambda''}) - \delta(  \omega - \omega_{\lambda'} + \omega_{\lambda''})] \\
&N_2^{(2)}(\mathbf{q}, \omega) =  \frac{1}{N} \sum_{\lambda'\lambda''}  \Delta(-\mathbf{q}+\mathbf{q}'+\mathbf{q}'') (n_{\lambda'}+ n_{\lambda''}+1) \delta( \omega - \omega_{\lambda'} - \omega_{\lambda''}),
\end{align}
\end{widetext}
where $n_\mathrm{a}$ in Eq.~(\ref{eq:defPeierls}) is the number of atoms in the primitive cell.

The computational cost of $\widetilde{\Gamma}_\lambda(\omega)$ may be
much smaller than that of $\Gamma_\lambda(\omega)$ if we can avoid the
computation of third-order force constants using a good approximation.
In Eq.~(\ref{eq:imag-selfenergy-ave}), $P_{\mathbf{q} j}$ represents the
averaged interaction experienced by phonon $\mathbf{q} j$ while
propagating in the medium. $N_2(\mathbf{q}, \omega)$ is a weighted JDOS
(w-JDOS) which measures the number of those interactions,
$N_2^{(1)}(\mathbf{q}, \omega)$ of them being of class 1, and
$N_2^{(2)}(\mathbf{q}, \omega)$ being of class 2.

In Fig.~\ref{fig:gamma-jdos-ZnTe-spectra} (c), $N_2(\mathbf{q}, \omega)$
at L point is presented at 300 K. Below $\omega \sim 2$ THz, the
divergence behavior in $D_2^{(1)}(\mathbf{q}, \omega)$ has been canceled
by the occupation functions in $N_2^{(1)}(\mathbf{q}, \omega)$ but the
intensity of the peak at $\omega \sim 10$ THz is still comparable to
those of the other peaks, which is a noticeable difference with
$2\Gamma_\lambda(\omega)$. However below $\omega \sim 6$ THz, where the
phonon modes do exist, $N_2(\mathbf{q}, \omega)$ reasonably reproduces
the shapes of the profiles of $2\Gamma_\lambda(\omega)$.

\begin{figure}[ht]
 \begin{center}
  \includegraphics[width=0.7\linewidth]{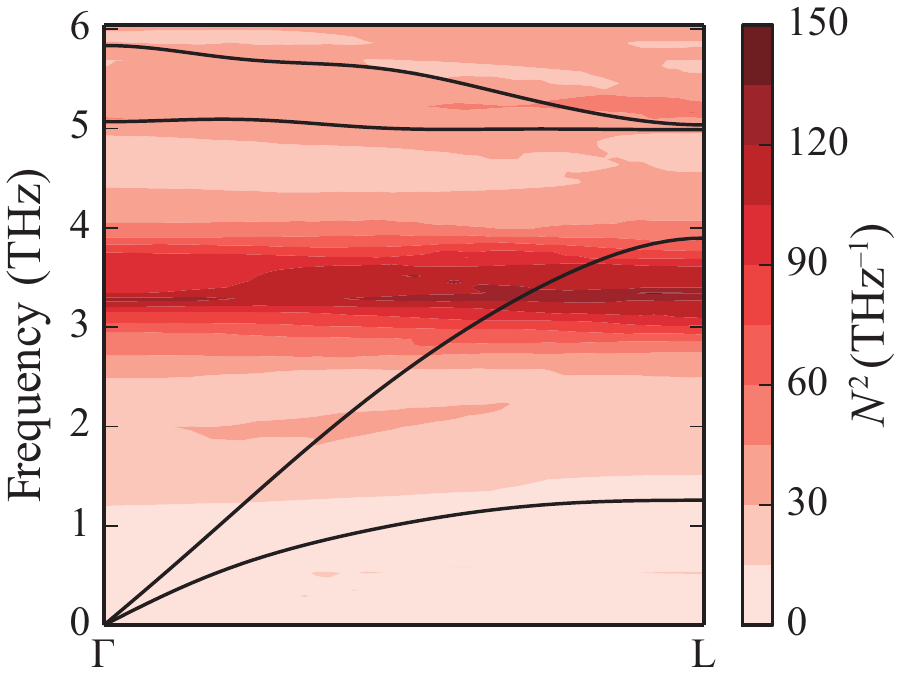} \caption{(color
  online) \label{fig:N2-ZnTe} w-JDOS ($N_2$) of ZnTe as functions of
  wave vector between $\Gamma$-L and phonon frequency at 300 K. The
  solid curves show phonon mode frequencies of the TA, LA, TO, and LO
  branches in ascending order along the vertical axis.}
 \end{center}
\end{figure}

One consequence is that $N_2(\mathbf{q}, \omega)$ may be used for a
screening of the lattice thermal conductivity of compounds. The
$(\mathbf{q}, \omega)$ map of $N_2(\mathbf{q}, \omega)$ for ZnTe is shown
in Fig.~\ref{fig:N2-ZnTe}. The phonon band structure is superimposed on
this $(\mathbf{q}, \omega)$ map. The intensity distribution is reasonably
similar to those of the imaginary parts of the self energies in
Fig.~\ref{fig:imag-self-energy-ZnTe}. Therefore low (high) lattice
thermal conductivity is expected if phonon band structure crosses
(avoids) the high intensity region with larger group velocity. Such a
discussion for the lattice thermal conductivity will be made elsewhere.

\subsection{Averaged linewidth}

\begin{figure*}[ht]
 \begin{center}
  \includegraphics[width=0.7\linewidth]{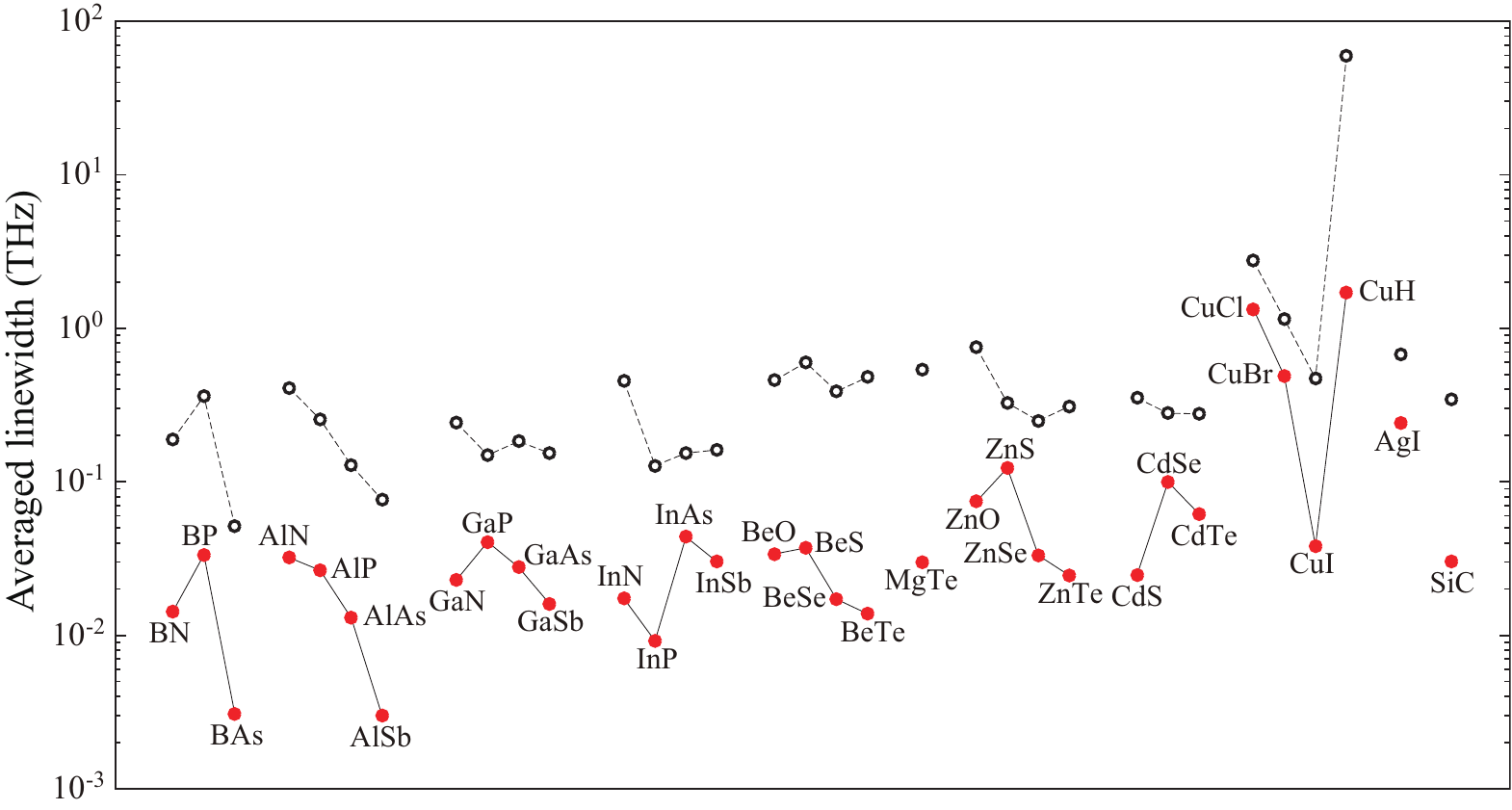}
  \caption{(color online) \label{fig:averaged-linewidths} Averaged
  linewidths $2\Gamma_\mathrm{ave}$ (filled circle) and
  $2\widetilde{\Gamma}_\mathrm{ave}$ (open circle) of the 33
  zincblende-type compounds. The solid and dashed lines are eye guide to
  group compounds.}
 \end{center}
\end{figure*}

To sort the linewidth distributions quantitatively, we define an
averaged linewidth over the Brillouin zone, $2\Gamma_\mathrm{ave}$, as a
temporal measure, with
\begin{equation}
 \label{eq:averaged-linewidth}
\Gamma_\mathrm{ave} = \frac{1}{3n_\mathrm{a}N}
\sum_{\lambda} \Gamma_\lambda(\omega_\lambda).
\end{equation}
Averaged linewidths of Cu$X$, Zn$X$, and Ga$X$ are listed in
Table~\ref{table:averaged-linewidth}. The averaged linewidths increase
from right to left and from lower to upper in
Table~\ref{table:averaged-linewidth}. Along the compounds with fixed
cations, relatively large differences of the averaged linewidths are
found between CuBr-CuI and ZnS-ZnSe. We may expect a qualitative change
of phonon collisions at these points. The averaged linewidths of the 33
zincblende-type compounds are summarized in
Fig.~\ref{fig:averaged-linewidths} (filled circles). In between BP-BAs,
AlAs-AlSb, InP-InAs, and CdS-CdSe, large differences of the averaged
linewidths are also found. These qualitative changes are also observed
in the linewidth distributions, which are shown in
Appendix~\ref{apx:bands}.

\begin{table}[ht]
 \caption{\label{table:averaged-linewidth} Averaged linewidths
 $2\Gamma_\mathrm{ave}$ of Cu$X$, Zn$X$, and Ga$X$ in THz. See
 Eq.~(\ref{eq:averaged-linewidth}) for the definition of
 $\Gamma_\mathrm{ave}$.}
 \begin{ruledtabular}
  \begin{tabular}{lcclcclc}
   CuCl & $1.3\times10^{0}$ && ZnS & $1.2\times10^{-1}$ && GaP & $4.1\times10^{-2}$ \\
   CuBr & $4.9\times10^{-1}$ & & ZnSe & $3.3\times10^{-2}$ & & GaAs &
   $2.8\times10^{-2}$ \\
   CuI & $3.8\times10^{-2}$ && ZnTe & $2.5\times10^{-2}$ && GaSb & $1.6\times10^{-2}$\\
  \end{tabular}
 \end{ruledtabular}
\end{table}

The averaged linewidths obtained with an averaged phonon-phonon
interaction, $2\widetilde{\Gamma}_\mathrm{ave}$ are also shown in
Fig.~\ref{fig:averaged-linewidths} (open
circles). $2\widetilde{\Gamma}_\mathrm{ave}$ tends to overestimate
$2\Gamma_\mathrm{ave}$, however the degree of the overestimation is not
systematic among the 33 zincblende-type compounds. This suggests that phonon-phonon
interaction strength is as complicated as three phonon selection rule
and their detailed combination is decisively important to determine
phonon linewidth quantitatively.

\section{Summary}
The lattice thermal conductivities of the zincblende- and wurtzite-type
compounds with 33 combinations of elements are calculated with RTA
and LBTE from first-principles anharmonic lattice dynamics
calculations. The zincblende- and wurtzite-type compounds with the same
compositions show similar lattice thermal conductivities, phonon
densities of states, and phonon linewidth distributions though their
stacking sequences are different. Focusing on the zincblende-type
compounds and selecting Cu$X$, Zn$X$, and Ga$X$, the phonon linewidth
distributions are investigated in detail. It is found that their phonon
linewidths are distributed non-smoothly in the Brillouin zones. This is
explained by the weak wave vector dependence and the strong frequency
dependence of the imaginary parts of the self energies. It is shown that
detailed combination of phonon-phonon interaction strength and three
phonon selection rule is critically important to determine imaginary
part of self energy. Therefore it is considered difficult to predict
phonon lifetime quantitatively without a full calculation using
anharmonic force constants. However we show a possibility that w-JDOS
may be used for a rough estimate of the lattice thermal conductivity and
thereby materials screening.

\section*{ACKNOWLEDGMENTS}
This work was supported by Grant-in-Aid for Scientific Research on
Innovative Areas ``Nano Informatics'' (Grant No. 25106005) from the
Japan Society for the Promotion of Science (JSPS), by MEXT Japan through
ESISM (Elements Strategy Initiative for Structural Materials) of Kyoto
University, by JSPS KAKENHI (Grant No. 26820284), and by a GENCI project
under Grant x2014097186.

\appendix
\clearpage

\section{Supercell approach and finite displacement approach}
\label{apx:supercell-approach}
We employed the supercell approach and the finite
displacement approach to calculate force constants.
The harmonic and cubic force constants are given by
\begin{equation}
 \Phi_{\alpha\beta}(l\kappa,l'\kappa') = \frac{\partial^2 \Phi} {\partial
  u_\alpha(l\kappa) \partial u_\beta(l'\kappa')},
\end{equation}
and
\begin{align}
 \Phi_{\alpha\beta\gamma}(l\kappa,l'\kappa',l''\kappa'') 
 = \frac{\partial^3 \Phi} {\partial u_\alpha(l\kappa) \partial
  u_\beta(l'\kappa') \partial u_\gamma(l''\kappa'')},
\end{align}
respectively. A set of finite atomic displacements is introduced to
supercells to calculate these force constants. Forces acting on atoms in
a supercell are obtained from the first-principles calculations and the
force constants are calculated from the finite difference method as
follows. The harmonic force constants may be approximately obtained by
\begin{align}
 \label{eq:harmonic-fc}
 \Phi_{\alpha\beta}(l\kappa,l'\kappa') \simeq -\frac{
  F_\beta[l'\kappa'; \mathbf{u}(l\kappa)]} {u_\alpha(l\kappa)},
\end{align}
where $\mathbf{F}[l'\kappa'; \mathbf{u}(l\kappa)]$ is defined as the
atomic force measured at $\mathbf{r}(l'\kappa')$ under an atomic
displacement $\mathbf{u}(l\kappa)$ in a supercell. The cubic force
constants may be similarly obtained by
\begin{align}
 \label{eq:cubic-fc}
 \Phi_{\alpha\beta\gamma}(l\kappa,&l'\kappa',l''\kappa'') 
 \simeq -\frac{F_\gamma[l''\kappa''; \mathbf{u}(l\kappa),
  \mathbf{u}(l'\kappa')]}{u_\alpha(l\kappa) 
  u_\beta(l'\kappa')},
\end{align}
where $\mathbf{F}[l''\kappa''; \mathbf{u}(l\kappa),
\mathbf{u}(l'\kappa')]$ is defined as the atomic force measured at
$\mathbf{r}(l''\kappa'')$ under a pair of atomic displacements
$\mathbf{u}(l\kappa)$ and $\mathbf{u}(l'\kappa')$ in a supercell. The
sets of Eqs.~(\ref{eq:harmonic-fc}) and (\ref{eq:cubic-fc}) forms
systems of linear equations, respectively, and they were solved using the
Moore-Penrose pseudoinverse. The computational details are
found in Ref.~\onlinecite{Laurent-phph-2011}.

Configurations of atomic displacements in supercells have to be chosen
to solve the linear equations. The number of the atomic displacement
configurations can be huge, however it may be reduced utilizing crystal
symmetry. Atoms that are independent with respect to space group
symmetry are selected to be displaced in supercells. The number of
displacement directions is reduced using site symmetry at the displaced
atom. For the case of cubic force constants, the second atom that is
displaced is determined from the symmetry lowered by the first atomic
displacement. To gain numerical stability, the opposite direction of a
chosen atomic displacement was introduced if they are symmetrically
inequivalent.
The crystal symmetry is again used to recover full elements of force
constants from the reduced set of supercell force calculations. Let a
site-symmetry operation $\mathrm{S}$ at the displaced atom
$\mathbf{r}(l\kappa)$ send an atomic point to the other point as
$\mathbf{r}(\tilde{l}'\tilde{\kappa}')\xrightarrow[\mathrm{S}]{}
\mathbf{r}(l'\kappa')$. With this site-symmetry operation and from
Eq.~(\ref{eq:harmonic-fc}), the first atomic displacement, forces on
atoms in the supercell, and harmonic force constants are related by
\begin{equation}
\sum_{\alpha\gamma}S_{\alpha\gamma}u_\gamma(l\kappa)
\Phi_{\alpha\beta}(l\kappa,l'\kappa')
 \simeq - \sum_\gamma S_{\beta\gamma} F_\gamma[\tilde{l}'\tilde{\kappa}';
 \mathbf{u}(l\kappa)].
\end{equation}
Similarly for the cubic force constants,
Eq.~(\ref{eq:cubic-fc}) is rewritten as
\begin{align}
 \sum_{\alpha\zeta}S_{\alpha\zeta}u_\zeta(l\kappa) & \Phi_{\alpha\beta\gamma}(l\kappa,
  l'\kappa', l''\kappa'') \simeq \nonumber \\
 & \sum_{\zeta\eta}
 S_{\beta\zeta}S_{\gamma\eta}\Delta\Phi_{\zeta\eta}
 [\tilde{l}'\tilde{\kappa}',\tilde{l}''\tilde{\kappa}'';
 \mathbf{u}(l\kappa)],
\end{align}
where
\begin{align}
\Delta\Phi_{\beta\gamma} &
 [{l}'{\kappa}',{l}''{\kappa}''
 ;\mathbf{u}(l\kappa)] \equiv \nonumber\\ &
 -\frac{F_\gamma[{l}''{\kappa}''; \mathbf{u}(l\kappa),
 \mathbf{u}(l'\kappa')]}{u_\beta({l}'{\kappa}')} -
 \Phi_{\beta\gamma}({l}'{\kappa}',
 {l}''{\kappa}'').
\end{align}

\section{Non-analytical term correction}
\label{apx:non-analytical-term-correction} Macroscopic electric field is
generated by atomic displacements of phonon modes. At long-wavelength
limit, this effect is incorporated as the non-analytical part of
harmonic force constants~\cite{Cochran-1962,Pick-1970,Giannozzi-1991}
given by
\begin{equation}
 \Phi_{\alpha\beta}^{\mathrm{na}}(\kappa\kappa', \mathbf{q}) =
  \frac{4\pi e^2}{V_0}
  \frac{\sum_{\gamma}q_{\gamma}Z^{*}_{\kappa,\gamma\alpha}
  \sum_{\gamma'}q_{\gamma'}Z^{*}_{\kappa',\gamma'\beta}}
  {\sum_{\gamma\gamma'}q_{\gamma}\epsilon_{\gamma\gamma'}^{\infty} q_{\gamma'}},
\end{equation}
where $e$ is the elemental charge, 
$Z^{*}_{\kappa}$ is the Born effective charge tensor for the atom
$\kappa$, and $\epsilon^{\infty}$ is the high frequency static
dielectric constant tensor. The non-analytical term is $\mathbf{q}$
dependent but only on its direction in reciprocal space and its
application is limited at $\mathbf{q}\rightarrow \mathbf{0}$. In the
supercell approach, the macroscopic electric field effect at general
$\mathbf{q}$-points may be included into the dynamical matrix using a
Fourier interpolation scheme presented in
Ref.~\onlinecite{Wang-2010}. The summation in
Eq.~(\ref{eq:dynamical-matrix-ti}) is made over lattice points in a
supercell and the non-analytical term is added to the harmonic supercell
force constants divided by the number of unit cells in the supercell,
$N_\mathrm{c}$, as
\begin{equation}
 \Phi^{\mathrm{NAC}}_{\alpha\beta}(l\kappa,l'\kappa', \mathbf{q}) =
  \Phi_{\alpha\beta}(l\kappa,l'\kappa') +
  \Phi_{\alpha\beta}^{\mathrm{na}}(\kappa\kappa', \mathbf{q}) /
  N_\mathrm{c},
\end{equation}
which is used instead of normal $\Phi_{\alpha\beta}(l\kappa,l'\kappa')$ to
 create dynamical matrices in Eq.~(\ref{eq:dynamical-matrix-ti}).

In Eq.~(\ref{eq:group-velocity}), first derivative of dynamical matrix
requires first derivative of the non-analytical term that is easily
obtained as
% \begin{widetext}
\begin{align} 
 & \frac{\partial
  \Phi_{\alpha\beta}^{\mathrm{na}}(\kappa\kappa', \mathbf{q})}
 {\partial q_\zeta}  = 
  \frac{4\pi e^2}{V_0} \times \nonumber \\
& \;\;\; \left[ \frac{Z^{*}_{\kappa,\zeta\alpha}\sum_{\gamma'}
  q_{\gamma'} Z^{*}_{\kappa',\gamma'\beta} +
  Z^{*}_{\kappa',\zeta\beta}\sum_\gamma q_\gamma
  Z^{*}_{\kappa,\gamma\alpha}}
  {\sum_{\gamma\gamma'}q_{\gamma}\epsilon_{\gamma\gamma'}^{\infty}
 q_{\gamma'}} \right.   \nonumber \\
& \;\;\;\;\; \left. - \frac{
  \sum_{\gamma}q_{\gamma}Z^{*}_{\kappa,\gamma\alpha}
  \sum_{\gamma'}q_{\gamma'}Z^{*}_{\kappa',\gamma'\beta}
  \sum_{\gamma''} q_{\gamma''} (\epsilon^{\infty}_{\zeta\gamma''} +
   \epsilon^{\infty}_{\gamma''\zeta})
 }
  {\left(\sum_{\gamma\gamma'}q_{\gamma}
    \epsilon_{\gamma\gamma'}^{\infty} q_{\gamma'}\right)^2} \right].
\end{align}
% \end{widetext}
Hence the first derivative of the dynamical matrix with the
non-analytical term correction becomes
\begin{align}
& \frac{\partial D^{\mathrm{NAC}}_{\alpha\beta}
 (\kappa\kappa', \mathbf{q})}{\partial q_\zeta} =
 \frac{1}{\sqrt{m_\kappa m_\kappa'}} \frac{1}{N_\mathrm{c}} \times \nonumber \\
& \;\;\sum_{ll'} \Bigg\{
 \frac{1}{N_\mathrm{c}}\frac{\partial \Phi^{\mathrm{na}}_{\alpha\beta}
 (\kappa\kappa',\mathbf{q})}{\partial q_\zeta} e^{i\mathbf{q}\cdot
 \left[\mathbf{r}(l'\kappa')-\mathbf{r}(l\kappa)\right]} + 
 \nonumber \\
& \;\; i
 \Phi^{\mathrm{NAC}}_{\alpha\beta}(l\kappa,l'\kappa',\mathbf{q})
 \left[r_\zeta(l'\kappa')-r_\zeta(l\kappa)\right]
 e^{i\mathbf{q}\cdot\left[\mathbf{r}(l'\kappa')-\mathbf{r}(l\kappa)\right]}
 \Bigg\}.
\end{align}

\section{Brillouin zone integration}
\label{apx:brillouin-zone-integration} 

Computations of phonon lifetimes were discretized using sampling meshes
in Brillouin zones. The calculation order of a phonon lifetime in
Eq.~(\ref{eq:imag-selfenergy}) is $O(n^6)$ with respect to the number of
mesh points along an axis. This is reduced using translation symmetry of
crystal to $O(n^3)$ by $\Delta(\mathbf{q} + \mathbf{q}' +
\mathbf{q}'')$.  In Eq.~(\ref{eq:imag-selfenergy}), it is supposed that
the leftmost index of the phonon-phonon interaction strength
(Eq.~\ref{eq:ph-ph-strength}) is fixed and the rest of the indices can
be interchanged as $\Phi_{-\lambda\lambda'\lambda''} =
\Phi_{-\lambda\lambda''\lambda'}$. Therefore the calculation order
becomes $O(n^3/2)$.  A set of $\mathbf{q}$-point triplets,
$\{(\mathbf{q}, \mathbf{q}', \mathbf{q}'')|\;\mathbf{q}, \mathbf{q}',
\mathbf{q}'' \in \mathrm{Brillouin\;zone}\}$ was thus reduced. The
number of the $\mathbf{q}$-point triplets was reduced further
considering crystallographic point group symmetry and time-reversal
symmetry.

We performed the Brillouin zone integration in
Eq.~(\ref{eq:imag-selfenergy}) using a linear tetrahedron
method.\cite{MacDonald-tetrahedron-1979,Blochl-tetrahedron-1994,Tadano-2014}
Smearing method is another popular method that is easier to implement in
computer programs. In a smearing method, the delta functions are simply
replaced by the Gaussian or Lorentzian functions that are normalized so
that those integrals become 1. The advantage of the tetrahedron method
over the smearing method is that it can be parameter-free. The
straightforward application of the tetrahedron method to
Eq.~(\ref{eq:imag-selfenergy}) requires the translational invariance
condition where Eq.~(\ref{eq:imag-selfenergy}) is reduced to a single
summation over Brillouin zone.\cite{Tadano-2014} By sorting the order of
integration of contributions from vertices of tetrahedra, each delta
function is simply replaced by respective integration weight that is
pre-calculated from phonon mode frequencies on a sampling
mesh.\cite{Blochl-tetrahedron-1994} Therefore after computing the
integration weights, the tetrahedron method and smearing method can be
treated in a computer program on equal footing. Generally the smearing
method with an appropriate smearing parameter underestimates phonon
lifetime, and thus lattice thermal conductivity as well. In both
methods, the integrations are necessary to be performed over the reduced
set of $\mathbf{q}$-points or triplets.

The Brillouin zone integration of Eq.~(\ref{eq:SMRT-kappa}) reduces to the
integration over an irreducible part of the Brillouin zone (IBZ):
\begin{align}
 \label{eq:SMRT-kappa-BZ} \kappa_{\alpha\beta} = \frac{1}{NV_0}
 & \sum_{\mathbf{q}\in \mathrm{IBZ}} 
 \frac{1}{p_\mathbf{q}}
 \sum_j C_{\mathbf{q} j} \tau_{\mathbf{q} j}  \nonumber \\
 \times & \sum_\mathrm{R} \sum_{\gamma\zeta}
 R_{\alpha\gamma} R_{\beta\zeta} v_\gamma(\mathbf{q} j)
  v_\zeta(\mathbf{q} j),
\end{align}
where $\mathrm{R}$ runs the point group operations and $p_\mathbf{q}$ is
the order of point-group of the $\mathbf{q}$-vector.

\section{Isotope scattering}
\label{apx:isotope}

The scattering rate for a phonon mode by randomly distributed isotopes
is given by the second-order perturbation theory
as~\cite{Tamura-isotope-lifetime-1983}
\begin{align}
 \label{eq:isotope-lifetime}
\frac{1}{\tau^\mathrm{iso}_\lambda(\omega)} =
 \frac{\pi}{2N}  \omega_\lambda^2&
 \sum_{\lambda'} \delta(\omega - \omega_{\lambda'}) 
 \nonumber \\
 \times & \sum_\kappa  g_\kappa \Big|
 \sum_\alpha W_\alpha(\kappa, \lambda)W_\alpha^*(\kappa, \lambda')
 \Big|^2,
\end{align}
where $g_\kappa$ is the mass variance parameter. $g_\kappa$ is defined
by
\begin{equation}
 g_\kappa=\sum_i f_i (1 - m_{i\kappa} / \bar{m}_\kappa)^2,
\end{equation}
where $f_i$ and $m_{i\kappa}$ are the mole fraction and relative atomic
mass of the $i$th isotope, respectively. $\bar{m}_\kappa$ is the average
mass given by $\bar{m}_\kappa = \sum_i f_i m_{i\kappa}$. When both of
phonon-phonon and phonon-isotope scattering effects were considered in the
calculation, we employed the scattering rate given as the sum of them:
\begin{equation}
 \label{eq:total-lifetime}
\frac{1}{\tau_\lambda^\mathrm{total}} =
\frac{1}{\tau_\lambda} + \frac{1}{\tau^\mathrm{iso}_\lambda(\omega_\lambda)}.
\end{equation}

\section{Direct solution of linearized phonon Boltzmann equation}
\label{apx:LBTE} Recently the method of the direct solution of the
linearized phonon Boltzmann equation was developed by
Chaput.\cite{Laurent-LBTE-2013} In this method, lattice thermal
conductivity is given by
\begin{align}
\kappa_{\alpha\beta} = \frac{\hbar^2}{4k_\mathrm{B}T^2NV_0}
 \sum_{\lambda\lambda'}
 \frac{\omega_\lambda v_\alpha(\lambda)}
 {\sinh\left(\frac{\hbar\omega_{\lambda}}{2k_\mathrm{B}T}\right)}
 \frac{\omega_{\lambda'} v_\beta(\lambda')}
 {\sinh\left(\frac{\hbar\omega_{\lambda'}}{2k_\mathrm{B}T}\right)}
 (\Omega^{\sim1})_{\lambda\lambda'},
\end{align}
where $\Omega^{\sim1}$ denotes the Moore-Penrose
inverse of the collision matrix $\Omega$ that is given by
\begin{align}
 \label{eq:collision-matrix}
 \Omega_{\lambda\lambda'} = &
  \delta_{\lambda\lambda'}
\frac{1}{\tau_{\lambda}} + \frac{\pi}{\hbar^2}
\sum_{\lambda''} \bigl|\Phi_{\lambda\lambda'\lambda''}\bigl|^2
 \frac{1}{\sinh\left(\frac{\hbar\omega_{\lambda''}}{2k_\mathrm{B}T}\right)}
 \nonumber \\
\times \left[ \right. & \delta(\omega_\lambda - \omega_\lambda' -
 \omega_\lambda'') \nonumber \\
 + & \delta(\omega_\lambda + \omega_\lambda' - \omega_\lambda'') 
+ \delta(\omega_\lambda - \omega_\lambda' + \omega_\lambda'')
  \left. \right].
\end{align}
To include the phonon-isotope scattering effect in
Appendix~\ref{apx:kappa}, $\tau_\lambda$ in
Eq.~(\ref{eq:collision-matrix}) was replaced by
$\tau_\lambda^\mathrm{total}$ of Eq.~(\ref{eq:total-lifetime}).

\section{Lattice parameters}
\label{apx:lattice-parameters}

Since phonon lifetime is sensitive to lattice parameters, the lattice
parameters and volumes of the (conventional) unit cells of the
zincblende- and wurtzite-type compounds with 33 combinations of chemical
elements used in the present study are listed in
Table~\ref{table:lattice-parameters}.
 
\begin{table}[ht]
 \caption{\label{table:lattice-parameters} Lattice parameters (\AA) and
 volumes (\AA$^3$) of conventional unit cells of the zincblende- and
 wurtzite-type structures used in the present study.}
 \begin{ruledtabular}
  \begin{tabular}{lccccccc}
   && \multicolumn{2}{c}{Zincblende-type} &&
   \multicolumn{3}{c}{Wurtzite-type} \\
   \cline{3-4} \cline{6-8}
   && $a$ & Volume && $a$ & $c$ & volume \\
   \hline
BN   && 3.63 &   47.6 && 2.56 & 4.22 &   23.9 \\
BP   && 4.55 &   94.0 && 3.20 & 5.30 &   47.0 \\
BAs  && 4.81 &  111.1 && 3.38 & 5.60 &   55.5 \\
AlN  && 4.40 &   85.2 && 3.13 & 5.02 &   42.5 \\
AlP  && 5.51 &  166.9 && 3.89 & 6.38 &   83.5 \\
AlAs && 5.73 &  187.8 && 4.04 & 6.64 &   93.9 \\
AlSb && 6.23 &  242.1 && 4.40 & 7.23 &  121.0 \\
GaN  && 4.55 &   94.1 && 3.22 & 5.24 &   47.0 \\
GaP  && 5.51 &  166.9 && 3.88 & 6.39 &   83.5 \\
GaAs && 5.75 &  189.9 && 4.05 & 6.67 &   95.0 \\
GaSb && 6.22 &  240.2 && 4.38 & 7.21 &  120.1 \\
InN  && 5.05 &  128.9 && 3.58 & 5.79 &   64.4 \\
InP  && 5.96 &  211.5 && 4.21 & 6.90 &  105.8 \\
InAs && 6.19 &  236.8 && 4.37 & 7.17 &  118.5 \\
InSb && 6.63 &  291.9 && 4.68 & 7.69 &  146.0 \\
BeO  && 3.83 &   56.1 && 2.71 & 4.41 &   28.1 \\
BeS  && 4.88 &  115.9 && 3.44 & 5.67 &   58.1 \\
BeSe && 5.18 &  139.2 && 3.65 & 6.02 &   69.7 \\
BeTe && 5.67 &  182.1 && 4.00 & 6.59 &   91.1 \\
MgTe && 6.51 &  276.3 && 4.61 & 7.50 &  138.2 \\
ZnO  && 4.63 &   99.2 && 3.29 & 5.30 &   49.6 \\
ZnS  && 5.45 &  161.7 && 3.85 & 6.31 &   80.9 \\
ZnSe && 5.74 &  189.2 && 4.05 & 6.65 &   94.7 \\
ZnTe && 6.18 &  236.5 && 4.36 & 7.18 &  118.4 \\
CdS  && 5.94 &  209.3 && 4.20 & 6.84 &  104.7 \\
CdSe && 6.21 &  239.2 && 4.39 & 7.17 &  119.7 \\
CdTe && 6.63 &  290.8 && 4.68 & 7.67 &  145.6 \\
CuCl && 5.43 &  159.7 && 3.82 & 6.32 &   79.9 \\
CuBr && 5.73 &  187.7 && 4.04 & 6.65 &   93.9 \\
CuI  && 6.08 &  224.8 && 4.29 & 7.06 &  112.4 \\
CuH  && 4.02 &   64.9 && 2.87 & 4.56 &   32.5 \\
AgI  && 6.61 &  288.8 && 4.67 & 7.64 &  144.3 \\
SiC  && 4.38 &   84.0 && 3.09 & 5.07 &   42.0 \\
  \end{tabular}
 \end{ruledtabular}
\end{table}

\section{Phonon band structures}
\label{apx:bands}

\begin{figure}[ht]
 \begin{center}
  \includegraphics[width=0.80\linewidth]{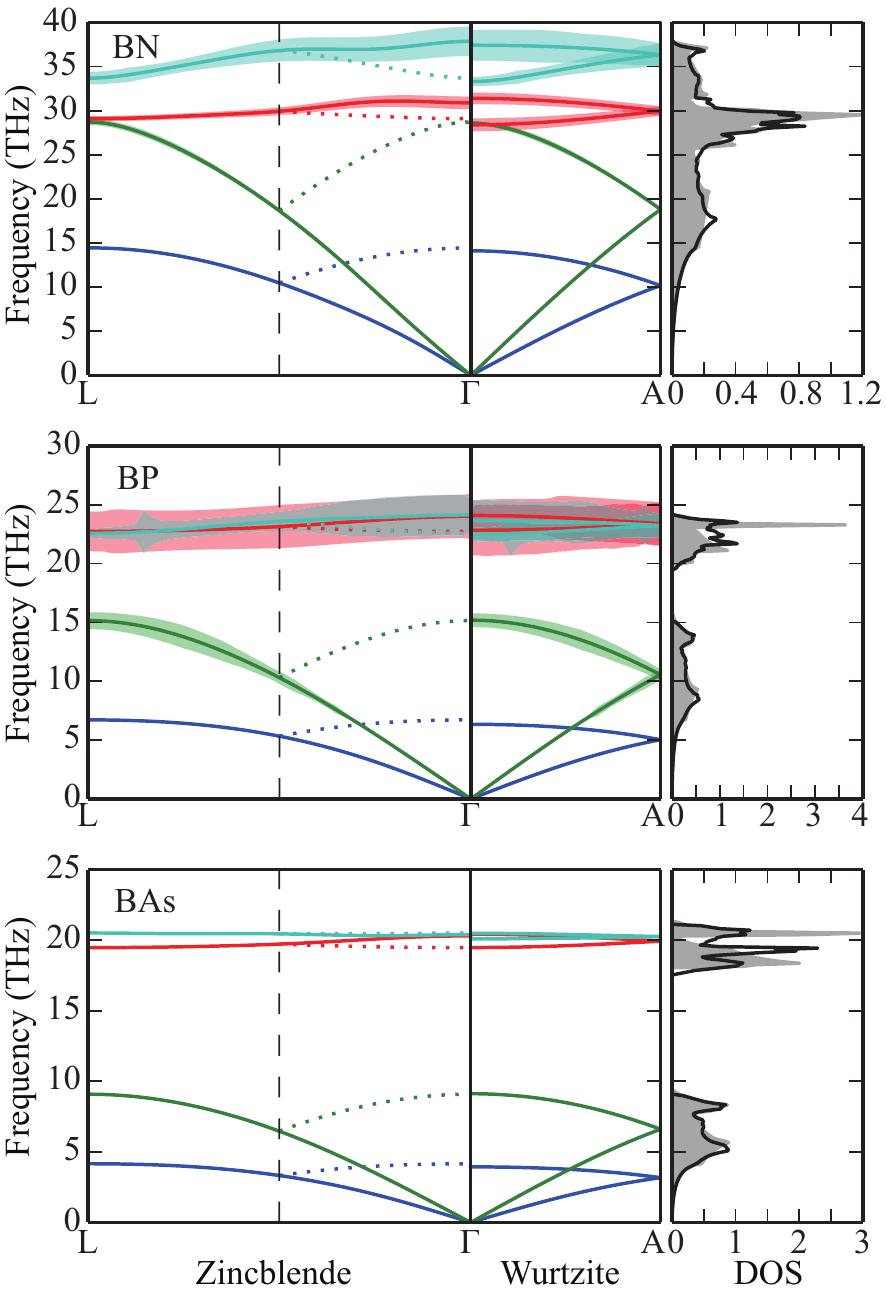}
  \caption{(color online) \label{fig:band-BX} Linewidth distributions
  and phonon densities of states of the zincblende- and wurtzite-type
   BN, BP, and BAs.  The linewidths are drawn by 50 times
  magnified.}
 \end{center}
\end{figure}

\begin{figure}[ht]
 \begin{center}
  \includegraphics[width=0.80\linewidth]{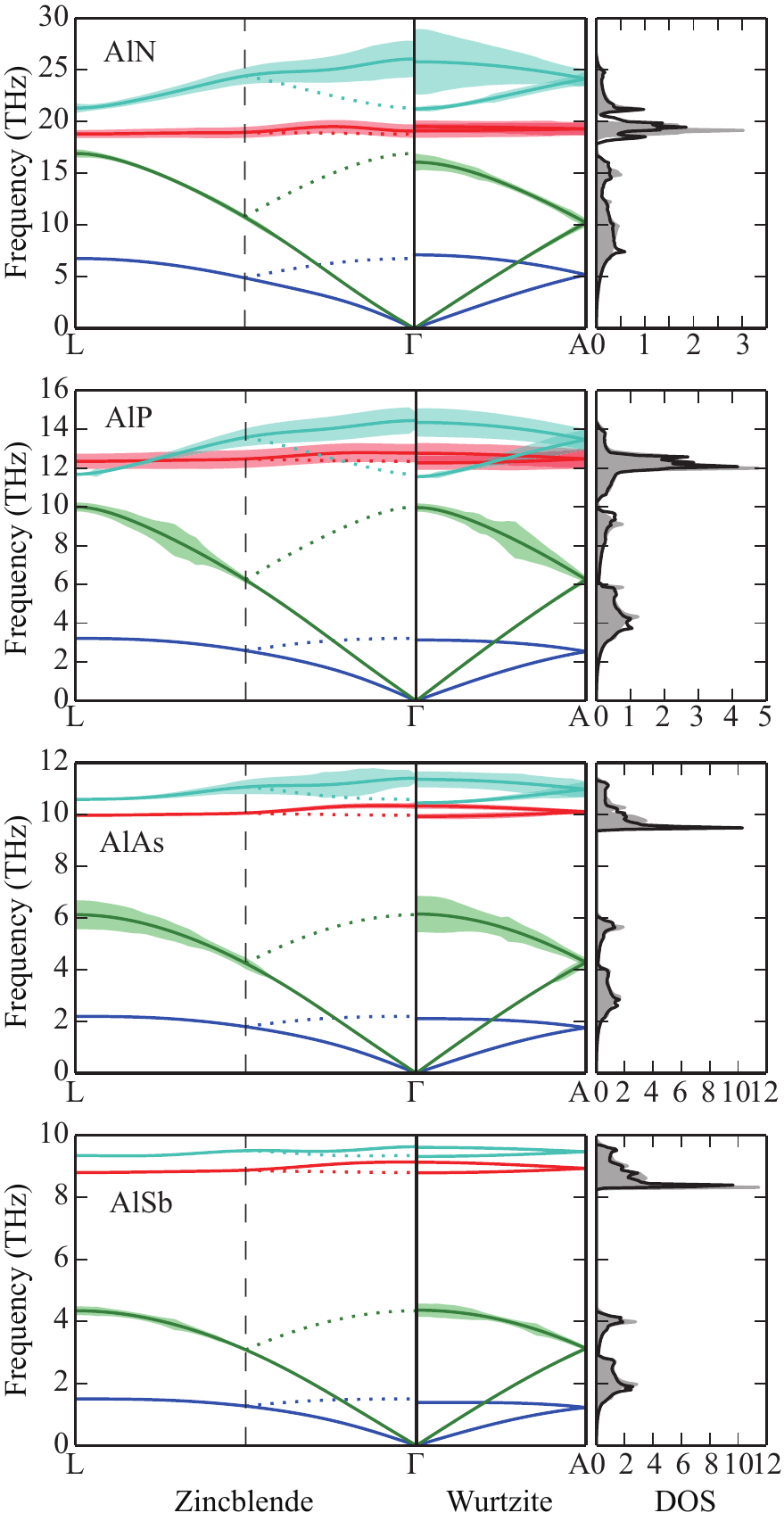}
  \caption{(color online) \label{fig:band-AlX} Linewidth distributions
  and phonon densities of states of the zincblende- and wurtzite-type
   AlN, AlP, AlAs, and AlSb. The linewidths are drawn
  by 20 times magnified.}
 \end{center}
\end{figure}

\begin{figure}[ht]
 \begin{center}
  \includegraphics[width=0.80\linewidth]{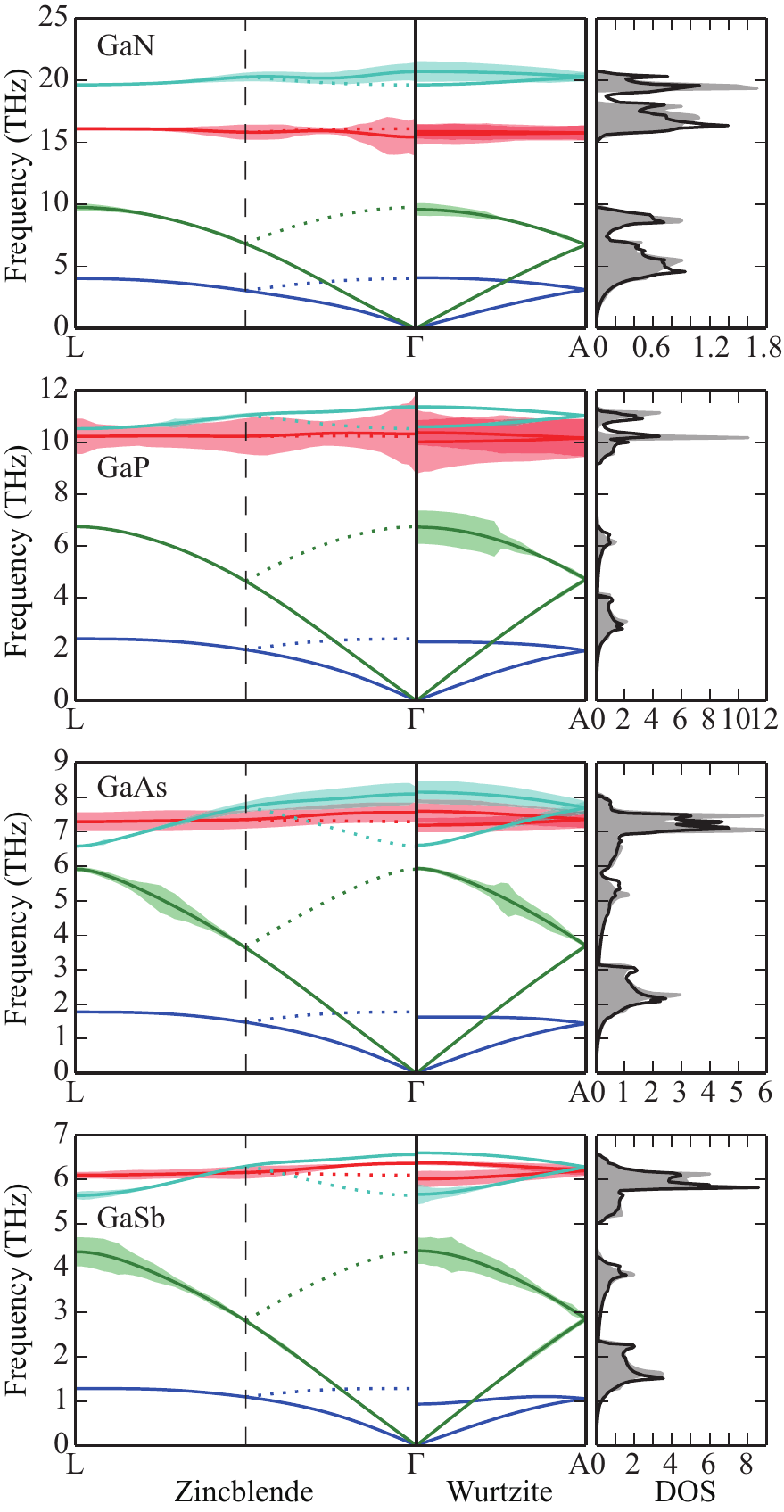}
  \caption{(color online) \label{fig:band-GaX} Linewidth distributions
  and phonon densities of states of the zincblende- and wurtzite-type
   GaN, GaP, GaAs, and GaSb. The linewidths are drawn
  by 10 times magnified.}
 \end{center}
\end{figure}

\begin{figure}[ht]
 \begin{center}
  \includegraphics[width=0.80\linewidth]{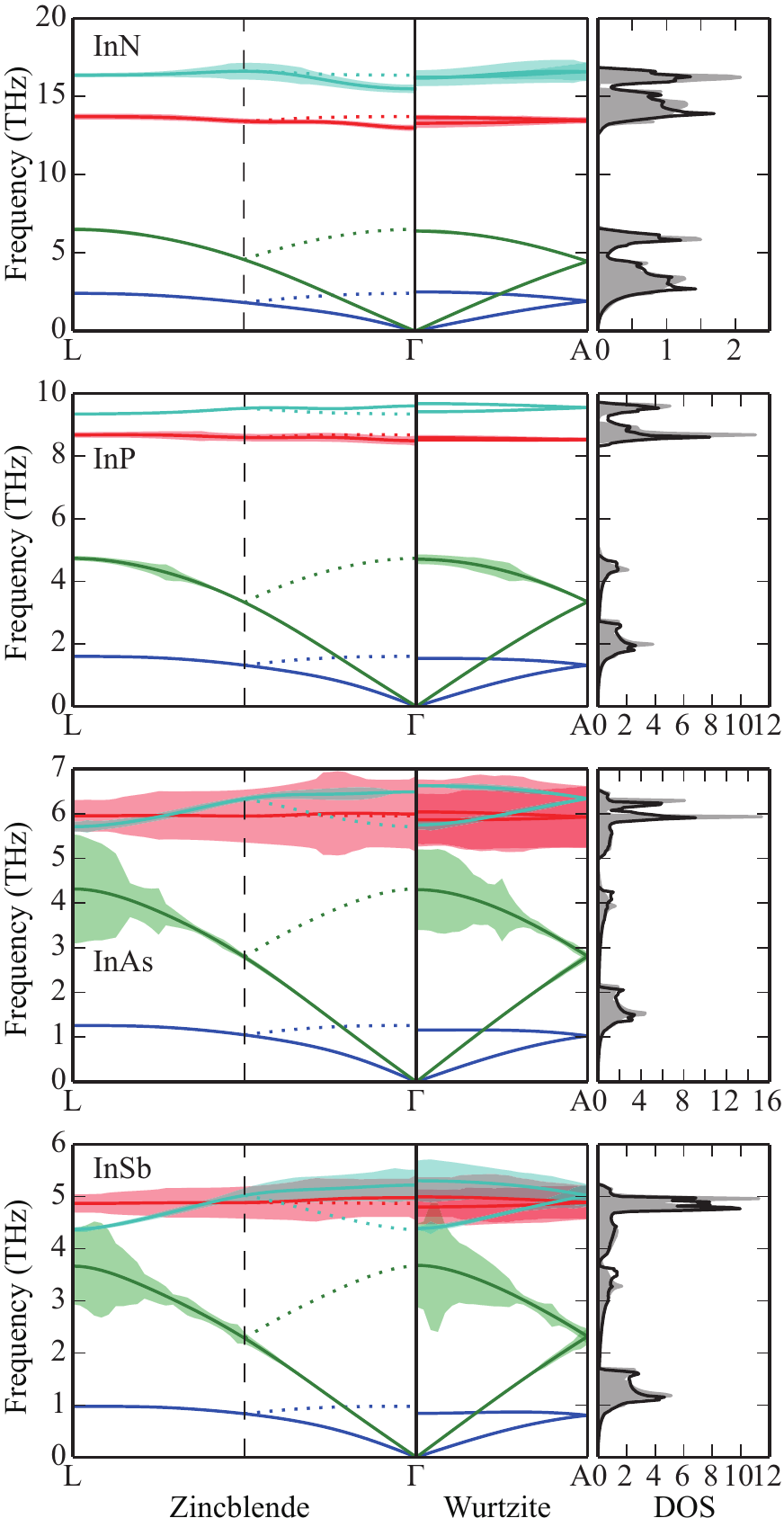}
  \caption{(color online) \label{fig:band-InX} Linewidth distributions
  and phonon densities of states of the zincblende- and wurtzite-type
   InN, InP, InAs, and InSb. The linewidths are drawn
  by 10 times magnified.}
 \end{center}
\end{figure}

\begin{figure}[ht]
 \begin{center}
  \includegraphics[width=0.80\linewidth]{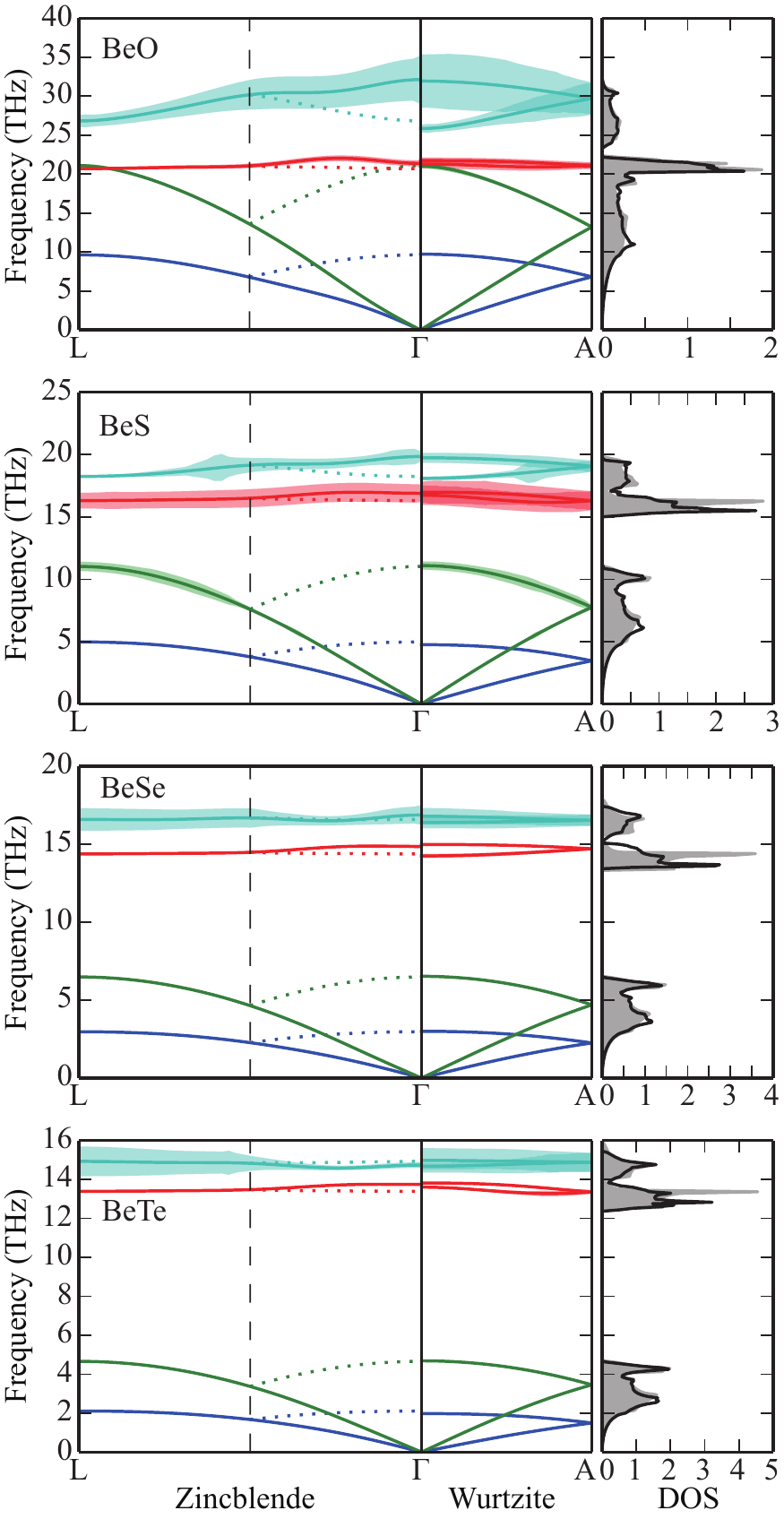}
  \caption{(color online) \label{fig:band-BeX} Linewidth distributions
  and phonon densities of states of the zincblende- and wurtzite-type
   BeO, BeS, BeSe, and BeTe. The linewidths are drawn
  by 20 times magnified.}
 \end{center}
\end{figure}

\begin{figure}[ht]
 \begin{center}
  \includegraphics[width=0.80\linewidth]{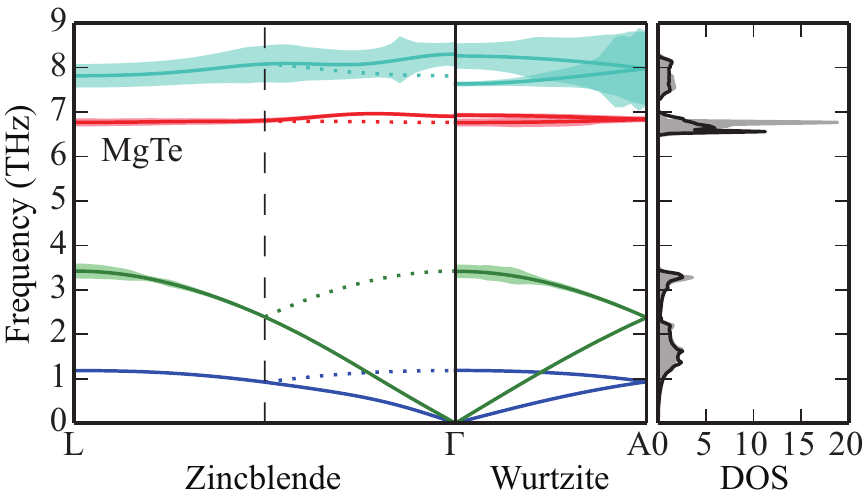}
  \caption{(color online) \label{fig:band-MgTe} Linewidth distributions
  and phonon densities of states of the zincblende- and wurtzite-type
   MgTe. The linewidths are drawn
  by 5 times magnified.}
 \end{center}
\end{figure}

\begin{figure}[ht]
 \begin{center}
  \includegraphics[width=0.80\linewidth]{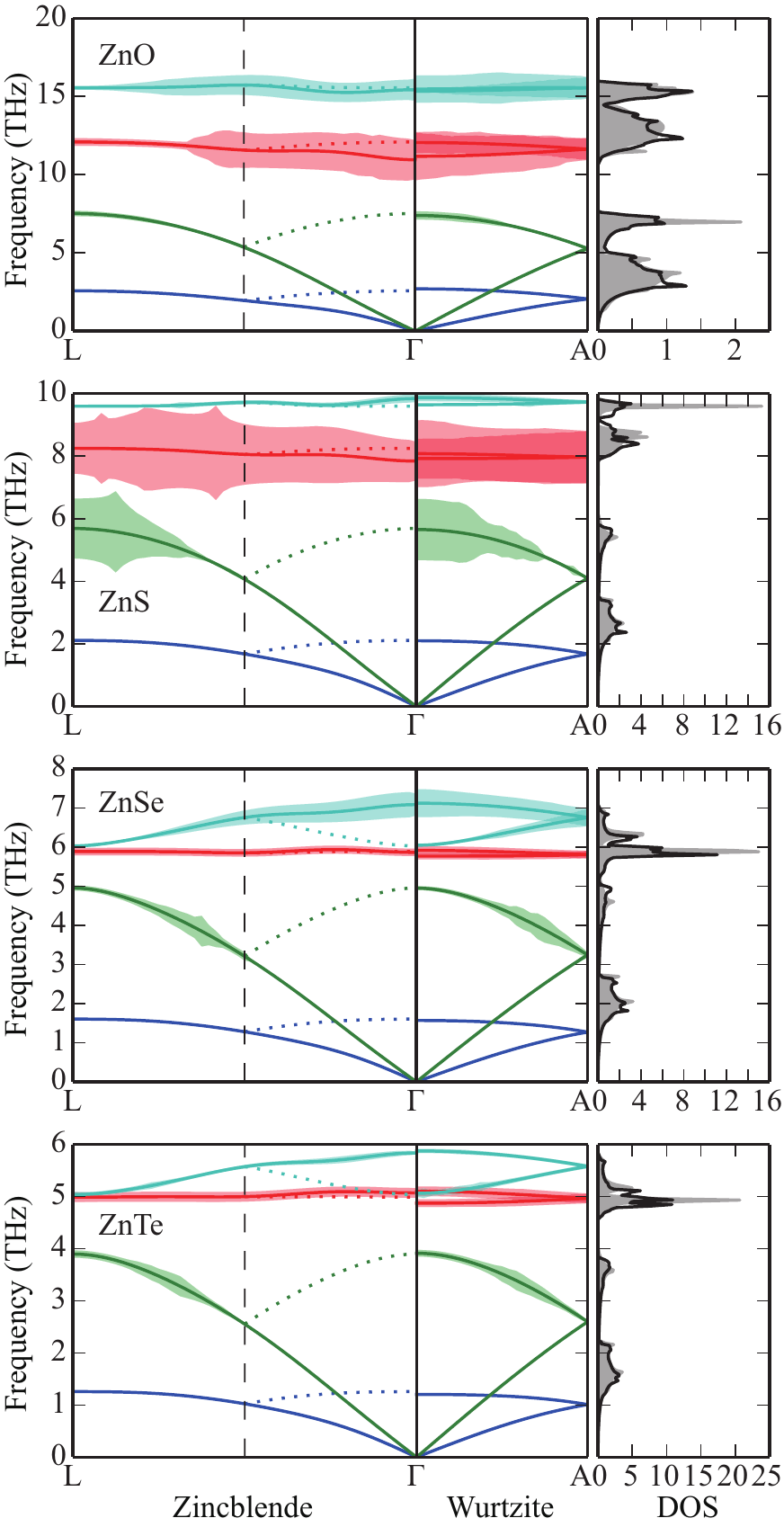}
  \caption{(color online) \label{fig:band-ZnX} Linewidth distributions
  and phonon densities of states of the zincblende- and wurtzite-type
   ZnO, ZnS, ZnSe, and ZnTe. The linewidths are drawn
  by 5 times magnified.}
 \end{center}
\end{figure}

\begin{figure}[ht]
 \begin{center}
  \includegraphics[width=0.80\linewidth]{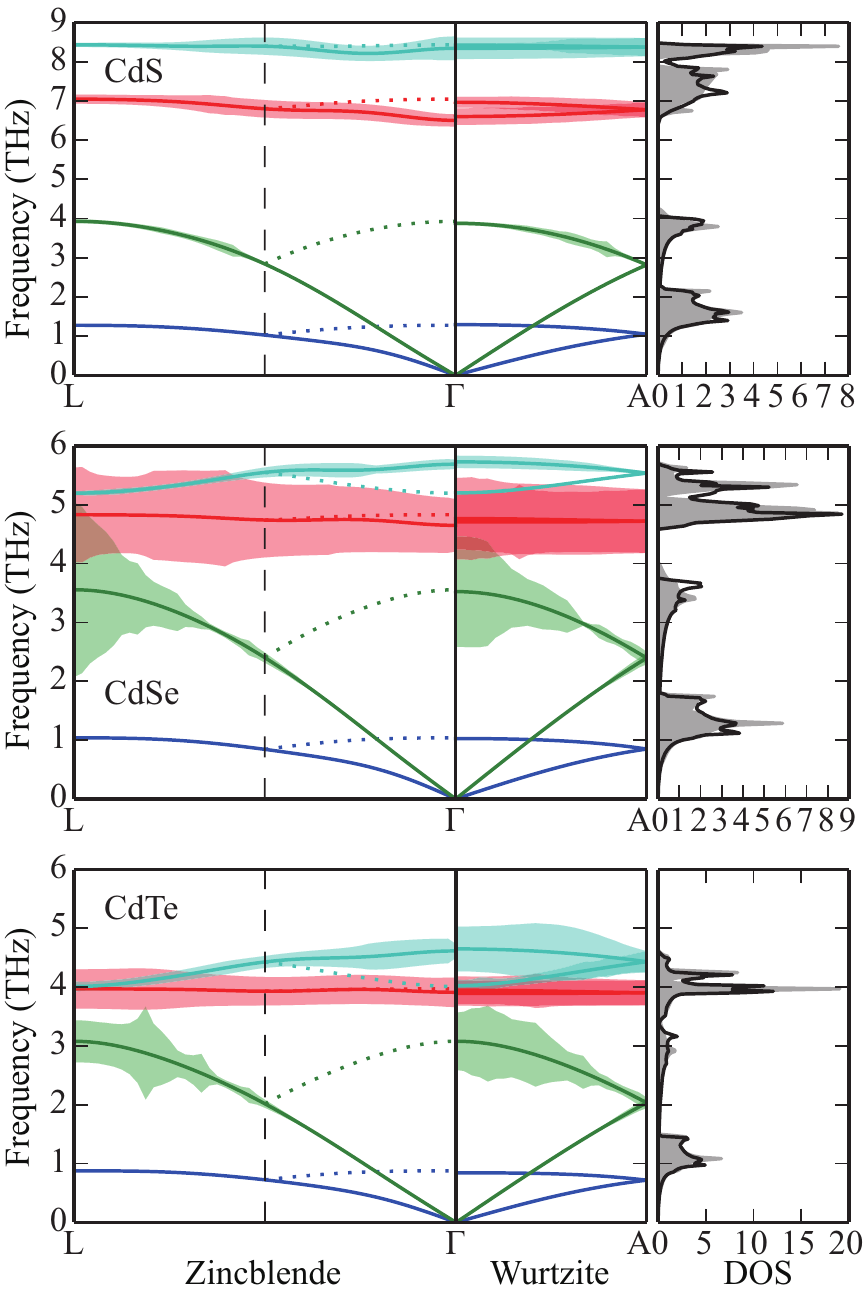}
  \caption{(color online) \label{fig:band-CdX} Linewidth distributions
  and phonon densities of states of the zincblende- and wurtzite-type
   CdS, CdSe, and CdTe. The linewidths are drawn
  by 5 times magnified.}
 \end{center}
\end{figure}

\begin{figure}[ht]
 \begin{center}
  \includegraphics[width=0.80\linewidth]{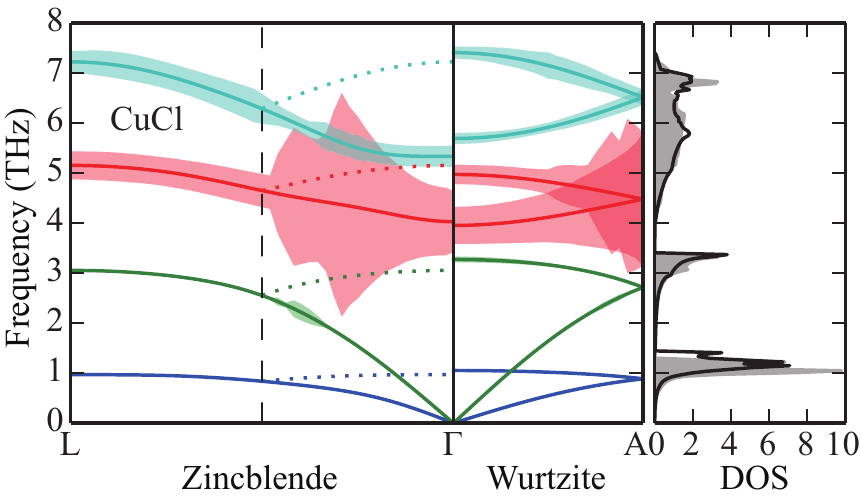}
  \caption{(color online) \label{fig:band-CuCl} Linewidth distributions
  and phonon densities of states of the zincblende- and wurtzite-type
   CuCl. The linewidths are drawn
  by 4 times reduced.}
 \end{center}
\end{figure}

\begin{figure}[ht]
 \begin{center}
  \includegraphics[width=0.80\linewidth]{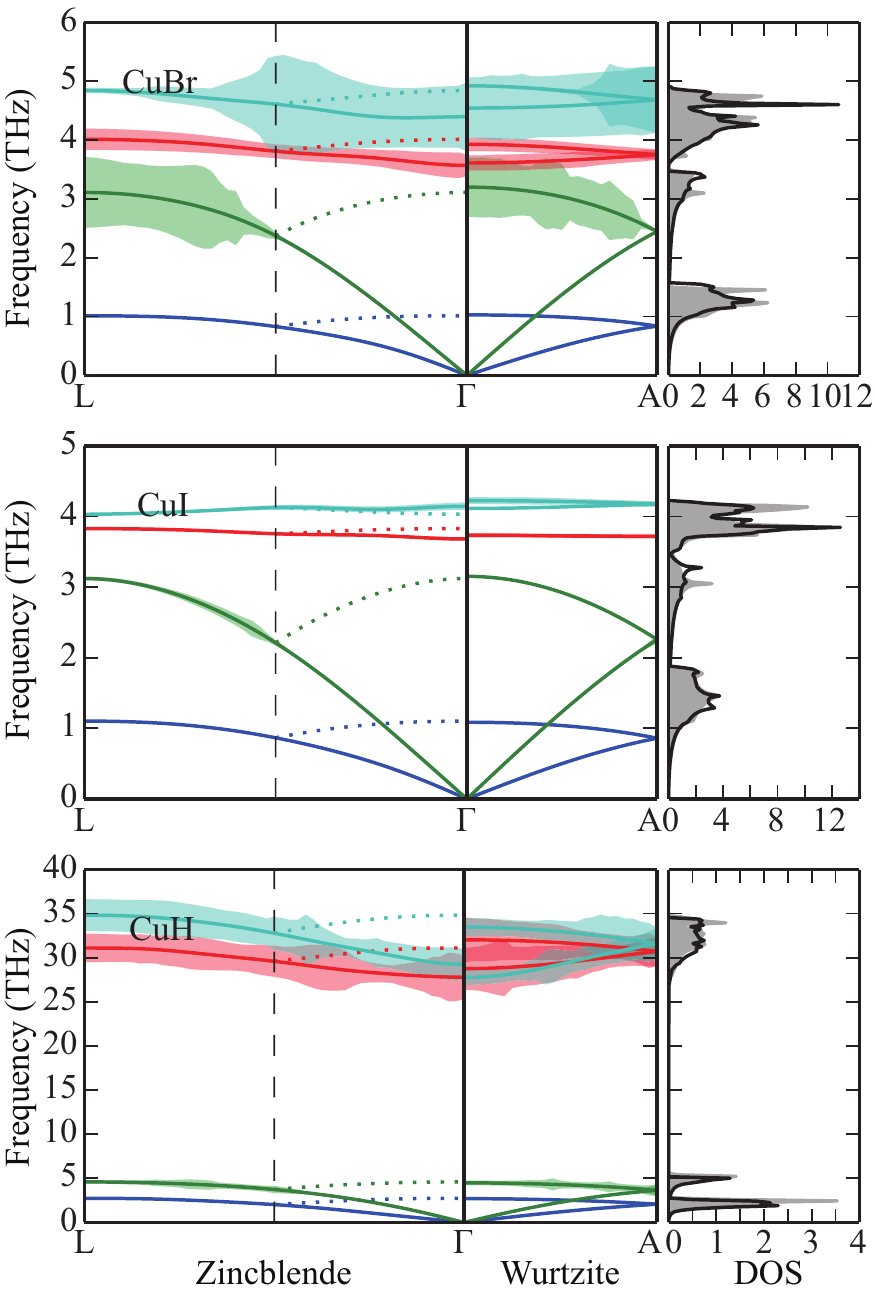}
  \caption{(color online) \label{fig:band-CuX} Linewidth distributions
  and phonon densities of states of the zincblende- and wurtzite-type
   CuBr, CuI, and CuH.}
 \end{center}
\end{figure}

\begin{figure}[ht]
 \begin{center}
  \includegraphics[width=0.80\linewidth]{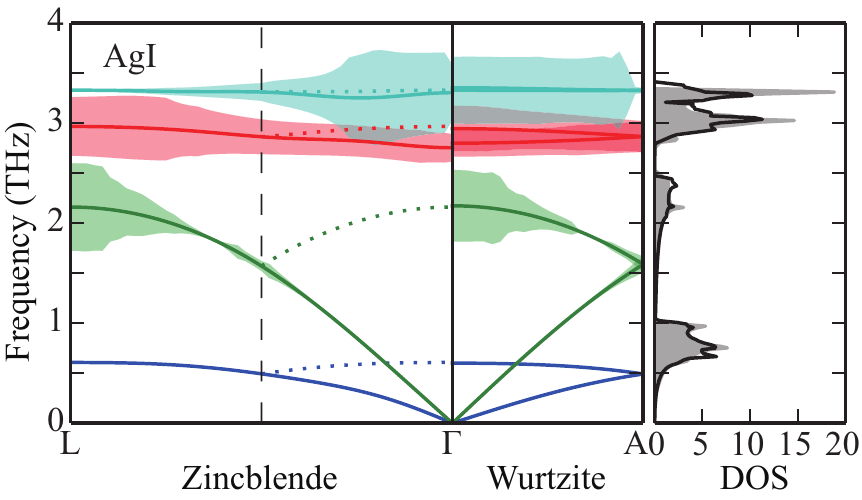}
  \caption{(color online) \label{fig:band-AgI} Linewidth distributions
  and phonon densities of states of the zincblende- and wurtzite-type
   AgI.}
 \end{center}
\end{figure}

\begin{figure}[ht]
 \begin{center}
  \includegraphics[width=0.75\linewidth]{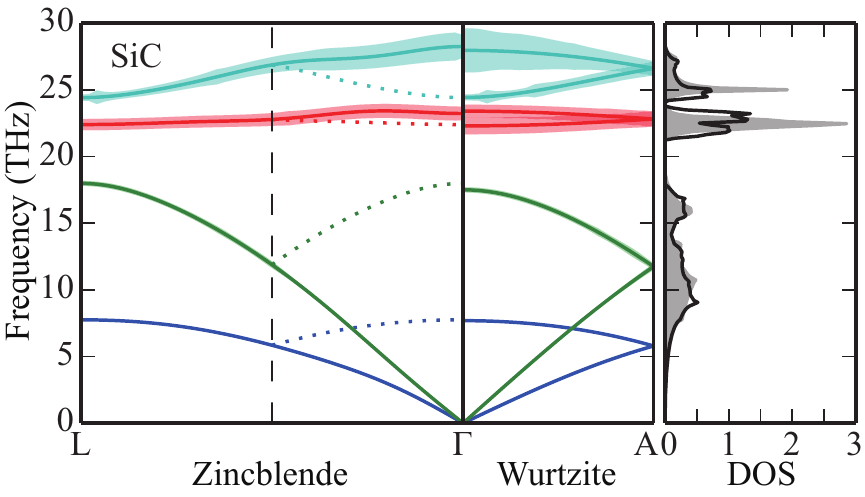}
  \caption{(color online) \label{fig:band-SiC} Linewidth distributions
  and phonon densities of states of the zincblende- and wurtzite-type
   SiC. The linewidths are drawn
  by 20 times magnified.}
 \end{center}
\end{figure}

Figures from \ref{fig:band-BX} to \ref{fig:band-SiC} show the phonon
band structures and distributions of phonon linewidths (left panels) and
phonon densities of states (DOS) (right panels) for the zincblende- and
wurtzite-type compounds with 33 combinations of elements used in the
present study. The reciprocal path of $\Gamma$-L of the zincblende-type
structure is around twice longer than that of $\Gamma$-A of the
wurtzite-type structure. The dotted curves in the left panel show the
phonon band structure of the zincblende-type compound folded at the
middle point between L and $\Gamma$ points to emphasize the similarity
between the phonon band structures of the zincblende- and wurtzite-type
compounds. The phonon DOS shaded and drawn by solid curves are those of
the zincblende- and wurtzite-type compounds, respectively.  \clearpage

\clearpage

\section{Imaginary parts of self energies}
\label{apx:self} 
\begin{figure}[ht]
 \begin{center}
  \includegraphics[height=0.80\textheight]{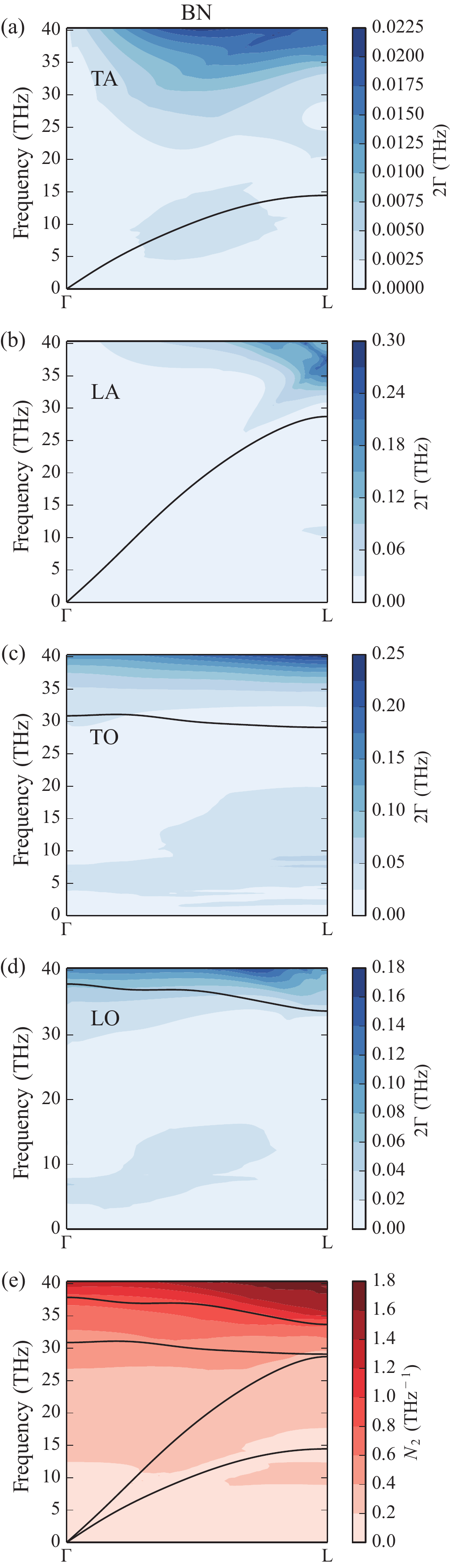}
  \caption{(color online) \label{fig:gamma-jdos-BN} $(\mathbf{q}, \omega)$ maps of imaginary parts of
  self energies 2$\Gamma_\lambda(\omega)$ and w-JDOS $N_2(\mathbf{q},\omega)$ of BN.}
 \end{center}
\end{figure}

\begin{figure}[ht]
 \begin{center}
  \includegraphics[height=0.80\textheight]{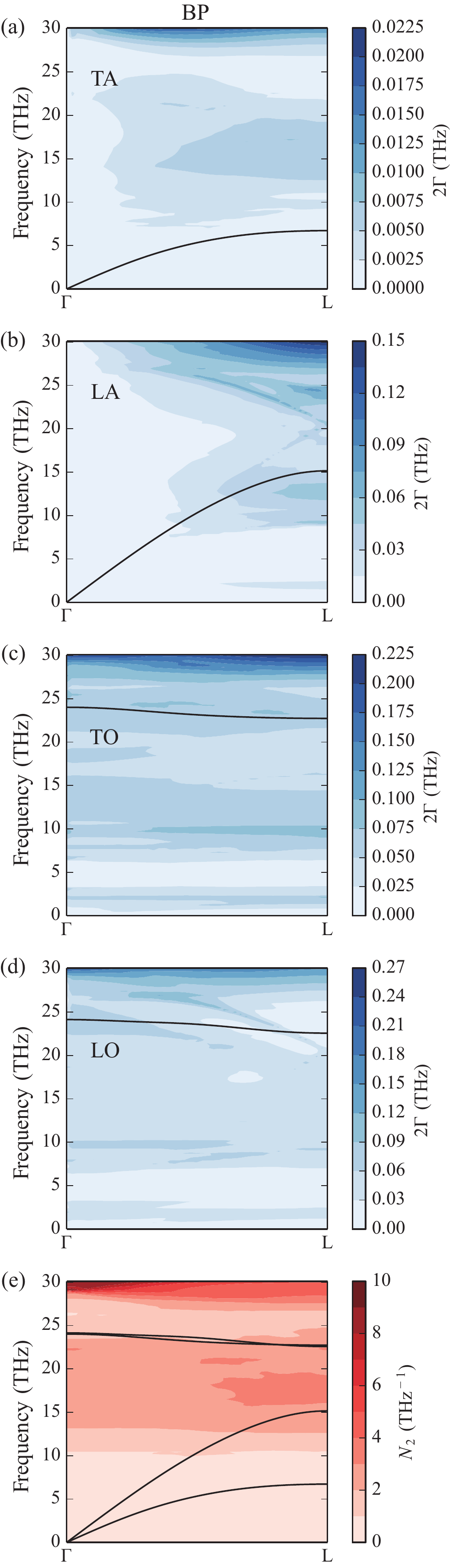}
  \caption{(color online) \label{fig:gamma-jdos-BP} $(\mathbf{q}, \omega)$ maps of imaginary parts of
  self energies 2$\Gamma_\lambda(\omega)$ and w-JDOS $N_2(\mathbf{q},\omega)$ of BP.}
 \end{center}
\end{figure}

\begin{figure}[ht]
 \begin{center}
  \includegraphics[height=0.80\textheight]{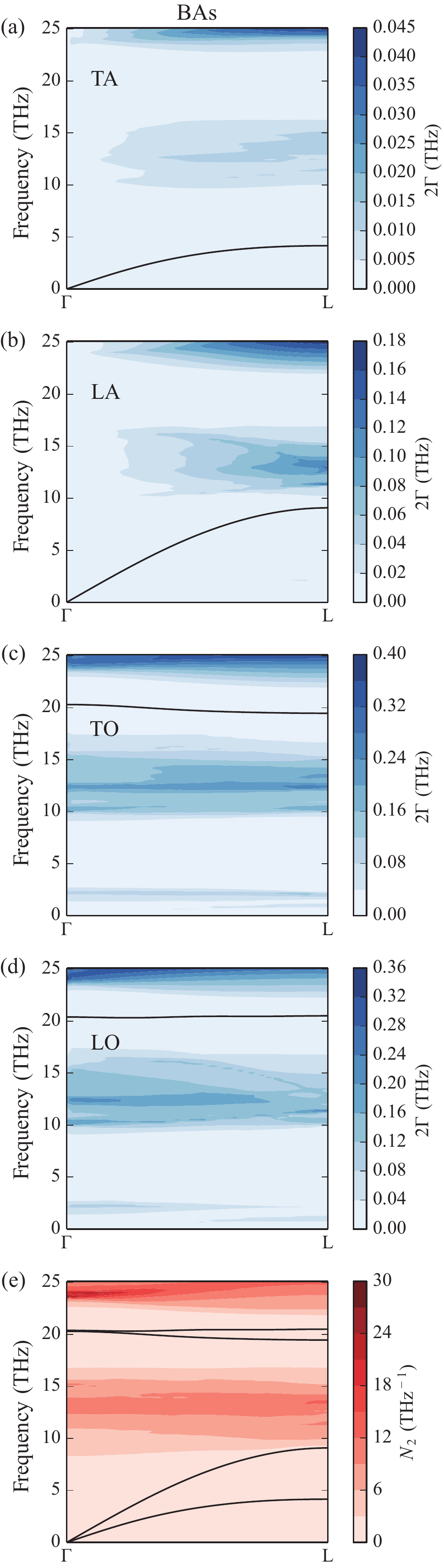}
  \caption{(color online) \label{fig:gamma-jdos-BAs} $(\mathbf{q}, \omega)$ maps of imaginary parts of
  self energies 2$\Gamma_\lambda(\omega)$ and w-JDOS $N_2(\mathbf{q},\omega)$ of BAs.}
 \end{center}
\end{figure}

\begin{figure}[ht]
 \begin{center}
  \includegraphics[height=0.80\textheight]{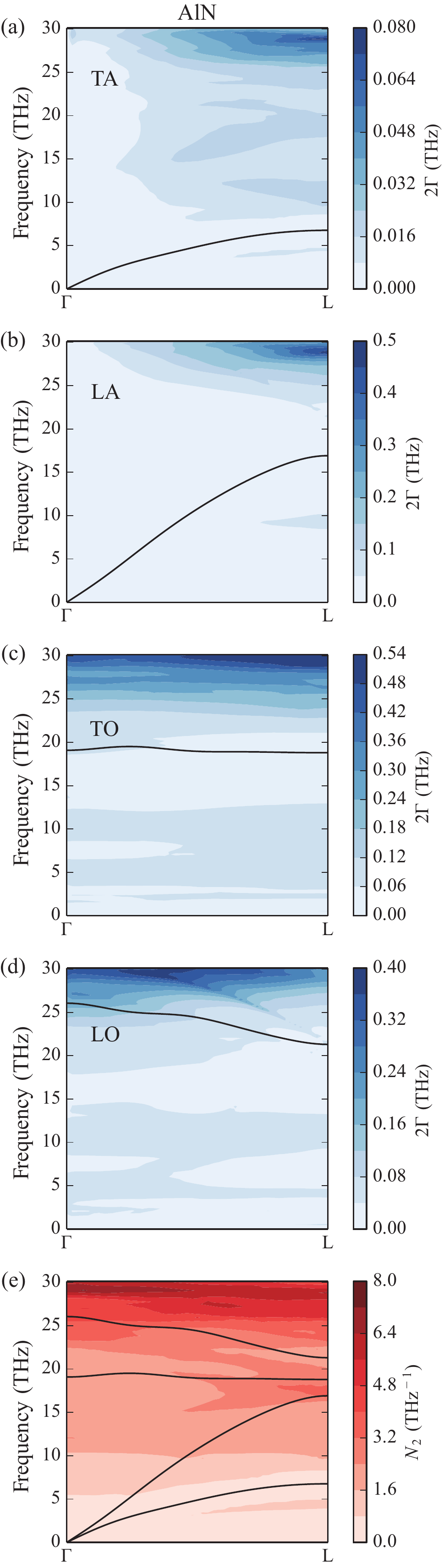}
  \caption{(color online) \label{fig:gamma-jdos-AlN} $(\mathbf{q}, \omega)$ maps of imaginary parts of
  self energies 2$\Gamma_\lambda(\omega)$ and w-JDOS $N_2(\mathbf{q},\omega)$ of AlN.}
 \end{center}
\end{figure}

\begin{figure}[ht]
 \begin{center}
  \includegraphics[height=0.80\textheight]{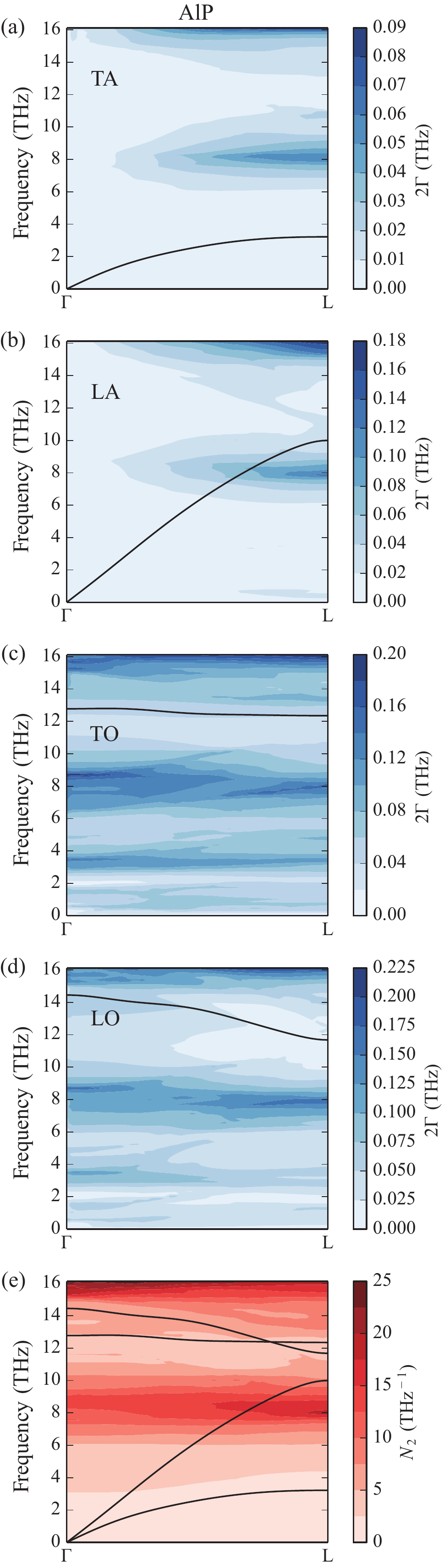}
  \caption{(color online) \label{fig:gamma-jdos-AlP} $(\mathbf{q}, \omega)$ maps of imaginary parts of
  self energies 2$\Gamma_\lambda(\omega)$ and w-JDOS $N_2(\mathbf{q},\omega)$ of AlP.}
 \end{center}
\end{figure}

\begin{figure}[ht]
 \begin{center}
  \includegraphics[height=0.80\textheight]{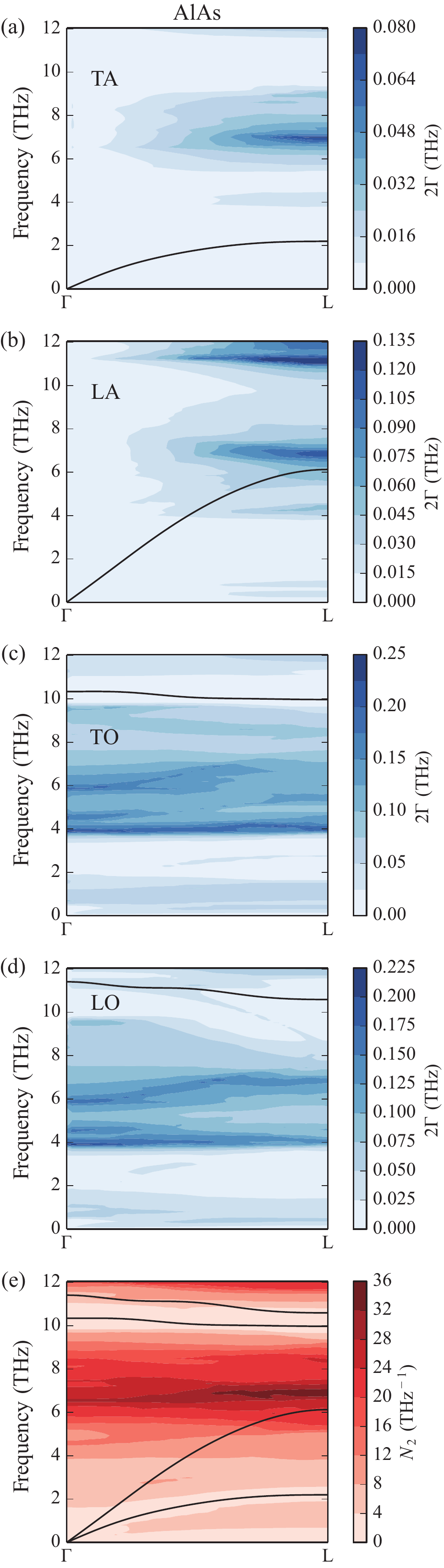}
  \caption{(color online) \label{fig:gamma-jdos-AlAs} $(\mathbf{q}, \omega)$ maps of imaginary parts of
  self energies 2$\Gamma_\lambda(\omega)$ and w-JDOS $N_2(\mathbf{q},\omega)$ of AlAs.}
 \end{center}
\end{figure}

\begin{figure}[ht]
 \begin{center}
  \includegraphics[height=0.80\textheight]{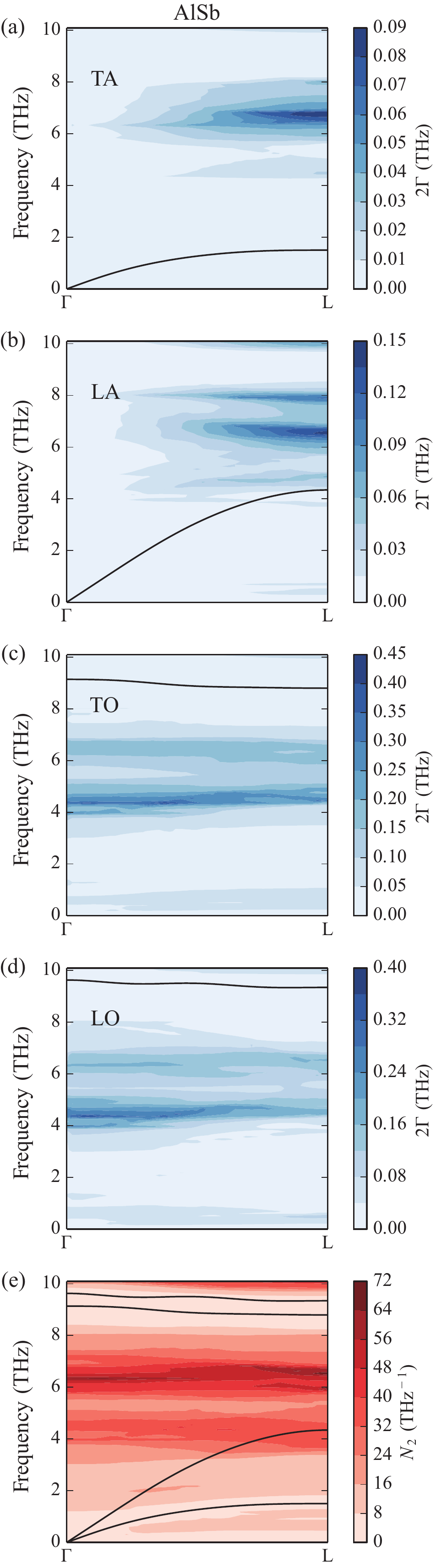}
  \caption{(color online) \label{fig:gamma-jdos-AlSb} $(\mathbf{q}, \omega)$ maps of imaginary parts of
  self energies 2$\Gamma_\lambda(\omega)$ and w-JDOS $N_2(\mathbf{q},\omega)$ of AlSb.}
 \end{center}
\end{figure}

\begin{figure}[ht]
 \begin{center}
  \includegraphics[height=0.80\textheight]{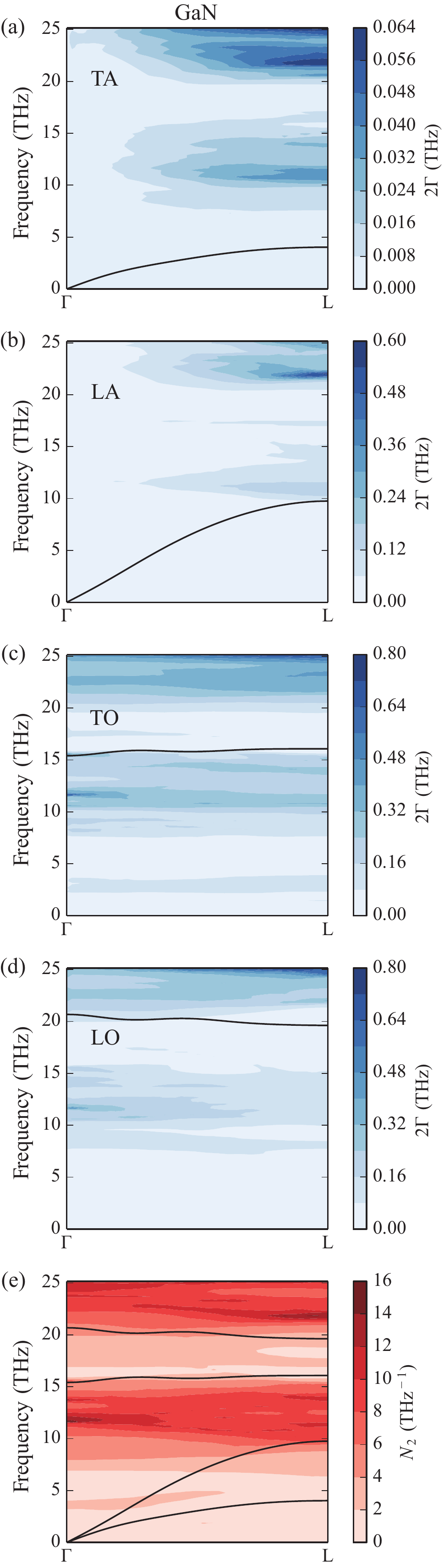}
  \caption{(color online) \label{fig:gamma-jdos-GaN} $(\mathbf{q}, \omega)$ maps of imaginary parts of
  self energies 2$\Gamma_\lambda(\omega)$ and w-JDOS $N_2(\mathbf{q},\omega)$ of GaN.}
 \end{center}
\end{figure}

\begin{figure}[ht]
 \begin{center}
  \includegraphics[height=0.80\textheight]{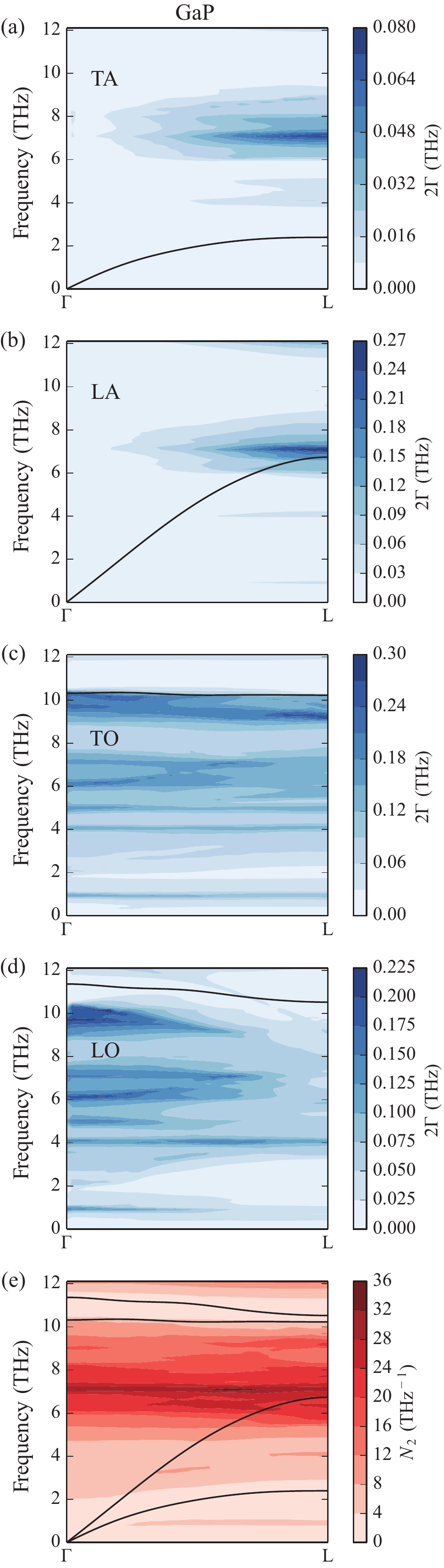}
  \caption{(color online) \label{fig:gamma-jdos-GaP} $(\mathbf{q}, \omega)$ maps of imaginary parts of
  self energies 2$\Gamma_\lambda(\omega)$ and w-JDOS $N_2(\mathbf{q},\omega)$ of GaP.}
 \end{center}
\end{figure}

\begin{figure}[ht]
 \begin{center}
  \includegraphics[height=0.80\textheight]{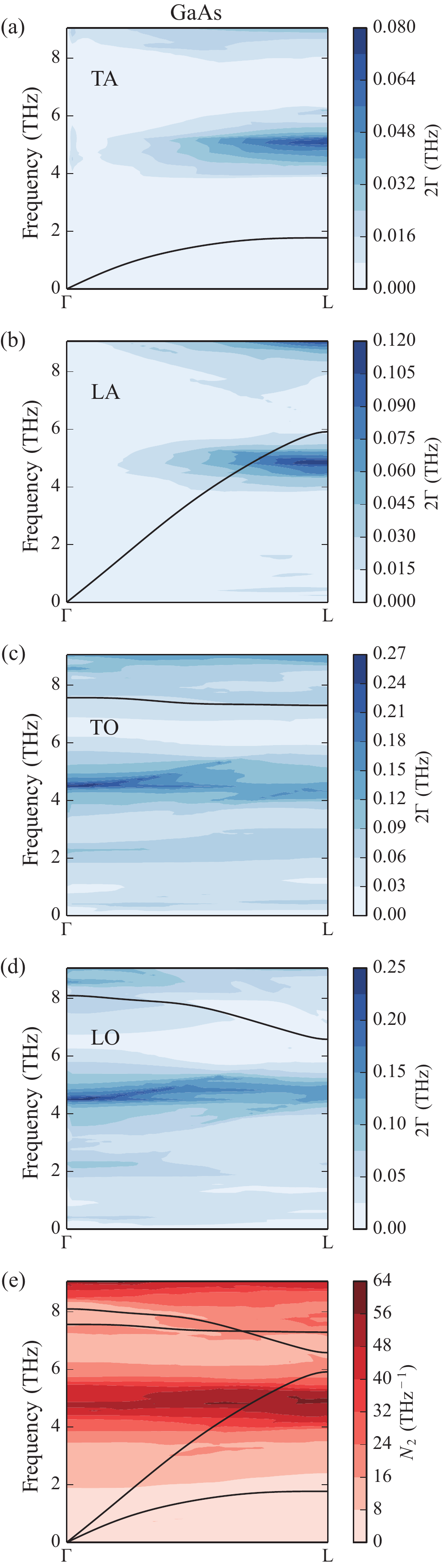}
  \caption{(color online) \label{fig:gamma-jdos-GaAs} $(\mathbf{q}, \omega)$ maps of imaginary parts of
  self energies 2$\Gamma_\lambda(\omega)$ and w-JDOS $N_2(\mathbf{q},\omega)$ of GaAs.}
 \end{center}
\end{figure}

\begin{figure}[ht]
 \begin{center}
  \includegraphics[height=0.80\textheight]{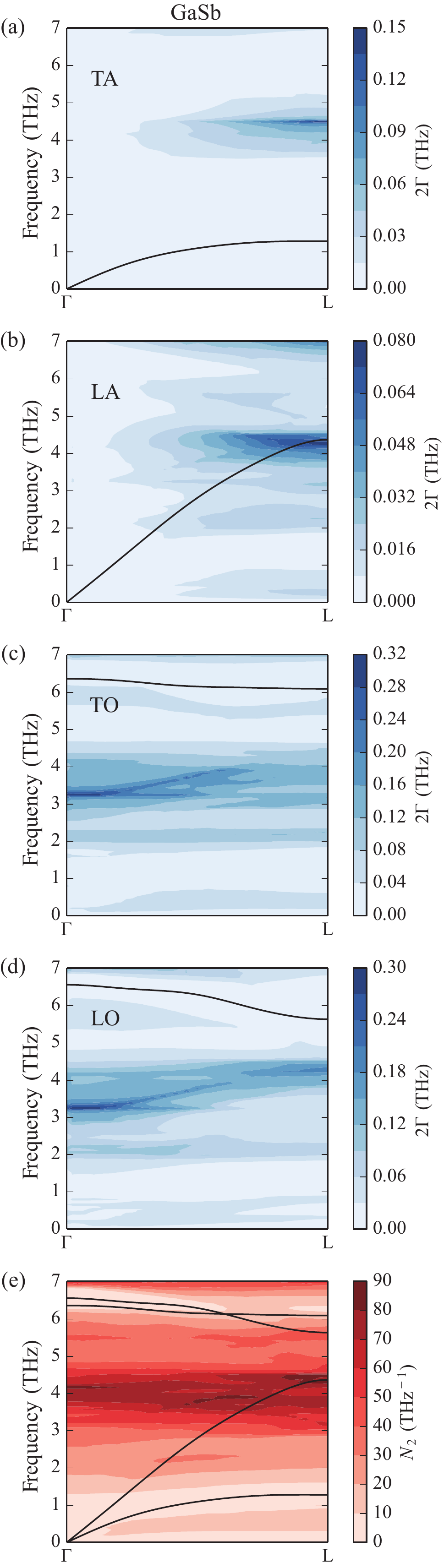}
  \caption{(color online) \label{fig:gamma-jdos-GaSb} $(\mathbf{q}, \omega)$ maps of imaginary parts of
  self energies 2$\Gamma_\lambda(\omega)$ and w-JDOS $N_2(\mathbf{q},\omega)$ of GaSb.}
 \end{center}
\end{figure}

\begin{figure}[ht]
 \begin{center}
  \includegraphics[height=0.80\textheight]{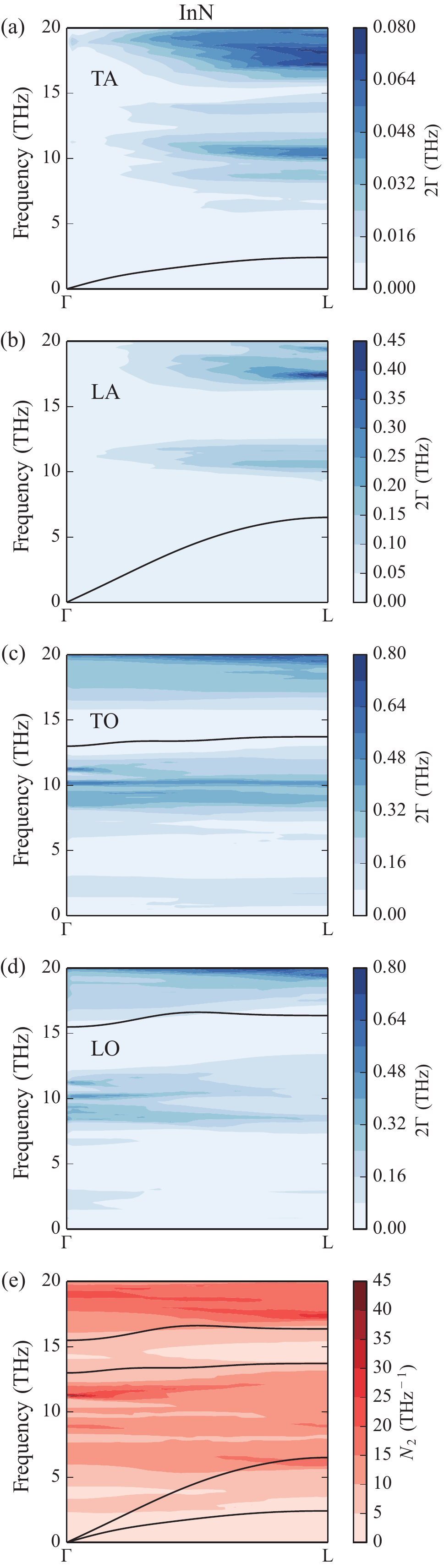}
  \caption{(color online) \label{fig:gamma-jdos-InN} $(\mathbf{q}, \omega)$ maps of imaginary parts of
  self energies 2$\Gamma_\lambda(\omega)$ and w-JDOS $N_2(\mathbf{q},\omega)$ of InN.}
 \end{center}
\end{figure}

\begin{figure}[ht]
 \begin{center}
  \includegraphics[height=0.80\textheight]{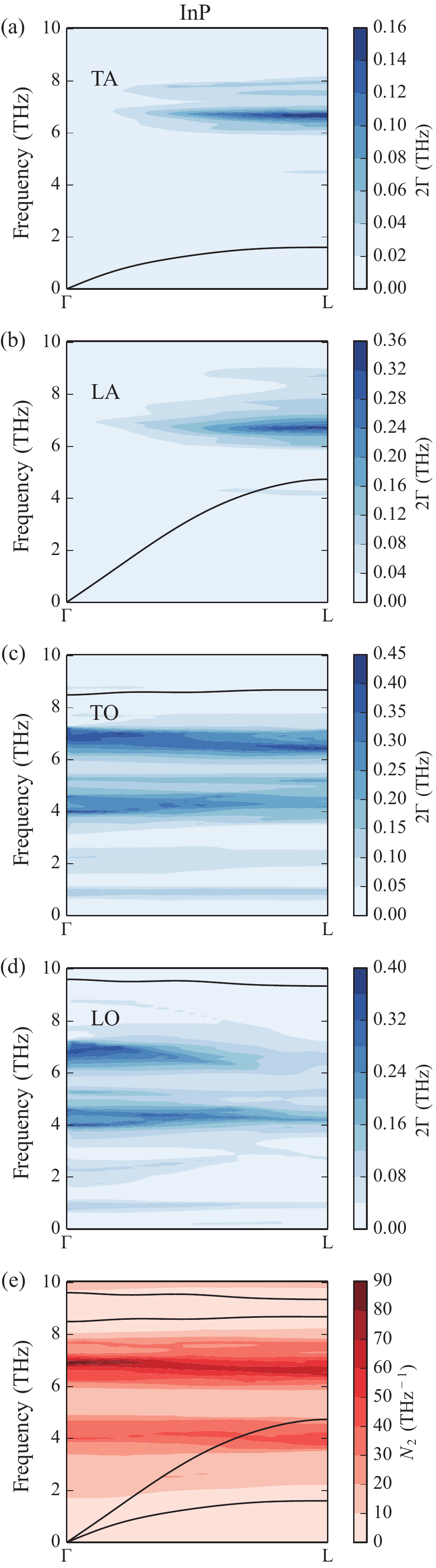}
  \caption{(color online) \label{fig:gamma-jdos-InP} $(\mathbf{q}, \omega)$ maps of imaginary parts of
  self energies 2$\Gamma_\lambda(\omega)$ and w-JDOS $N_2(\mathbf{q},\omega)$ of InP.}
 \end{center}
\end{figure}

\begin{figure}[ht]
 \begin{center}
  \includegraphics[height=0.80\textheight]{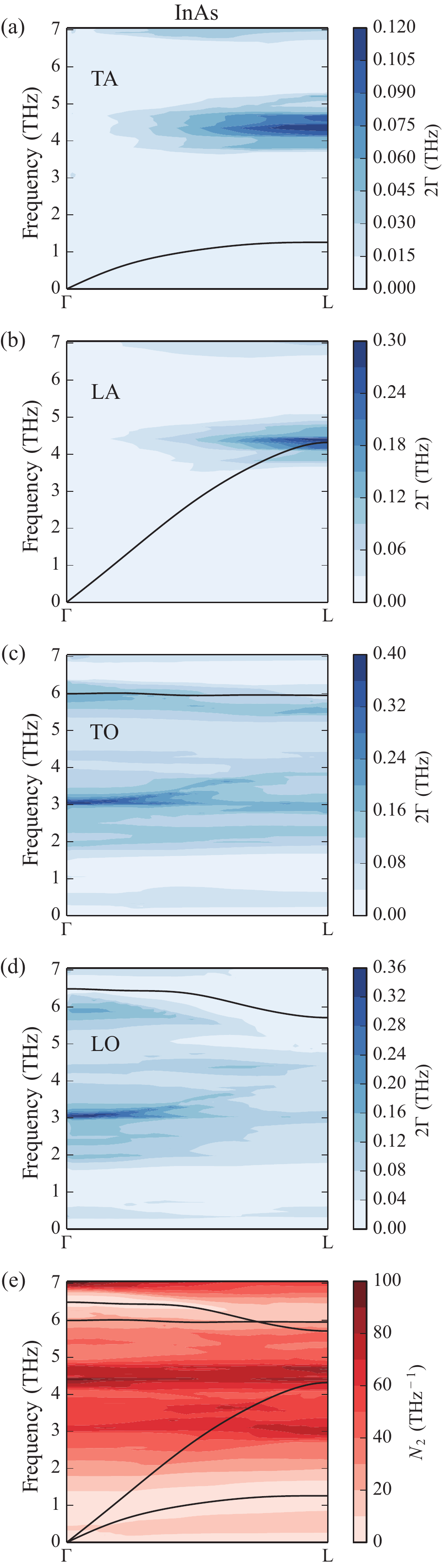}
  \caption{(color online) \label{fig:gamma-jdos-InAs} $(\mathbf{q}, \omega)$ maps of imaginary parts of
  self energies 2$\Gamma_\lambda(\omega)$ and w-JDOS $N_2(\mathbf{q},\omega)$ of InAs.}
 \end{center}
\end{figure}

\begin{figure}[ht]
 \begin{center}
  \includegraphics[height=0.80\textheight]{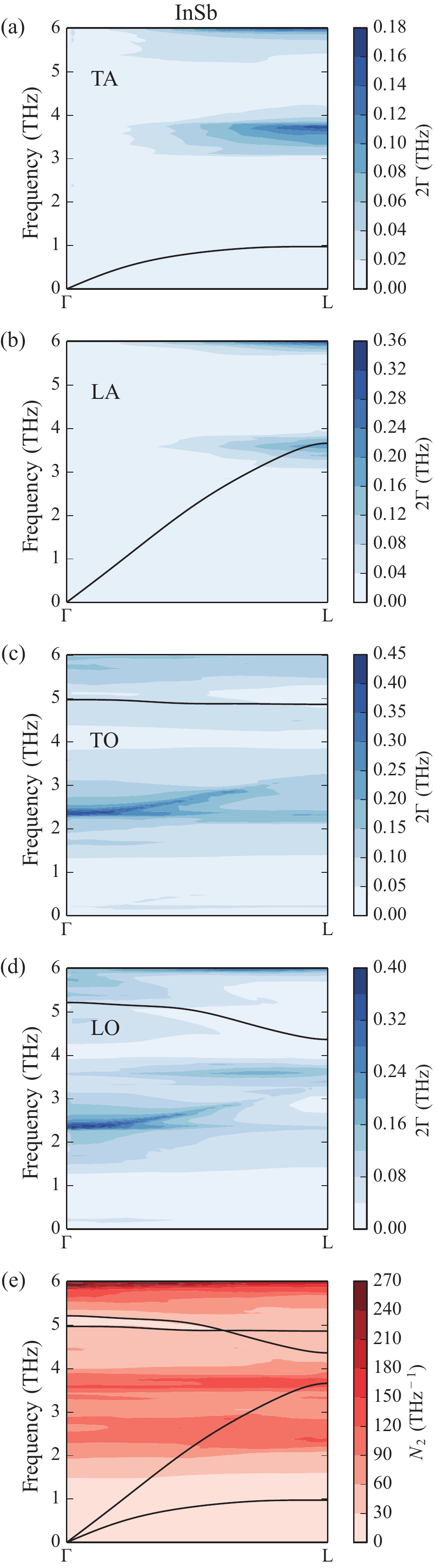}
  \caption{(color online) \label{fig:gamma-jdos-InSb} $(\mathbf{q}, \omega)$ maps of imaginary parts of
  self energies 2$\Gamma_\lambda(\omega)$ and w-JDOS $N_2(\mathbf{q},\omega)$ of InSb.}
 \end{center}
\end{figure}

\begin{figure}[ht]
 \begin{center}
  \includegraphics[height=0.80\textheight]{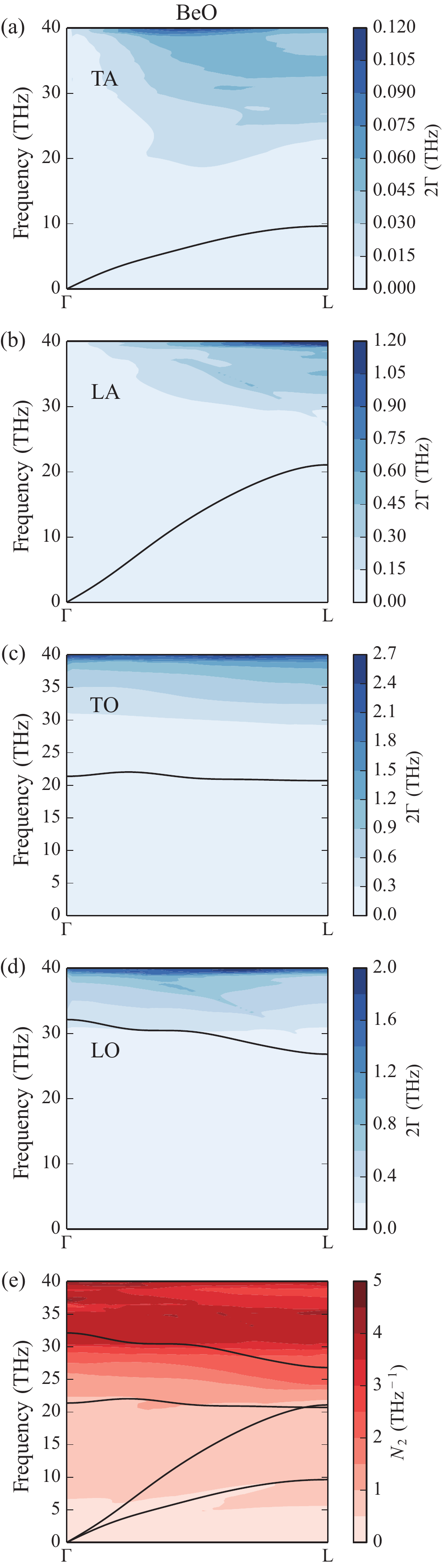}
  \caption{(color online) \label{fig:gamma-jdos-BeO} $(\mathbf{q}, \omega)$ maps of imaginary parts of
  self energies 2$\Gamma_\lambda(\omega)$ and w-JDOS $N_2(\mathbf{q},\omega)$ of BeO.}
 \end{center}
\end{figure}

\begin{figure}[ht]
 \begin{center}
  \includegraphics[height=0.80\textheight]{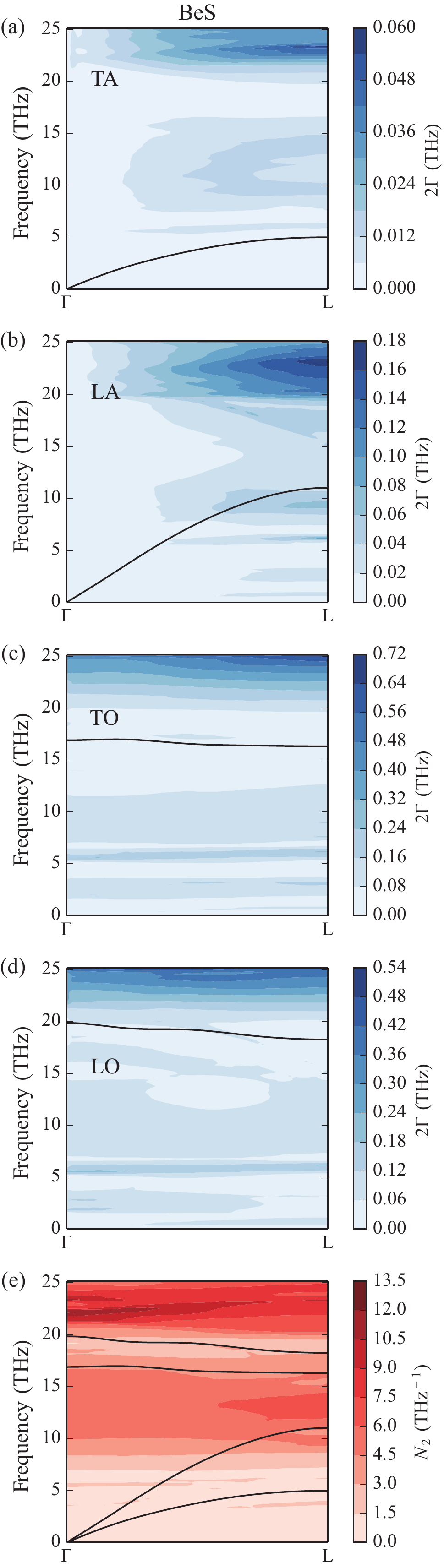}
  \caption{(color online) \label{fig:gamma-jdos-BeS} $(\mathbf{q}, \omega)$ maps of imaginary parts of
  self energies 2$\Gamma_\lambda(\omega)$ and w-JDOS $N_2(\mathbf{q},\omega)$ of BeS.}
 \end{center}
\end{figure}

\begin{figure}[ht]
 \begin{center}
  \includegraphics[height=0.80\textheight]{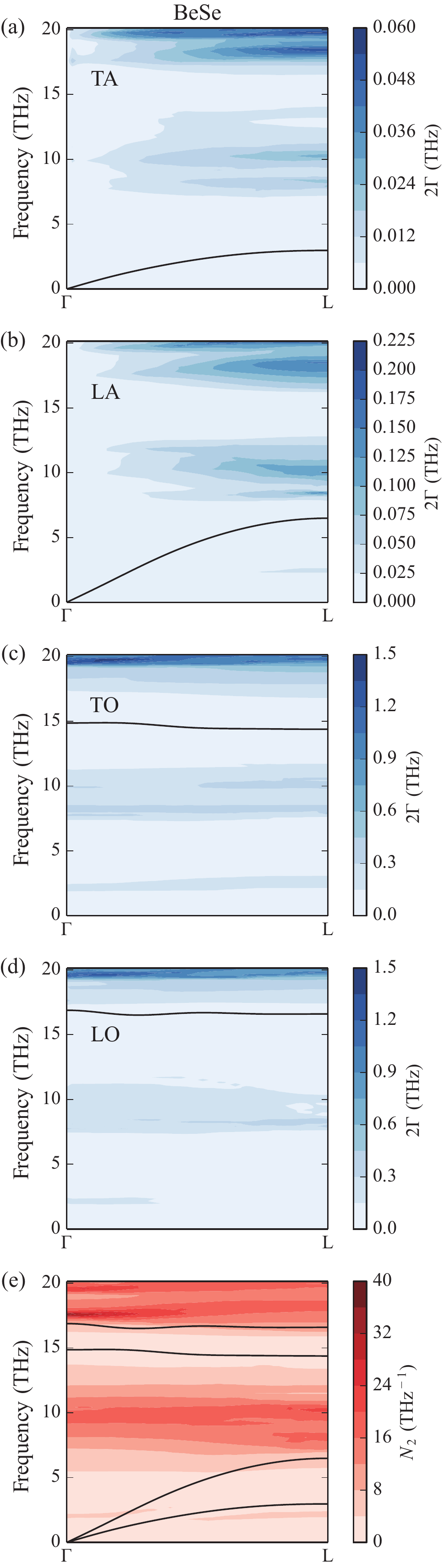}
  \caption{(color online) \label{fig:gamma-jdos-BeSe} $(\mathbf{q}, \omega)$ maps of imaginary parts of
  self energies 2$\Gamma_\lambda(\omega)$ and w-JDOS $N_2(\mathbf{q},\omega)$ of BeSe.}
 \end{center}
\end{figure}

\begin{figure}[ht]
 \begin{center}
  \includegraphics[height=0.80\textheight]{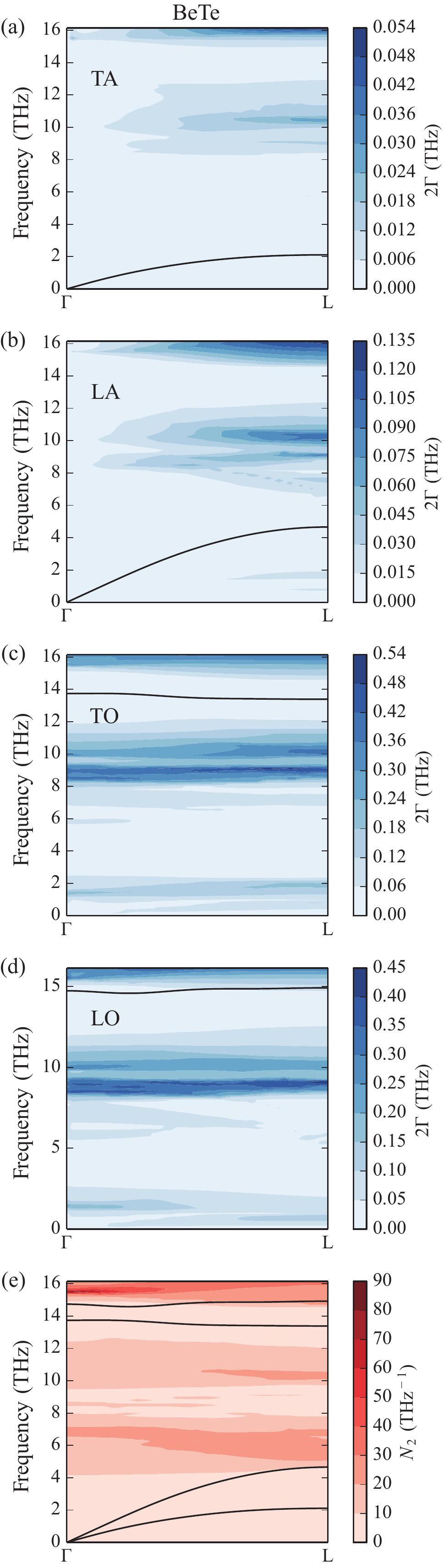}
  \caption{(color online) \label{fig:gamma-jdos-BeTe} $(\mathbf{q}, \omega)$ maps of imaginary parts of
  self energies 2$\Gamma_\lambda(\omega)$ and w-JDOS $N_2(\mathbf{q},\omega)$ of BeTe.}
 \end{center}
\end{figure}

\begin{figure}[ht]
 \begin{center}
  \includegraphics[height=0.80\textheight]{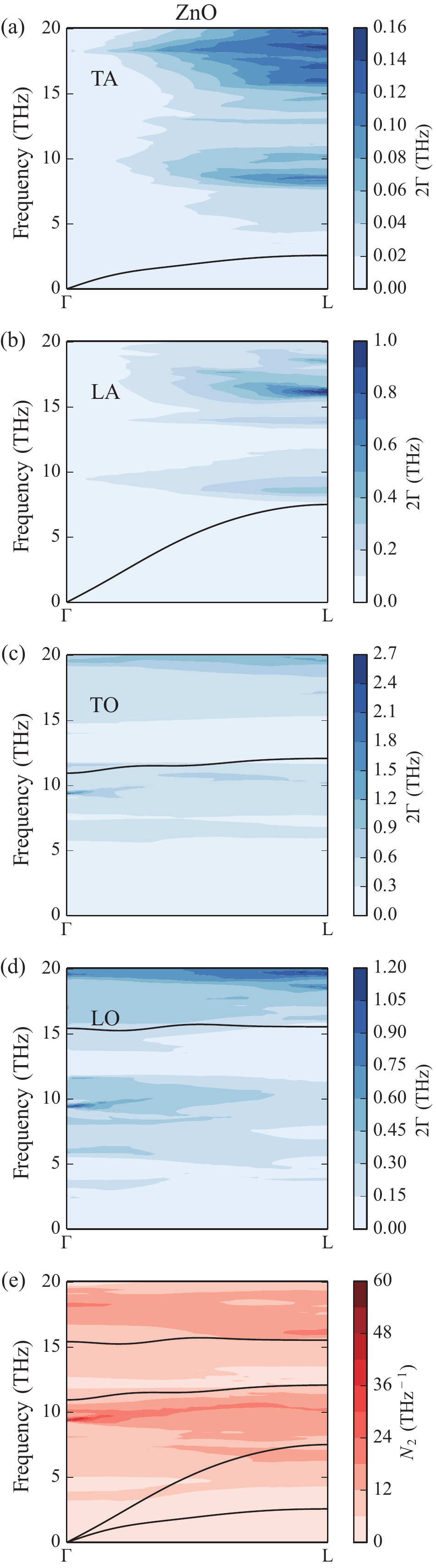}
  \caption{(color online) \label{fig:gamma-jdos-ZnO} $(\mathbf{q}, \omega)$ maps of imaginary parts of
  self energies 2$\Gamma_\lambda(\omega)$ and w-JDOS $N_2(\mathbf{q},\omega)$ of ZnO.}
 \end{center}
\end{figure}

\begin{figure}[ht]
 \begin{center}
  \includegraphics[height=0.80\textheight]{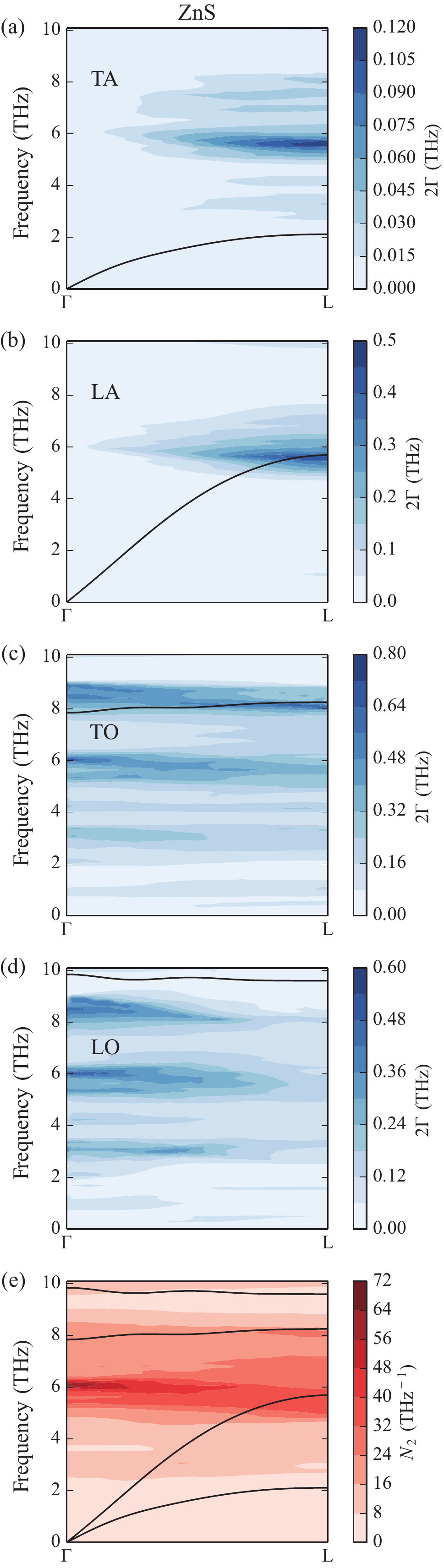}
  \caption{(color online) \label{fig:gamma-jdos-ZnS} $(\mathbf{q}, \omega)$ maps of imaginary parts of
  self energies 2$\Gamma_\lambda(\omega)$ and w-JDOS $N_2(\mathbf{q},\omega)$ of ZnS.}
 \end{center}
\end{figure}

\begin{figure}[ht]
 \begin{center}
  \includegraphics[height=0.80\textheight]{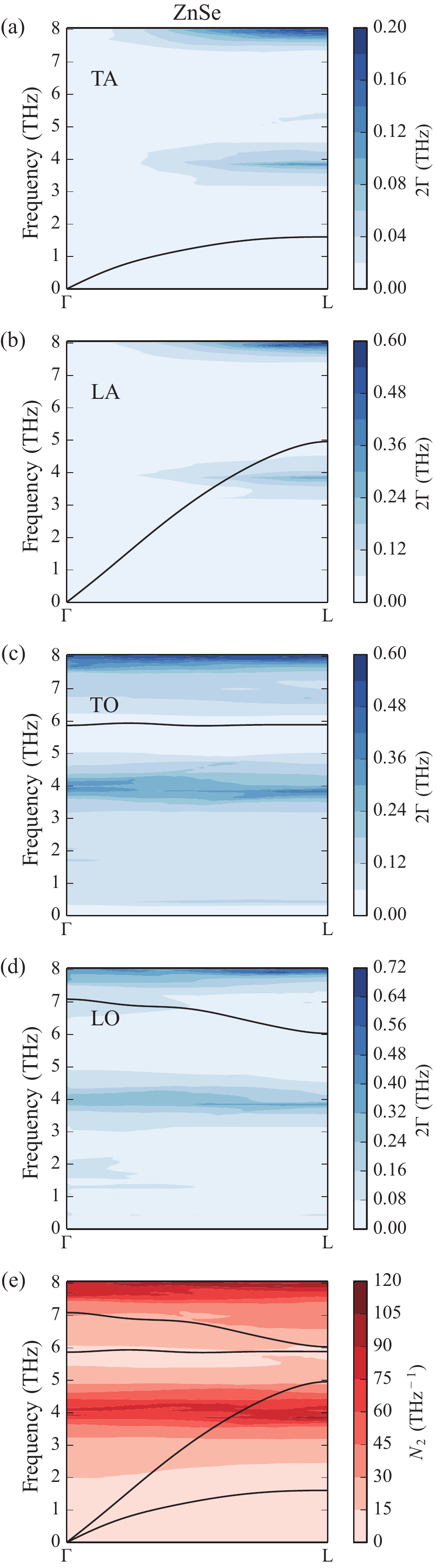}
  \caption{(color online) \label{fig:gamma-jdos-ZnSe} $(\mathbf{q}, \omega)$ maps of imaginary parts of
  self energies 2$\Gamma_\lambda(\omega)$ and w-JDOS $N_2(\mathbf{q},\omega)$ of ZnSe.}
 \end{center}
\end{figure}

\begin{figure}[ht]
 \begin{center}
  \includegraphics[height=0.80\textheight]{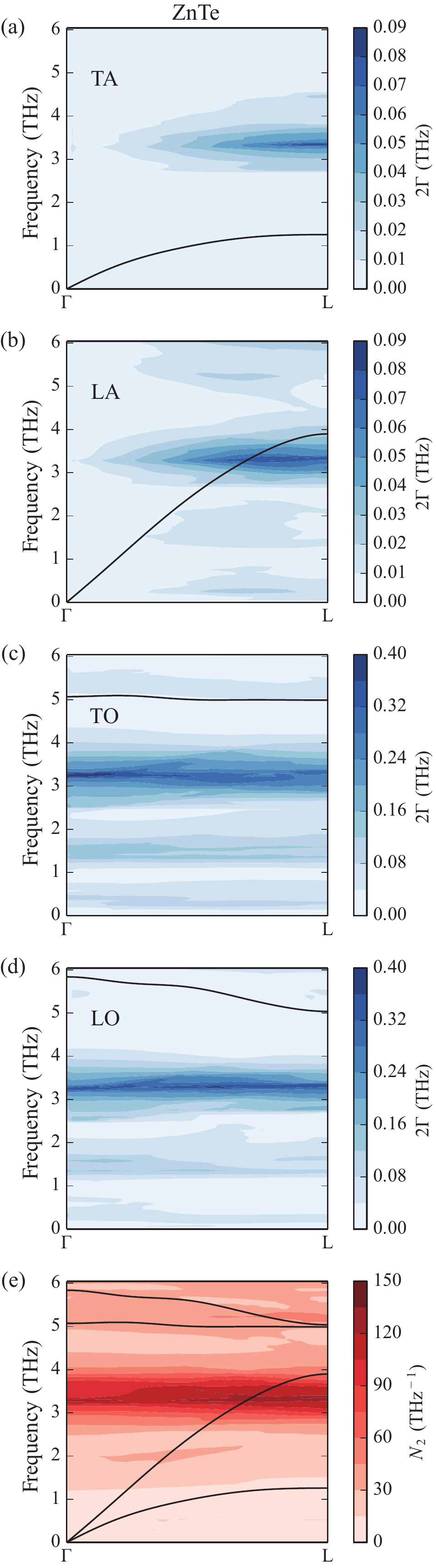}
  \caption{(color online) \label{fig:gamma-jdos-ZnTe} $(\mathbf{q}, \omega)$ maps of imaginary parts of
  self energies 2$\Gamma_\lambda(\omega)$ and w-JDOS $N_2(\mathbf{q},\omega)$ of ZnTe.}
 \end{center}
\end{figure}

\begin{figure}[ht]
 \begin{center}
  \includegraphics[height=0.80\textheight]{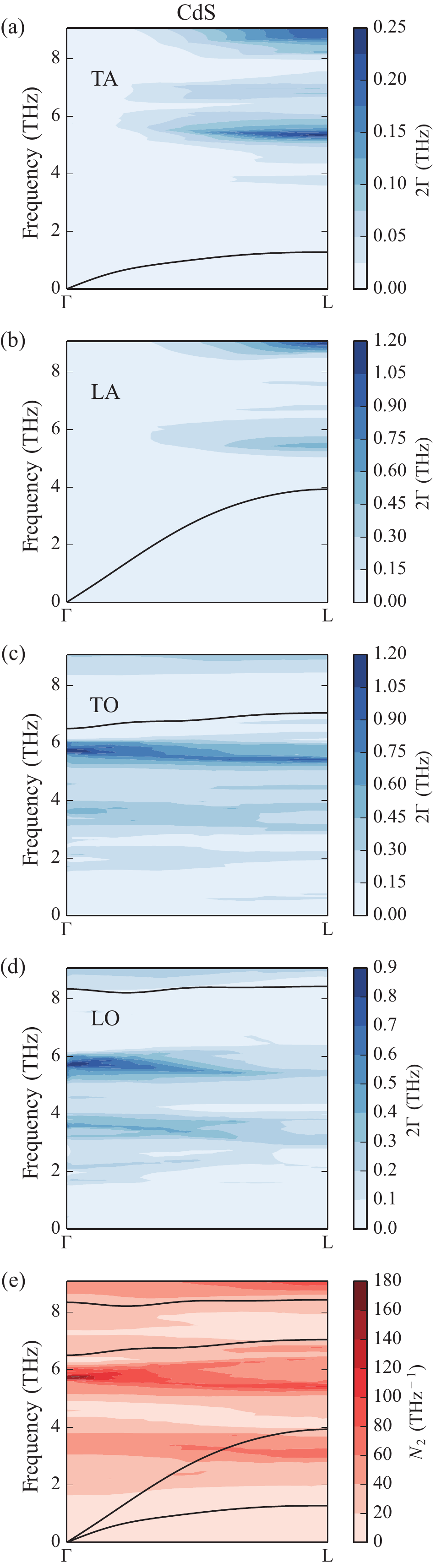}
  \caption{(color online) \label{fig:gamma-jdos-CdS} $(\mathbf{q}, \omega)$ maps of imaginary parts of
  self energies 2$\Gamma_\lambda(\omega)$ and w-JDOS $N_2(\mathbf{q},\omega)$ of CdS.}
 \end{center}
\end{figure}

\begin{figure}[ht]
 \begin{center}
  \includegraphics[height=0.80\textheight]{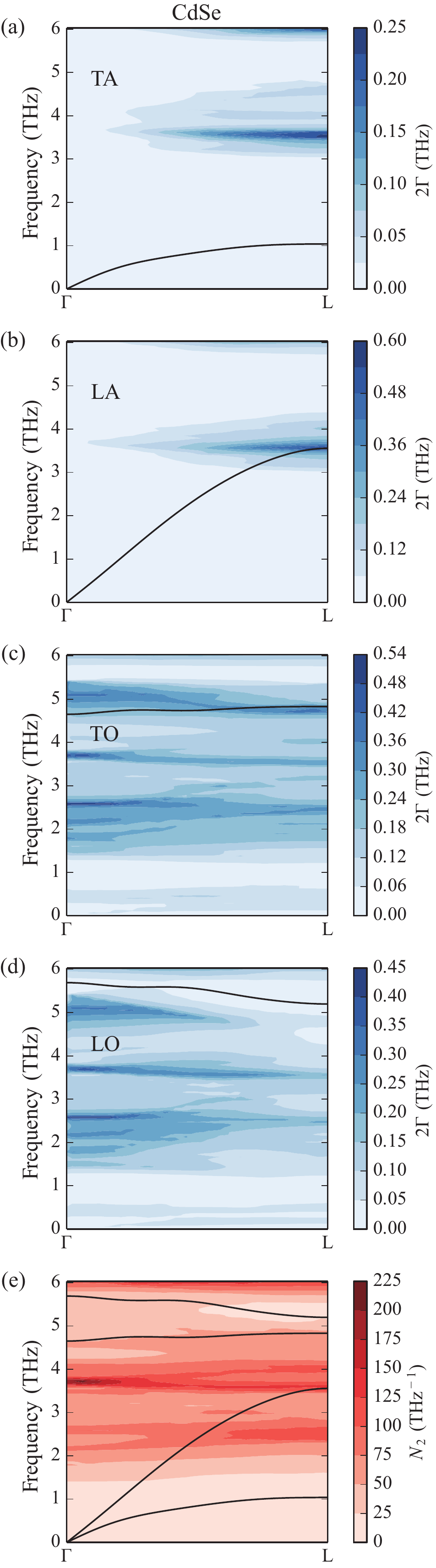}
  \caption{(color online) \label{fig:gamma-jdos-CdSe} $(\mathbf{q}, \omega)$ maps of imaginary parts of
  self energies 2$\Gamma_\lambda(\omega)$ and w-JDOS $N_2(\mathbf{q},\omega)$ of CdSe.}
 \end{center}
\end{figure}

\begin{figure}[ht]
 \begin{center}
  \includegraphics[height=0.80\textheight]{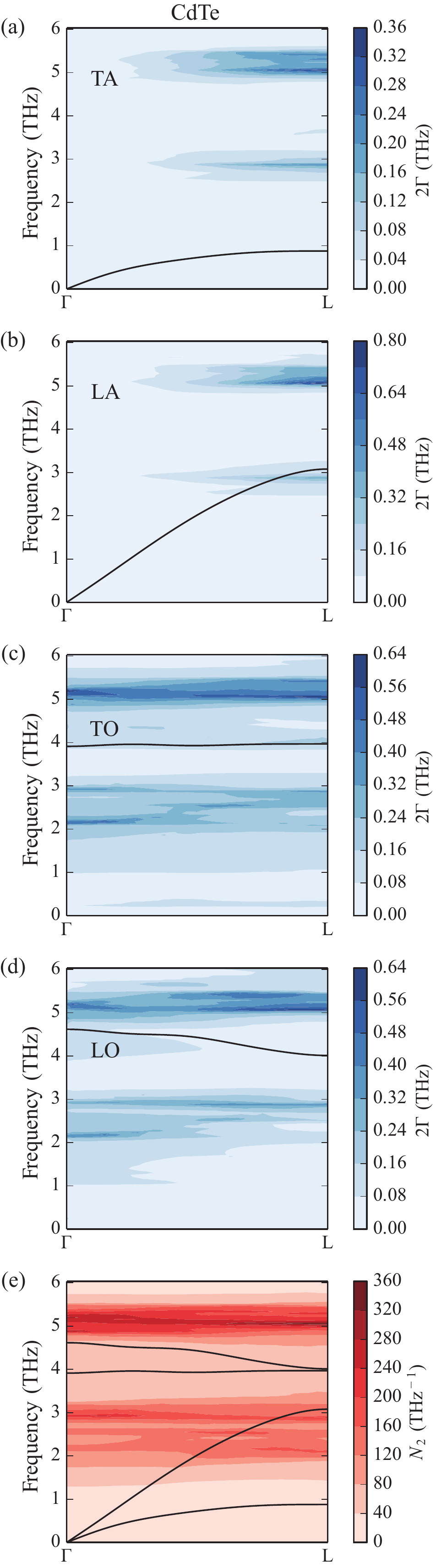}
  \caption{(color online) \label{fig:gamma-jdos-CdTe} $(\mathbf{q}, \omega)$ maps of imaginary parts of
  self energies 2$\Gamma_\lambda(\omega)$ and w-JDOS $N_2(\mathbf{q},\omega)$ of CdTe.}
 \end{center}
\end{figure}

\begin{figure}[ht]
 \begin{center}
  \includegraphics[height=0.80\textheight]{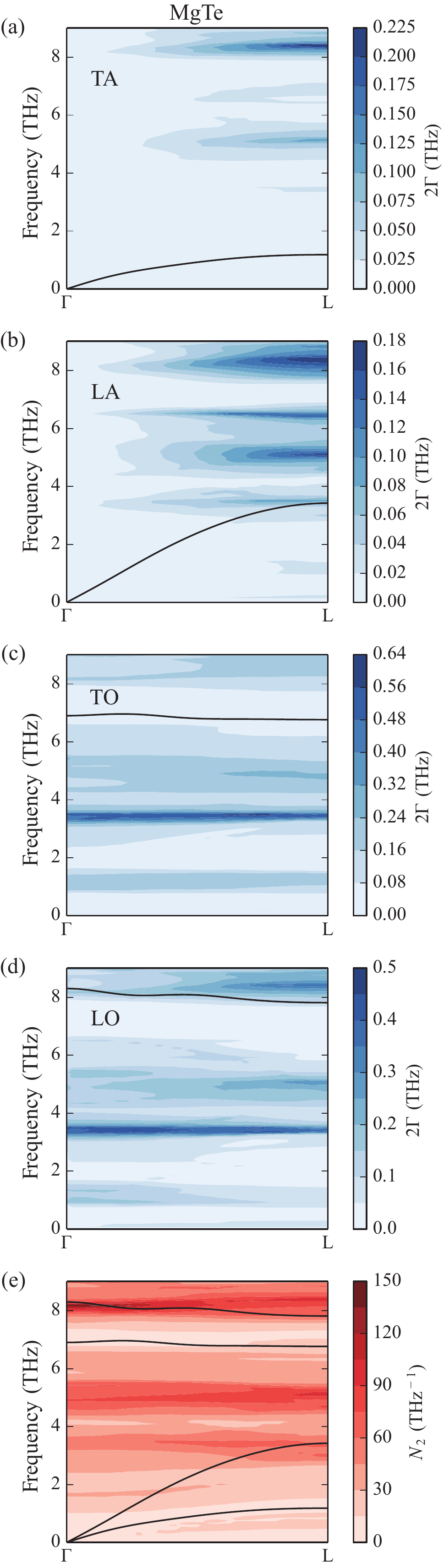}
  \caption{(color online) \label{fig:gamma-jdos-MgTe} $(\mathbf{q}, \omega)$ maps of imaginary parts of
  self energies 2$\Gamma_\lambda(\omega)$ and w-JDOS $N_2(\mathbf{q},\omega)$ of MgTe.}
 \end{center}
\end{figure}

\begin{figure}[ht]
 \begin{center}
  \includegraphics[height=0.80\textheight]{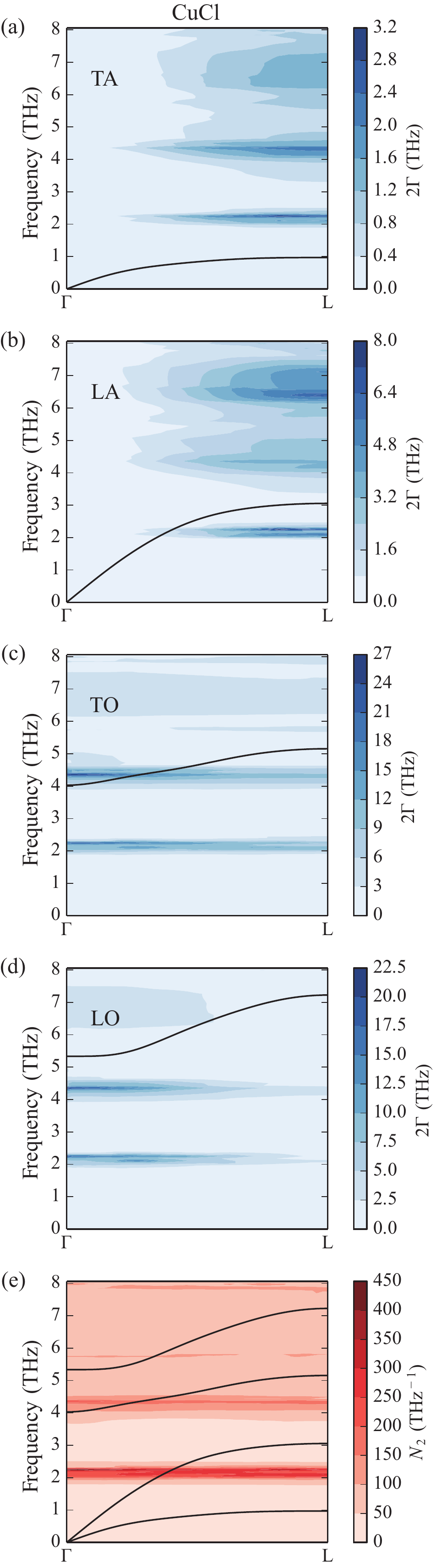}
  \caption{(color online) \label{fig:gamma-jdos-CuCl} $(\mathbf{q}, \omega)$ maps of imaginary parts of
  self energies 2$\Gamma_\lambda(\omega)$ and w-JDOS $N_2(\mathbf{q},\omega)$ of CuCl.}
 \end{center}
\end{figure}

\begin{figure}[ht]
 \begin{center}
  \includegraphics[height=0.80\textheight]{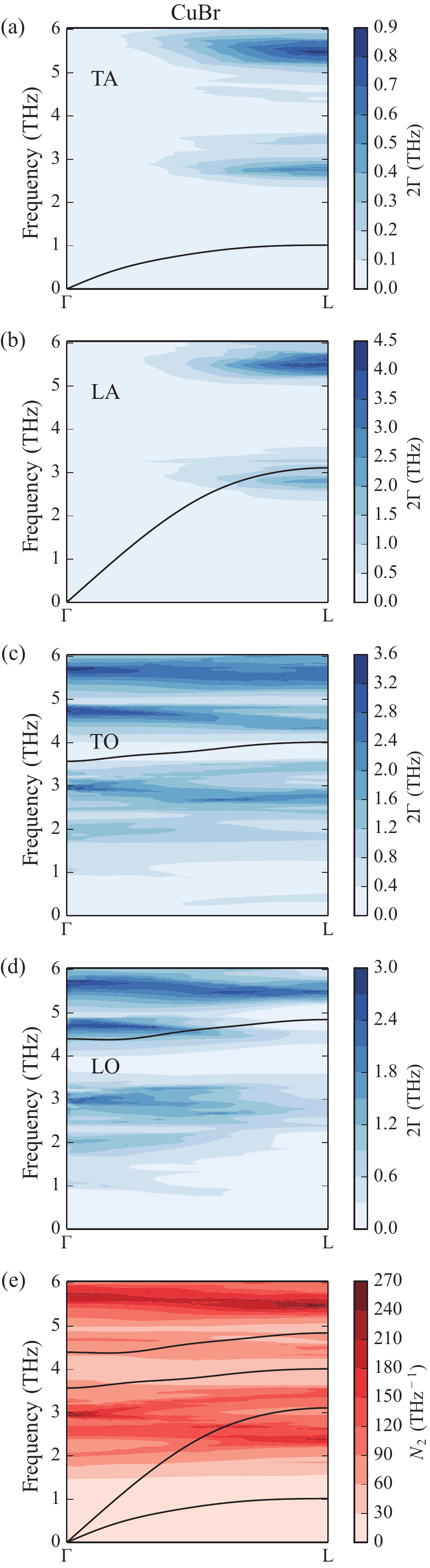}
  \caption{(color online) \label{fig:gamma-jdos-CuBr} $(\mathbf{q}, \omega)$ maps of imaginary parts of
  self energies 2$\Gamma_\lambda(\omega)$ and w-JDOS $N_2(\mathbf{q},\omega)$ of CuBr.}
 \end{center}
\end{figure}

\begin{figure}[ht]
 \begin{center}
  \includegraphics[height=0.80\textheight]{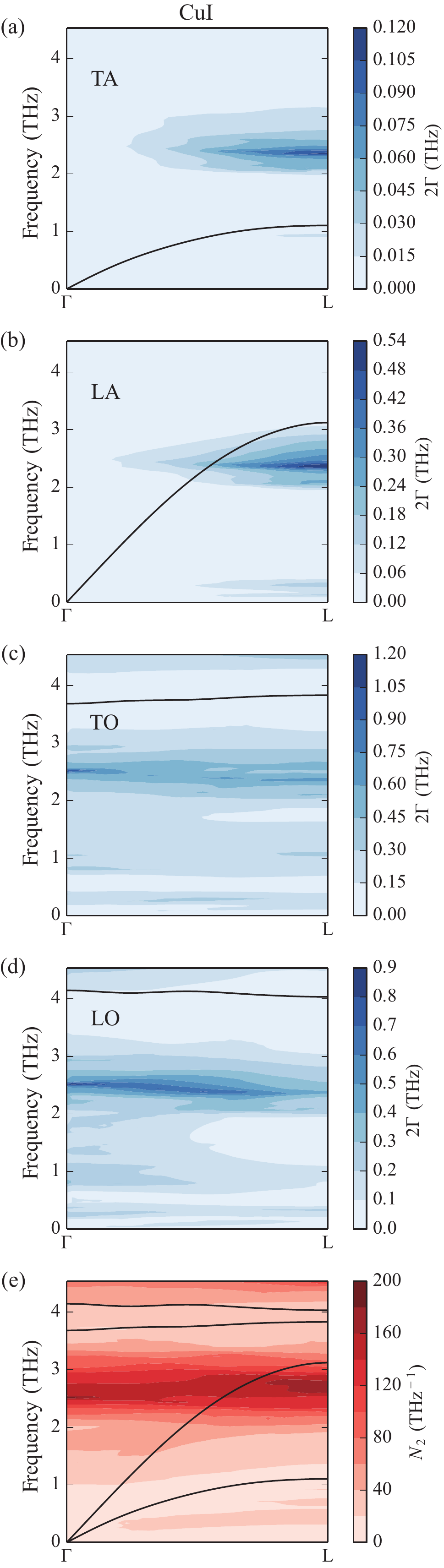}
  \caption{(color online) \label{fig:gamma-jdos-CuI} $(\mathbf{q}, \omega)$ maps of imaginary parts of
  self energies 2$\Gamma_\lambda(\omega)$ and w-JDOS $N_2(\mathbf{q},\omega)$ of CuI.}
 \end{center}
\end{figure}

\begin{figure}[ht]
 \begin{center}
  \includegraphics[height=0.80\textheight]{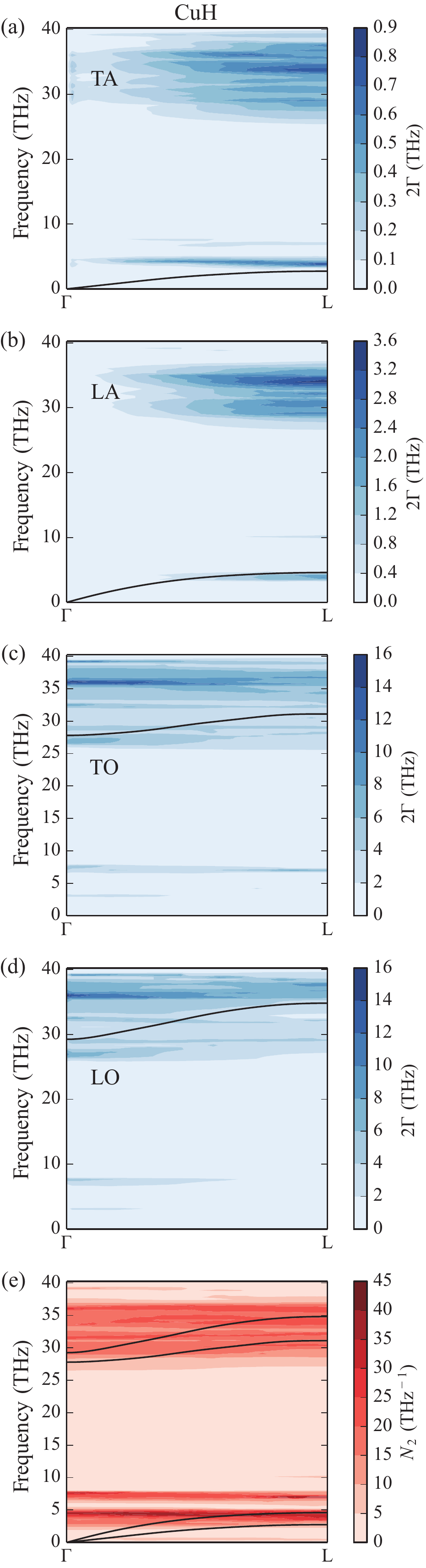}
  \caption{(color online) \label{fig:gamma-jdos-CuH} $(\mathbf{q}, \omega)$ maps of imaginary parts of
  self energies 2$\Gamma_\lambda(\omega)$ and w-JDOS $N_2(\mathbf{q},\omega)$ of CuH.}
 \end{center}
\end{figure}

\begin{figure}[ht]
 \begin{center}
  \includegraphics[height=0.80\textheight]{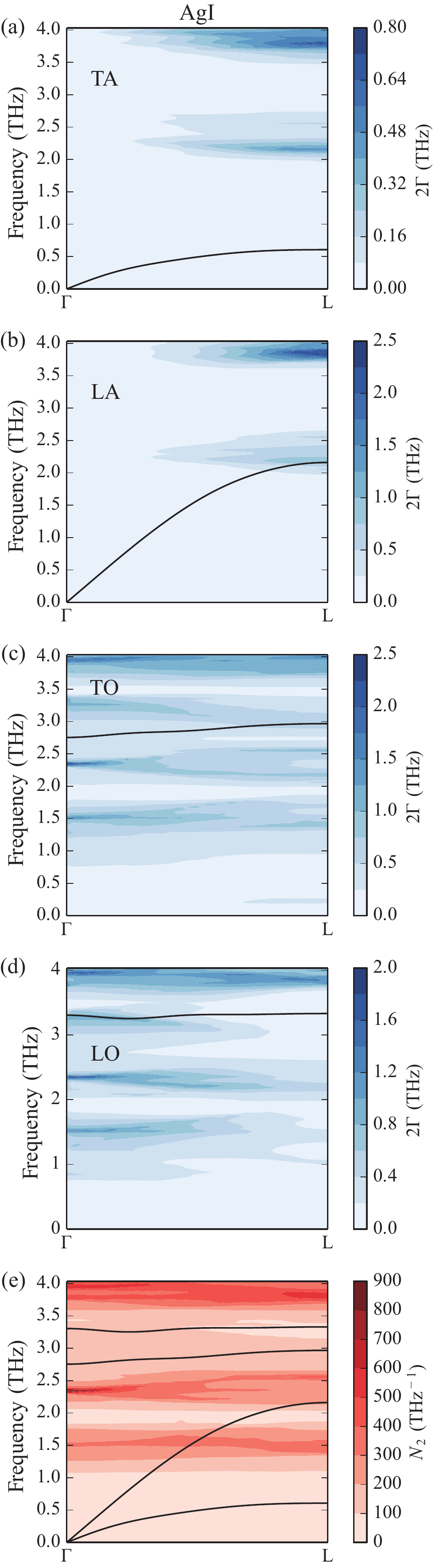}
  \caption{(color online) \label{fig:gamma-jdos-AgI} $(\mathbf{q}, \omega)$ maps of imaginary parts of
  self energies 2$\Gamma_\lambda(\omega)$ and w-JDOS $N_2(\mathbf{q},\omega)$ of AgI.}
 \end{center}
\end{figure}

\begin{figure}[ht]
 \begin{center}
  \includegraphics[height=0.80\textheight]{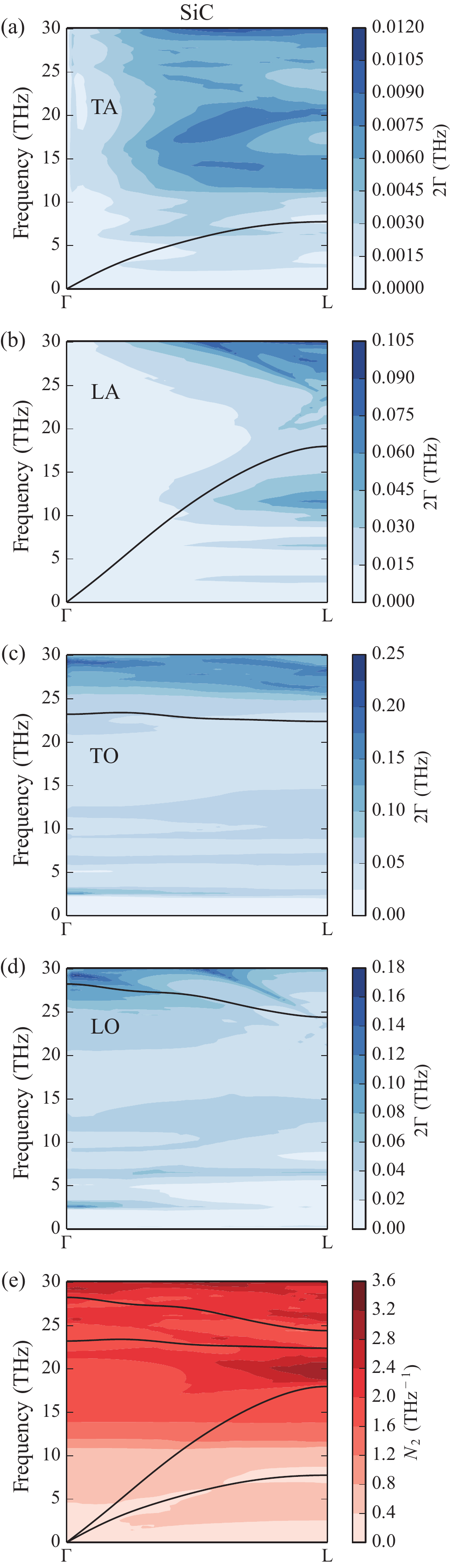}
  \caption{(color online) \label{fig:gamma-jdos-SiC} $(\mathbf{q}, \omega)$ maps of imaginary parts of
  self energies 2$\Gamma_\lambda(\omega)$ and w-JDOS $N_2(\mathbf{q},\omega)$ of SiC.}
 \end{center}
\end{figure}

Figures from \ref{fig:gamma-jdos-BN} to
\ref{fig:gamma-jdos-SiC} present $(\mathbf{q}, \omega)$ maps of
imaginary parts of self energies and w-JDOS of the 33 zincblende-type
compounds used in the present study. Their phonon band structures are
superimposed on the $(\mathbf{q}, \omega)$ maps.

\clearpage

\clearpage

\clearpage

\clearpage

\clearpage

\clearpage

\clearpage

\clearpage

\clearpage

\clearpage

\clearpage

\clearpage

\clearpage

\clearpage

\clearpage

\onecolumngrid
\section{Lattice thermal conductivities}
\label{apx:kappa} Lattice thermal conductivities of the 33 zincblende-
and wurtzite-type compounds calculated with RTA and the method of the
direct solution of LBTE~\cite{Laurent-LBTE-2013} are listed in
Tables~\ref{table:kappa} and \ref{table:kappa-isotope}. Scattering by
the isotopes with the natural
abundance~\cite{Laeter-natunal-abundance-2003} is considered in
Table~\ref{table:kappa-isotope}. Available experimental values in
reports are shown in Table~\ref{table:kappa-isotope} and comparisons
between the experiments and calculations are graphically shown in
Fig.~\ref{fig:kappa-exp-calc}.  The methods to include isotope effect
and the direct solution of LBTE are described in Appendices
\ref{apx:isotope} and \ref{apx:LBTE}, respectively.

\begin{table*}[ht]
 \caption{\label{table:kappa} Calculated lattice thermal conductivities of the
 33 zincblende- and wurtzite-type compounds at 300 K in W/m-K.}
 \begin{ruledtabular}
  \begin{tabular}{lccccccc}
   & \multicolumn{2}{c}{Zincblende-type} && \multicolumn{4}{c}{Wurtzite-type} \\
   \cline{2-3} \cline{5-8}
   & \multicolumn{1}{c}{RTA} & \multicolumn{1}{c}{LBTE} && \multicolumn{2}{c}{RTA} & \multicolumn{2}{c}{LBTE} \\
   \hline
   & $xx$ & $xx$ && $xx$ & $zz$ & $xx$ & $zz$ \\
   \hline
BN   & 1.17$\times 10^3$ & 1.80$\times 10^3$ && $9.30\times10^2$ & $9.24\times10^2$ & $1.23\times10^3$ & $1.04\times10^3$ \\
BP   & 4.27$\times 10^2$ & 4.64$\times 10^2$ && $4.18\times10^2$ & $3.19\times10^2$ & $4.31\times10^2$ & $3.28\times10^2$ \\
BAs  & 1.59$\times 10^3$ & 5.61$\times 10^3$ && $1.44\times10^3$ & $1.09\times10^3$ & $2.38\times10^3$ & $1.21\times10^3$ \\
AlN  & 2.12$\times 10^2$ & 2.35$\times 10^2$ && $2.24\times10^2$ & $2.07\times10^2$ & $2.40\times10^2$ & $2.12\times10^2$ \\
AlP  & 8.30$\times 10^1$ & 8.40$\times 10^1$ && $7.48\times10^1$ & $7.14\times10^1$ & $7.46\times10^1$ & $7.17\times10^1$ \\
AlAs & 8.56$\times 10^1$ & 8.68$\times 10^1$ && $7.23\times10^1$ & $6.96\times10^1$ & $7.39\times10^1$ & $7.10\times10^1$ \\
AlSb & 8.92$\times 10^1$ & 9.48$\times 10^1$ && $6.18\times10^1$ & $5.84\times10^1$ & $6.35\times10^1$ & $5.82\times10^1$ \\
GaN  & 2.31$\times 10^2$ & 3.37$\times 10^2$ && $2.56\times10^2$ & $2.38\times10^2$ & $2.73\times10^2$ & $2.66\times10^2$ \\
GaP  & 9.62$\times 10^1$ & 1.22$\times 10^2$ && $1.04\times10^2$ & $9.08\times10^1$ & $1.11\times10^2$ & $9.63\times10^1$ \\
GaAs & 3.18$\times 10^1$ & 3.31$\times 10^1$ && $2.85\times10^1$ & $2.61\times10^1$ & $2.86\times10^1$ & $2.66\times10^1$ \\
GaSb & 3.30$\times 10^1$ & 3.39$\times 10^1$ && $2.42\times10^1$ & $1.97\times10^1$ & $2.48\times10^1$ & $2.02\times10^1$ \\
InN  & 9.54$\times 10^1$ & 1.08$\times 10^2$ && $1.10\times10^2$ & $1.13\times10^2$ & $1.18\times10^2$ & $1.19\times10^2$ \\
InP  & 7.51$\times 10^1$ & 8.71$\times 10^1$ && $6.87\times10^1$ & $6.57\times10^1$ & $7.02\times10^1$ & $6.93\times10^1$ \\
InAs & 2.05$\times 10^1$ & 2.53$\times 10^1$ && $1.83\times10^1$ & $1.71\times10^1$ & $1.86\times10^1$ & $1.80\times10^1$ \\
InSb & 1.27$\times 10^1$ & 1.40$\times 10^1$ && $9.71\times10^0$ & $9.63\times10^0$ & $9.83\times10^0$ & $9.93\times10^0$ \\
BeO  & 3.58$\times 10^2$ & 4.64$\times 10^2$ && $2.86\times10^2$ & $2.98\times10^2$ & $3.36\times10^2$ & $3.20\times10^2$ \\
BeS  & 1.57$\times 10^2$ & 1.74$\times 10^2$ && $1.29\times10^2$ & $1.21\times10^2$ & $1.37\times10^2$ & $1.27\times10^2$ \\
BeSe & 3.89$\times 10^2$ & 6.75$\times 10^2$ && $3.21\times10^2$ & $3.12\times10^2$ & $4.08\times10^2$ & $3.36\times10^2$ \\
BeTe & 2.86$\times 10^2$ & 5.03$\times 10^2$ && $1.83\times10^2$ & $1.82\times10^2$ & $2.36\times10^2$ & $1.93\times10^2$ \\
MgTe & 1.57$\times 10^1$ & 1.57$\times 10^1$ && $1.22\times10^1$ & $1.36\times10^1$ & $1.24\times10^1$ & $1.35\times10^1$ \\
ZnO  & 5.82$\times 10^1$ & 8.93$\times 10^1$ && $4.63\times10^1$ & $6.24\times10^1$ & $4.66\times10^1$ & $6.48\times10^1$ \\
ZnS  & 3.72$\times 10^1$ & 4.72$\times 10^1$ && $3.73\times10^1$ & $4.03\times10^1$ & $3.89\times10^1$ & $4.69\times10^1$ \\
ZnSe & 1.63$\times 10^1$ & 1.72$\times 10^1$ && $1.50\times10^1$ & $1.53\times10^1$ & $1.51\times10^1$ & $1.57\times10^1$ \\
ZnTe & 1.96$\times 10^1$ & 2.04$\times 10^1$ && $1.45\times10^1$ & $1.55\times10^1$ & $1.45\times10^1$ & $1.58\times10^1$ \\
CdS  & 2.13$\times 10^1$ & 2.30$\times 10^1$ && $1.79\times10^1$ & $2.01\times10^1$ & $1.88\times10^1$ & $2.05\times10^1$ \\
CdSe & 1.10$\times 10^1$ & 1.24$\times 10^1$ && $8.86\times10^0$ & $9.58\times10^0$ & $8.81\times10^0$ & $1.04\times10^1$ \\
CdTe & 6.51$\times 10^0$ & 6.95$\times 10^0$ && $5.03\times10^0$ & $5.67\times10^0$ & $5.01\times10^0$ & $5.87\times10^0$ \\
CuCl & 1.48$\times 10^0$ & 1.11$\times 10^0$ && $1.01\times10^0$ & $1.75\times10^0$ & $1.04\times10^0$ & $2.02\times10^0$ \\
CuBr & 2.55$\times 10^0$ & 2.79$\times 10^0$ && $1.77\times10^0$ & $2.88\times10^0$ & $1.75\times10^0$ & $3.04\times10^0$ \\
CuI  & 6.53$\times 10^0$ & 7.22$\times 10^0$ && $5.89\times10^0$ & $6.94\times10^0$ & $6.07\times10^0$ & $7.32\times10^0$ \\
CuH  & 1.44$\times 10^1$ & 1.50$\times 10^1$ && $8.17\times10^0$ & $6.47\times10^0$ & $8.62\times10^0$ & $8.12\times10^0$ \\
AgI  & 2.02$\times 10^0$ & 2.47$\times 10^0$ && $1.39\times10^0$ & $1.78\times10^0$ & $1.38\times10^0$ & $1.90\times10^0$ \\
SiC  & 4.15$\times 10^2$ & 4.42$\times 10^2$ && $4.05\times10^2$ & $3.44\times10^2$ & $4.20\times10^2$ & $3.48\times10^2$ \\    
  \end{tabular}
 \end{ruledtabular}
\end{table*}

\begin{table*}[ht]
 \caption{\label{table:kappa-isotope} Calculated lattice thermal conductivities of the 33
 zincblende- and wurtzite-type compounds with isotope effect at 300 K in
 W/m-K. Experimental values at room temperature are also presented.}
 \begin{ruledtabular}
  \begin{tabular}{lccccccccc}
   & \multicolumn{3}{c}{Zincblende-type} && \multicolumn{5}{c}{Wurtzite-type} \\
   \cline{2-4} \cline{6-10}
   & \multicolumn{1}{c}{RTA} & \multicolumn{1}{c}{LBTE} & Exp. && \multicolumn{2}{c}{RTA} & \multicolumn{2}{c}{LBTE} & Exp. \\
   \hline
   & $xx$ & $xx$ & && $xx$ & $zz$ & $xx$ & $zz$ & \\
   \hline
BN   & 6.48$\times 10^2$ & 7.26$\times 10^2$ & 7.6$\times 10^2$ \footnotemark[1] && $5.61\times10^2$    & $5.42\times10^2$ & $6.02\times10^2$ & $5.73\times10^2$ & -                \\
BP   & 3.96$\times 10^2$ & 4.20$\times 10^2$ & 3.5$\times 10^2$ \footnotemark[1] && $3.89\times10^2$    & $3.01\times10^2$ & $3.95\times10^2$ & $3.07\times10^2$ & -                \\
BAs  & 1.23$\times 10^3$ & 2.94$\times 10^3$ & -                && $1.15\times10^3$    & $9.07\times10^2$ & $1.73\times10^3$ & $9.94\times10^2$ & -                \\
AlN  & 2.11$\times 10^2$ & 2.35$\times 10^2$ & -                && $2.24\times10^2$    & $2.07\times10^2$ & $2.39\times10^2$ & $2.12\times10^2$ & 3.5$\times 10^2$ \footnotemark[1] \\
AlP  & 8.30$\times 10^1$ & 8.40$\times 10^1$ & 9$\times 10^1$    \footnotemark[2]&& $7.48\times10^1$    & $7.14\times10^1$ & $7.46\times10^1$ & $7.17\times10^1$ & -                \\
AlAs & 8.56$\times 10^1$ & 8.68$\times 10^1$ & 9.8$\times 10^1$ \footnotemark[1] && $7.23\times10^1$    & $6.96\times10^1$ & $7.39\times10^1$ & $7.10\times10^1$ & -                \\
AlSb & 7.51$\times 10^1$ & 7.74$\times 10^1$ & 5.6$\times 10^1$ \footnotemark[1] && $5.71\times10^1$    & $5.44\times10^1$ & $5.82\times10^1$ & $5.40\times10^1$ & -                \\
GaN  & 1.54$\times 10^2$ & 1.81$\times 10^2$ & -                && $1.65\times10^2$    & $1.66\times10^2$ & $1.71\times10^2$ & $1.72\times10^2$ & 2.1$\times 10^2$ \footnotemark[1] \\
GaP  & 8.55$\times 10^1$ & 1.04$\times 10^2$ & 1.00$\times 10^2$ \footnotemark[1]&& $9.19\times10^1$    & $8.16\times10^1$ & $9.65\times10^1$ & $8.54\times10^1$ & -                \\
GaAs & 3.10$\times 10^1$ & 3.21$\times 10^1$ & 4.5$\times 10^1$ \footnotemark[1] && $2.78\times10^1$    & $2.55\times10^1$ & $2.78\times10^1$ & $2.59\times10^1$ & -                \\
GaSb & 3.18$\times 10^1$ & 3.25$\times 10^1$ & 4$\times 10^1$   \footnotemark[1] && $2.34\times10^1$    & $1.92\times10^1$ & $2.39\times10^1$ & $1.96\times10^1$ & -                \\
InN  & 9.31$\times 10^1$ & 1.05$\times 10^2$ & -                && $1.06\times10^2$    & $1.10\times10^2$ & $1.14\times10^2$ & $1.16\times10^2$ & -                \\
InP  & 7.38$\times 10^1$ & 8.52$\times 10^1$ & 9.3$\times 10^1$ \footnotemark[1] && $6.78\times10^1$    & $6.49\times10^1$ & $6.93\times10^1$ & $6.82\times10^1$ & -                \\
InAs & 2.04$\times 10^1$ & 2.52$\times 10^1$ & 3$\times 10^1$   \footnotemark[1] && $1.83\times10^1$    & $1.70\times10^1$ & $1.85\times10^1$ & $1.80\times10^1$ & -                \\
InSb & 1.26$\times 10^1$ & 1.38$\times 10^1$ & 2$\times 10^1$   \footnotemark[1] && $9.64\times10^0$    & $9.56\times10^0$ & $9.76\times10^0$ & $9.86\times10^0$ & -                \\
BeO  & 3.50$\times 10^2$ & 4.47$\times 10^2$ & -                && $2.80\times10^2$    & $2.93\times10^2$ & $3.27\times10^2$ & $3.14\times10^2$ & 3.7$\times 10^2$ \footnotemark[1] \\
BeS  & 1.43$\times 10^2$ & 1.57$\times 10^2$ & -                && $1.19\times10^2$    & $1.12\times10^2$ & $1.26\times10^2$ & $1.18\times10^2$ & -                \\
BeSe & 9.50$\times 10^1$ & 1.08$\times 10^2$ & -                && $7.57\times10^1$    & $8.02\times10^1$ & $8.19\times10^1$ & $8.31\times10^1$ & -                \\
BeTe & 7.75$\times 10^1$ & 9.04$\times 10^1$ & -                && $6.41\times10^1$    & $6.07\times10^1$ & $7.10\times10^1$ & $6.29\times10^1$ & -                \\
MgTe & 1.48$\times 10^1$ & 1.47$\times 10^1$ & -                && $1.14\times10^1$    & $1.27\times10^1$ & $1.16\times10^1$ & $1.27\times10^1$ & -                \\
ZnO  & 5.04$\times 10^1$ & 6.45$\times 10^1$ & -                && $4.10\times10^1$    & $5.56\times10^1$ & $4.13\times10^1$ & $5.73\times10^1$ & 6$\times 10^1$   \footnotemark[1] \\
ZnS  & 3.10$\times 10^1$ & 3.65$\times 10^1$ & 2.7$\times 10^1$ \footnotemark[1] && $3.15\times10^1$    & $3.40\times10^1$ & $3.24\times10^1$ & $3.83\times10^1$ & -                \\
ZnSe & 1.49$\times 10^1$ & 1.56$\times 10^1$ & 1.9$\times 10^1$ \footnotemark[1] && $1.37\times10^1$    & $1.42\times10^1$ & $1.38\times10^1$ & $1.45\times10^1$ & -                \\
ZnTe & 1.84$\times 10^1$ & 1.89$\times 10^1$ & 1.8$\times 10^1$ \footnotemark[1] && $1.38\times10^1$    & $1.48\times10^1$ & $1.38\times10^1$ & $1.51\times10^1$ & -                \\
CdS  & 1.90$\times 10^1$ & 1.99$\times 10^1$ & -                && $1.61\times10^1$    & $1.83\times10^1$ & $1.66\times10^1$ & $1.85\times10^1$ & 1.6$\times 10^1$ \footnotemark[1] \\
CdSe & 1.03$\times 10^1$ & 1.14$\times 10^1$ & -                && $8.42\times10^0$    & $9.16\times10^0$ & $8.37\times10^0$ & $9.87\times10^0$ & -                \\
CdTe & 6.28$\times 10^0$ & 6.67$\times 10^0$ & 7.5$\times 10^0$ \footnotemark[1] && $4.90\times10^0$    & $5.52\times10^0$ & $4.87\times10^0$ & $5.71\times10^0$ & -                \\
CuCl & 1.44$\times 10^0$ & 1.26$\times 10^0$ & -                && $9.95\times10^{-1}$ & $1.71\times10^0$ & $1.02\times10^0$ & $1.97\times10^0$ & 8.4$\times 10^{-1}$ \footnotemark[3] \\
CuBr & 2.52$\times 10^0$ & 2.75$\times 10^0$ & -                && $1.76\times10^0$    & $2.83\times10^0$ & $1.74\times10^0$ & $2.99\times10^0$ & 1.25$\times 10^0$ \footnotemark[3] \\
CuI  & 6.45$\times 10^0$ & 7.10$\times 10^0$ & -                && $5.82\times10^0$    & $6.85\times10^0$ & $5.99\times10^0$ & $7.22\times10^0$ & 1.68$\times 10^0$ \footnotemark[3] \\
CuH  & 1.39$\times 10^1$ & 1.44$\times 10^1$ & -                && $7.81\times10^0$    & $6.16\times10^0$ & $8.22\times10^0$ & $7.60\times10^0$ & -                \\
AgI  & 2.01$\times 10^0$ & 2.44$\times 10^0$ & -                && $1.38\times10^0$    & $1.78\times10^0$ & $1.37\times10^0$ & $1.89\times10^0$ & 1.03$\times 10^0$ \footnotemark[4] \\
SiC  & 3.55$\times 10^2$ & 3.72$\times 10^2$ & -                && $3.55\times10^2$    & $3.11\times10^2$ & $3.67\times10^2$ & $3.14\times10^2$ & 4.9$\times 10^2$ \footnotemark[1] \\
   \footnotetext[1]{Ref.~\onlinecite{High-thermal-conductivity-materials}.}
   \footnotetext[2]{Ref.~\onlinecite{Steigmeier-kappas-1966}.}
   \footnotetext[3]{Ref.~\onlinecite{CRC10}.}
   \footnotetext[4]{Ref.~\onlinecite{Goetz-AgI-1982}.}
  \end{tabular}
 \end{ruledtabular}
\end{table*}

\clearpage
\twocolumngrid

\bibliography{zb-wz-kappa}
\end{document}